\documentclass[aps,prl,onecolumn,superscriptaddress,floatfix,showpacs,showkeys]{revtex4}
 \usepackage{graphicx}
\usepackage{afterpage} \usepackage{dcolumn} \usepackage{bm}
\usepackage{amsmath}
\usepackage{amssymb}
\usepackage{epsfig}
\usepackage{array}

\begin{document}
\title{Optical probes of electron correlations in solids.}
\author{E.\ van Heumen and D. van der Marel}
\affiliation{D\'epartement de Physique de la Mati\`ere
Condens\'ee, Universit\'e de Gen\`eve, quai Ernest-Ansermet 24,
CH1211 , Gen\`eve 4, Switzerland}

\date{\today }

\begin{abstract}
Classically the interaction between light and matter is given by
the Maxwell relations. These are briefly reviewed and will be used
as a basis to discuss several techniques that are used in optical
spectroscopy. We then discuss the quantum mechanical description
of the optical conductivity based on the Kubo formalism. This is
used as a basis to understand how strong correlation effects can
be observed using optical techniques. We will discuss the use of
sum rules in the interpretation of optical experiments. Finally,
we describe the effect of including interactions between
electronic and collective degrees of freedom on optical spectra.
\end{abstract}



\maketitle

\section{Introduction}
We will discuss the physics of correlated electron systems from an
experimental viewpoint, focussing on optical spectroscopy. The
interaction of light and matter will be discussed first from a
classical point of view, based on the Maxwell equations. This
review will be the basis for a discussion of optical techniques
that are most commonly used. We will then continue with a
discussion of the quantum mechanical description of the
interaction between light and matter, using the Kubo-formalism. We
finally discuss the application of sum rules to correlated systems
and what happens when interactions, like the electron-phonon
interaction, become important. The first part of our review is not
meant to be complete. Readers with interest for further details
are referred to references \citep{Wooten} and \citep{Dressel}. In
the following all fields, currents, charge densities etc. are
implied to be position and time dependent if not written
explicitly. Bold quantities imply vectors or matrices.
\newpage
\section{Electromagnetism and Matter}
\subsection{Maxwell's equations}
We start this review with the microscopic Maxwell equations,
\begin{eqnarray}
\nabla\cdot \mathbf{e} &=& 4\pi\rho_{micro},\label{micromax1} \\
\nabla\times \mathbf{e} &=& -\frac{1}{c}\frac{\partial}{\partial t}\mathbf{b},\label{micromax2} \\
\nabla\cdot \mathbf{b} &=& 0, \label{micromax3}\\
\nabla\times \mathbf{b} &=& \frac{1}{c}\frac{\partial}{\partial
t}\mathbf{e}+\frac{4\pi}{c}\mathbf{j}_{micro}.\label{micromax4}
\end{eqnarray}
Here \textbf{e} and \textbf{b} are the microscopic electric and
magnetic fields respectively. $\rho_{micro}$ is the total
microscopic charge distribution and $j_{micro}$ the total
microscopic current distribution (i.e. due to internal and
external sources). Note that these equations are written in the
C.G.S. system of units. To convert them to S.I. units, simply
replace $4\pi$ by $1/\varepsilon_{0}$. The charge distribution for
a collection of point sources with charge $q_{i}$ can be written
classically,
\begin{equation}
\rho_{micro}=\sum_{i}q_{i}\delta(\mathbf{r}-\mathbf{r_{i}}),
\end{equation}
or quantum mechanically as,
\begin{equation}
\rho_{micro}=-e\Psi^{*}(\mathbf{r})\Psi(\mathbf{r}).
\end{equation}
Equations (\ref{micromax1}-\ref{micromax4}) are not very practical
to work with. As a first step we rewrite them in a more familiar
form. To do this we average the fields, charge and current
distributions over a volume $\Delta V$,
\begin{eqnarray}
\rho_{total}(\mathbf{r})=\frac{1}{\Delta V}\int_{\Delta
V}\rho_{micro}(\mathbf{r}+\mathbf{r'})d^{3}\mathbf{r'},\\
\mathbf{J}_{total}(\mathbf{r})=\frac{1}{\Delta V}\int_{\Delta
V}\mathbf{J}_{micro}(\mathbf{r}+\mathbf{r'})d^{3}\mathbf{r'},
\end{eqnarray}
and similarly for $\mathbf{e}$ and $\mathbf{b}$. This is a
sensible procedure under the condition that $a_{0}\ll\Delta
V\ll(2\pi c/\omega)^{3}$ where $a_{0}$ is the Bohr radius. Using
these averaged distributions we arrive at the standard Maxwell
equations in free space,
\begin{eqnarray}
\nabla\cdot \mathbf{E}_{total}(\mathbf{r},t) &=& 4\pi\rho_{total}(\mathbf{r},t),\label{max1} \\
\nabla\times \mathbf{E}_{total}(\mathbf{r},t) &=& -\frac{1}{c}\frac{\partial}{\partial t}\mathbf{B}_{total}(\mathbf{r},t),\label{max2} \\
\nabla\cdot \mathbf{B}_{total}(\mathbf{r},t) &=& 0, \label{max3}\\
\nabla\times \mathbf{B}_{total}(\mathbf{r},t) &=&
\frac{1}{c}\frac{\partial}{\partial
t}\mathbf{E}_{total}(\mathbf{r},t)+\frac{4\pi}{c}\mathbf{J}_{total}(\mathbf{r},t).\label{max4}
\end{eqnarray}
In order to see how matter interacts with propagating
electromagnetic waves we have to distinguish between induced and
external sources. We write $\mathbf{J}_{total}\equiv
\mathbf{J}_{ext}+\mathbf{J}_{ind}$ and $\rho_{total}\equiv
\rho_{ext}+\rho_{ind}$. Both the induced and external charge and
current distributions have to obey the continuity equations
separately,
\begin{equation}
\nabla\cdot \mathbf{J}_{ind/ext}+\frac{\partial}{\partial
t}\rho_{ind/ext}=0.
\end{equation}
We can distinguish three different types of macroscopic internal
sources,
\begin{equation}\label{Jind}
\mathbf{J}_{ind}=\mathbf{J}_{cond}+\frac{\partial
\mathbf{P}}{\partial t}+c\nabla\times \mathbf{M}.
\end{equation}
The first term on the right hand side, $\mathbf{J}_{cond}$,
corresponds to the response of the free charges. The second term
is the current due to changes in the polarization state of the
system. Finally, we include a term representing a current due to
magnetization. Note that this last term is purely transversal (the
divergence of a rotation is always zero) and so is easy to
distinguish from the other two terms. Since the induced free
charge current due to photons is necessarily transversal,
$\nabla\cdot \mathbf{J}_{cond}=0$, we can use the continuity
equations to show that the induced free charge density has to be
zero  and as a consequence that the total induced charge density
\begin{equation}\label{rho_ind}
\rho_{ind}=-\nabla\cdot P.
\end{equation}
It is convenient to introduce new fields
\begin{eqnarray}
\mathbf{D}(\mathbf{r},t)\equiv \mathbf{E}_{ext}(\mathbf{r},t)\equiv\mathbf{E}(\mathbf{r},t)+4\pi\mathbf{P}(\mathbf{r},t),\label{Dfield}\\
\mathbf{H}(\mathbf{r},t)\equiv\mathbf{B}(\mathbf{r},t)-4\pi\mathbf{M}(\mathbf{r},t),\label{Hfield}
\end{eqnarray}
so that using equations (\ref{Jind}-\ref{Hfield}) in equations
(\ref{max1}) and (\ref{max4}) we find,
\begin{eqnarray}
\nabla\cdot \mathbf{D}(\mathbf{r},t) &=& 4\pi\rho_{ext}(\mathbf{r},t),\label{mattermax1} \\
\nabla\times \mathbf{H}(\mathbf{r},t) &=&
\frac{1}{c}\frac{\partial}{\partial
t}\mathbf{D}(\mathbf{r},t)+\frac{4\pi}{c}\mathbf{J}_{ext}(\mathbf{r},t)+\frac{4\pi}{c}\mathbf{J}_{cond}(\mathbf{r},t).\label{mattermax4}
\end{eqnarray}

\subsection{Linear Response Theory}
In the spirit of linear response theory we assume that the
response of polarization, magnetization or current are linear in
the applied fields:
\begin{eqnarray}
\mathbf{P}=\chi_{e}\mathbf{E},\\
\mathbf{M}=\chi_{m}\mathbf{H},\label{Mag}\\
\mathbf{J}=\sigma\mathbf{E}.
\end{eqnarray}
The electric and magnetic susceptibilities can be expressed in
terms of a dielectric function
$\mathbf{\varepsilon'}=1+4\pi\chi_{e}$ and magnetic permittivity
$\mathbf{\mu'}=1+4\pi\chi_{m}$. The dielectric function is a
response function that connects the external field
$\textbf{E}_{ext}$ at position \textbf{r} and time $t$ with the
field $\textbf{E}$ at all other times and positions. So in
general,
\begin{equation}
\mathbf{E}_{ext}(\mathbf{r},t)=\int_{-\infty}^{t}\int\varepsilon'(\mathbf{r},\mathbf{r'},t,t')\mathbf{E}(\mathbf{r'},t')d^{3}r'dt'.
\end{equation}
We will be mainly interested in the Fourier transform of
$\varepsilon'\equiv\varepsilon(\textbf{q},\omega)$ however. It is
an easy exercise to express the Maxwell equations in terms of
$\textbf{q}$ and $\omega$ which we leave to the reader. We can use
these definitions to once again rewrite the Maxwell equations in
the following form,
\begin{eqnarray}
\nabla\cdot(\varepsilon'\mathbf{E}) &=& 4\pi\rho_{ext},\label{linresmax1} \\
\nabla\times \mathbf{E} &=& -\frac{1}{c}\frac{\partial}{\partial t}(\mu')\mathbf{H},\label{linresmax2} \\
\nabla\cdot\mu'\mathbf{H} &=& 0, \label{linresmax3}\\
\nabla\times \mathbf{H} &=& \frac{1}{c}\frac{\partial}{\partial
t}(\varepsilon'\mathbf{E})+\frac{4\pi\sigma}{c}\mathbf{E}.\label{linresmax4}
\end{eqnarray}
We are now in a position to study the response of a solid to an
externally applied field or light wave. For simplicity we assume
that our solid is homogeneous so that $\nabla\varepsilon'=0$ and
$\nabla\mu'=0$. We can describe light waves by plane waves, i.e.
\begin{eqnarray}
\mathbf{E}(\mathbf{r},t)=\mathbf{E_{0}}e^{i(\mathbf{q\centerdot
r}-\omega t)},\label{E}\\
\mathbf{B}(\mathbf{r},t)=\mathbf{B_{0}}e^{i(\mathbf{q\centerdot
r}-\omega t)}\label{B}.
\end{eqnarray}
Using (\ref{B}) on the right hand side of Faraday's equation
(\ref{max2}) and rearranging we find,
\begin{equation}
\mathbf{B}=\frac{c}{i\omega}\nabla\times\mathbf{E}.
\end{equation}
If we now take the curl of this equation and use the fact that we
can express $\textbf{M}$ in terms of $\textbf{B}$ as (see
equations (\ref{Hfield}) and (\ref{Mag})),
\begin{equation}
\mathbf{M}=\frac{\mu'^{-1}-1}{4\pi}\mathbf{B}.
\end{equation}
we find that
\begin{eqnarray}
\nabla\times\mathbf{M}&=&\frac{\mu'^{-1}-1}{4\pi}\nabla\times\mathbf{B}=\frac{c}{i\omega}\frac{\mu'^{-1}-1}{4\pi}\nabla\times\nabla\times\mathbf{E}\nonumber\\
&=&\frac{c}{i\omega}\frac{\mu'^{-1}-1}{4\pi}(\nabla^{2}\mathbf{E}-\nabla(\nabla\cdot\mathbf{E})=\frac{cq^{2}}{i\omega}\frac{\mu'^{-1}-1}{4\pi}\mathbf{E}^{T}.
\end{eqnarray}
Note that in this equation we are left with only the transversal
field since the curl of a curl is transverse. We give two further
identities for completeness,
\begin{equation}
\frac{\partial\mathbf{P}}{\partial
t}=-i\omega\frac{1-\varepsilon'}{4\pi}\mathbf{E},
\end{equation}
and,
\begin{equation}
J_{cond}=\sigma\mathbf{E}.
\end{equation}
Finally, we note that inside the solid $\rho_{ext}=J_{ext}=0$.
With this we have all the ingredients to express equation
(\ref{mattermax4}) in terms of $\textbf{E}$ and $\textbf{J}$. We
split this equation in transversal and longitudinal parts to find,
\begin{eqnarray}
\frac{\mathbf{J}_{ind}^{T}(\mathbf{q},\omega)}{\mathbf{E}^{T}(\mathbf{q},\omega)}&\equiv&\frac{i\omega}{4\pi}\{1-\varepsilon'(\mathbf{q},\omega)-
\frac{i4\pi}{\omega}\sigma(\mathbf{q},\omega)-\frac{c^{2}q^{2}}{\omega^{2}}(1-\frac{1}{\mu
(\mathbf{q},\omega)})\},\\
\frac{\mathbf{J}_{ind}^{L}(\mathbf{q},\omega)}{\mathbf{E}^{L}(\mathbf{q},\omega)}&\equiv&\frac{i\omega}{4\pi}\{1-\varepsilon'(\mathbf{q},\omega)-
\frac{i4\pi}{\omega}\sigma(\mathbf{q},\omega)\}.
\end{eqnarray}
We can define a new dielectric function with longitudinal and
transverse components and write the previous equation in a more
compact form,
\begin{equation}
\frac{\mathbf{J}_{ind}^{L,T}(\mathbf{q},\omega)}{\mathbf{E}^{L,T}(\mathbf{q},\omega)}\equiv\frac{i\omega}{4\pi}\{1-\varepsilon^{L,T}(\mathbf{q},\omega)\}.
\end{equation}
This new dielectric function $\varepsilon$ is now a complex
quantity:
$\varepsilon\equiv\varepsilon'+i\varepsilon''=\varepsilon'+i4\pi\sigma/\omega$.
Using the last relation we can also define a complex conductivity
$\hat{\sigma}\equiv\sigma'+i\sigma''$ and it is related to the
dielectric function by,
\begin{equation}
\hat{\sigma}=\frac{i\omega}{4\pi}(1-\varepsilon).
\end{equation}
The real part of $\varepsilon$ is often called the reactive part
and the imaginary part the dissipative part. The real and
imaginary parts are also indicated with a subscript 1 and 2
instead of (') and ('').

\subsection{Kramers-Kronig relations}\label{KK}
\begin{figure}[tbh]
\includegraphics[width=8.5 cm]{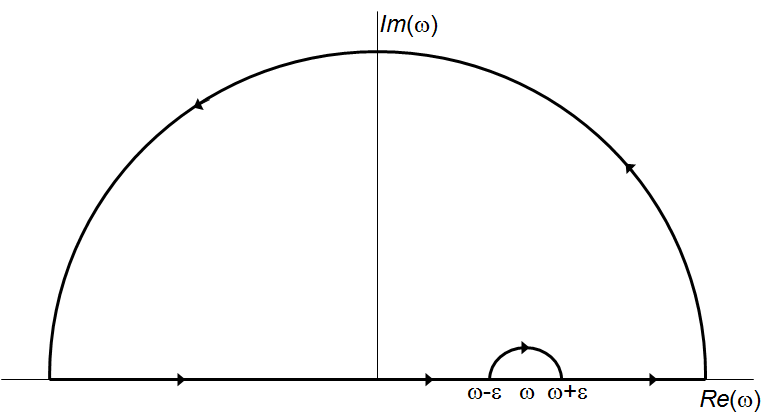}
\caption{\label{contour}Contour used to derive the KK-relations.}
\end{figure}

A fundamental principle in physics is the principle of causality:
an effect cannot precede its cause. This principle provides us
with a very useful relation between the real and imaginary parts
of a response function like the optical conductivity as we now
show. First we express the induced current due to an electric
field in terms of a memory function,
\begin{equation}\label{memfunc}
j(t)=\int^{t}_{-\infty}M(t-t')E(t')dt'.
\end{equation}
The memory function has the property that $M(\tau<0)=0$. This is
simply a restatement of the causality principle: we switch on a
driving force at time $\tau=0$ so before that time there can be no
current. We define the Fourier transform of $M$ in equation
(\ref{memfunc}) as,
\begin{equation}\label{FTmem}
\hat{\sigma}(\omega)=\int_{0}^{\infty}M(\tau)e^{i\omega\tau}d\tau.
\end{equation}
To do the integral we change to the complex frequency plane,
$\omega\to z=\omega_{1}+i\omega_{2}$. The exponential in Eq.
(\ref{FTmem}) now splits in a complex and real part,
\begin{equation}\label{FTmem2}
\hat{\sigma}(\omega)=\int_{0}^{\infty}M(\tau)e^{i\omega_{1}\tau}e^{-\omega_{2}\tau}d\tau.
\end{equation}
The second exponent in this integral is bounded in the upper half
plane for $\tau>0$ and in the lower half plane for $\tau<0$, so
that we can evaluate the integral in the upper half plane since
$M(\tau<0)=0$. We use the contour shown in figure \ref{contour}.
Since all poles occur on the real axis, the complete contour is
zero,
\begin{equation}
\oint dz \frac{\hat{\sigma}(z)}{z-\omega}=0.
\end{equation}
The integral along the large semi circle is also zero. So we are
left with,
\begin{equation}
\int_{-\infty}^{\omega-\varepsilon} dz
\frac{\hat{\sigma}(z)}{z-\omega} +
\int^{\infty}_{\omega+\varepsilon} dz
\frac{\hat{\sigma}(z)}{z-\omega} + \int_{\pi}^{0}d(\omega+\epsilon
e^{i\phi})\frac{\hat{\sigma}(\omega+\epsilon e^{i\phi})}{\epsilon
e^{i\phi}}=0.
\end{equation}
The first two integrals give the principle value of the integral
for $\epsilon\to 0$,
\begin{equation}
\mathcal{P}\int_{-\infty}^{\infty} d\omega'
\frac{\hat{\sigma}(\omega')}{\omega'-\omega}-\pi
i\hat{\sigma}(\omega')=0.
\end{equation}
From which the Kramers-Kronig relations follow,
\begin{eqnarray}
\sigma_{1}(\omega)&=&\frac{1}{\pi}\mathcal{P}\int_{-\infty}^{\infty}\frac{\sigma_{2}(\omega')}{\omega'-\omega}d\omega',\label{KK1}\\
\sigma_{2}(\omega)&=&-\frac{1}{\pi}\mathcal{P}\int_{-\infty}^{\infty}\frac{\sigma_{1}(\omega')}{\omega'-\omega}d\omega'.\label{KK2}
\end{eqnarray}
Using $Im(M(\tau))=0$ we see that
$\hat{\sigma}(-\omega)=\hat{\sigma}^{*}(\omega)$, which implies
that $\sigma_{1}(-\omega)=\sigma_{1}(\omega)$ and
$\sigma_{2}(-\omega)=-\sigma_{2}(\omega)$. These relations can be
used to rewrite equations (\ref{KK1}) and (\ref{KK2}),
\begin{eqnarray}
\sigma_{1}(\omega)&=&\frac{2}{\pi}\mathcal{P}\int_{0}^{\infty}\frac{\omega'\sigma_{2}(\omega')}{\omega'^{2}-\omega^{2}}d\omega',\\
\sigma_{2}(\omega)&=&-\frac{2\omega}{\pi}\mathcal{P}\int_{0}^{\infty}\frac{\sigma_{1}(\omega')}{\omega'^{2}-\omega^{2}}d\omega'.
\end{eqnarray}
The relations (\ref{KK1}) and (\ref{KK2}) between the real and
imaginary parts of the optical conductivity are examples of the
general relations between real and imaginary parts of causal
response functions and they are referred to as Kramers-Kronig (KK)
relations.

\subsection{Polaritons}\label{polariton}
In this section we discuss the properties of electromagnetic waves
propagating through solids. Such a wave is called a polariton. A
polariton is a photon dressed up with the excitations that exist
inside solids. For example one can have phonon-polaritons which
are photons dressed up with phonons. Although the solutions of the
Maxwell equations, i.e. the fields $\textbf{E}$ and $\textbf{B}$,
have the same form as before (Eq. (\ref{E}) and (\ref{B})) they
obey different dispersion relations as we will now see. As before
we assume that $\nabla\varepsilon'=\nabla\mu'=0$ and that
$\rho_{ext}=J_{ext}=0$. Taking the curl of Eq. (\ref{linresmax2})
we obtain for the left-hand side,
\begin{equation}
\nabla\times\nabla\times\mathbf{E}=-\nabla^{2}\mathbf{E}^{T},
\end{equation}
where the T indicates that we are left with a purely transverse
field. We then use Eq. (\ref{linresmax4}) to work out the
right-hand side of Eq. (\ref{linresmax2}) and we obtain the wave
equation,
\begin{equation}
\nabla^{2}\mathbf{E}^{T}=\frac{\varepsilon'\mu}{c^{2}}\frac{\partial^{2}\mathbf{E}^{T}}{\partial
t^{2}}+\frac{4\pi\sigma\mu}{c^{2}}\frac{\partial\mathbf{E}^{T}}{\partial
t}.
\end{equation}
From this wave equation we easily obtain the dispersion relation
for polaritons travelling through a solid by substituting Eq.
(\ref{E}),
\begin{equation}\label{poldisp}
\mu(\mathbf{q},\omega)\{\varepsilon'(\mathbf{q},\omega)+i\frac{4\pi\sigma(\mathbf{q},\omega)}{\omega}\}\omega^{2}=\mu\varepsilon^{L}\omega^{2}=\mathbf{q}^{2}c^{2}.
\end{equation}
The dispersion relation for longitudinal waves can be found by
observing that for longitudinal waves $\nabla\times\textbf{E}=0$
and hence the dispersion relation is simply,
\begin{equation}
\mu(\mathbf{q},\omega)\{\varepsilon'(\mathbf{q},\omega)+i\frac{4\pi\sigma(\mathbf{q},\omega)}{\omega}\}=0.
\end{equation}
The polariton solutions to Eq. (\ref{poldisp}) are of the form
\begin{equation}
\mathbf{E}(\mathbf{r},t)=\mathbf{E_{0}}e^{i(\mathbf{q\centerdot
r}-\omega t)},
\end{equation}
with
\begin{equation}
|q|=\frac{\sqrt{\mu\varepsilon}\omega}{c}.
\end{equation}
We now define the refractive index,
\begin{equation}
\hat{n}(\mathbf{q},\omega)=n+ik\equiv\sqrt{\mu\varepsilon}.
\end{equation}
In all cases considered here $n>0$ and $k>0$. We also note that
$Im(\varepsilon)\geq 0$ but it is possible to have
$Re(\varepsilon)<0$. If $k>0$ the wave travelling through the
solid gets attenuated,
\begin{equation}
\mathbf{E}(\mathbf{r},t)=\mathbf{E_{0}}e^{i\omega(nr/c-t)-r/\delta}.
\end{equation}
The extinction of the wave occurs over a characteristic length
scale $\delta$ called the skin depth,
\begin{equation}\label{skin}
\delta=\frac{c}{\omega k}=\frac{c}{\omega
Im\sqrt{\mu\varepsilon_{1}+i4\pi\mu\sigma_{1}/\omega}}.
\end{equation}
\begin{figure}[hbt]
\begin{minipage}{8cm}
\includegraphics[width=8cm]{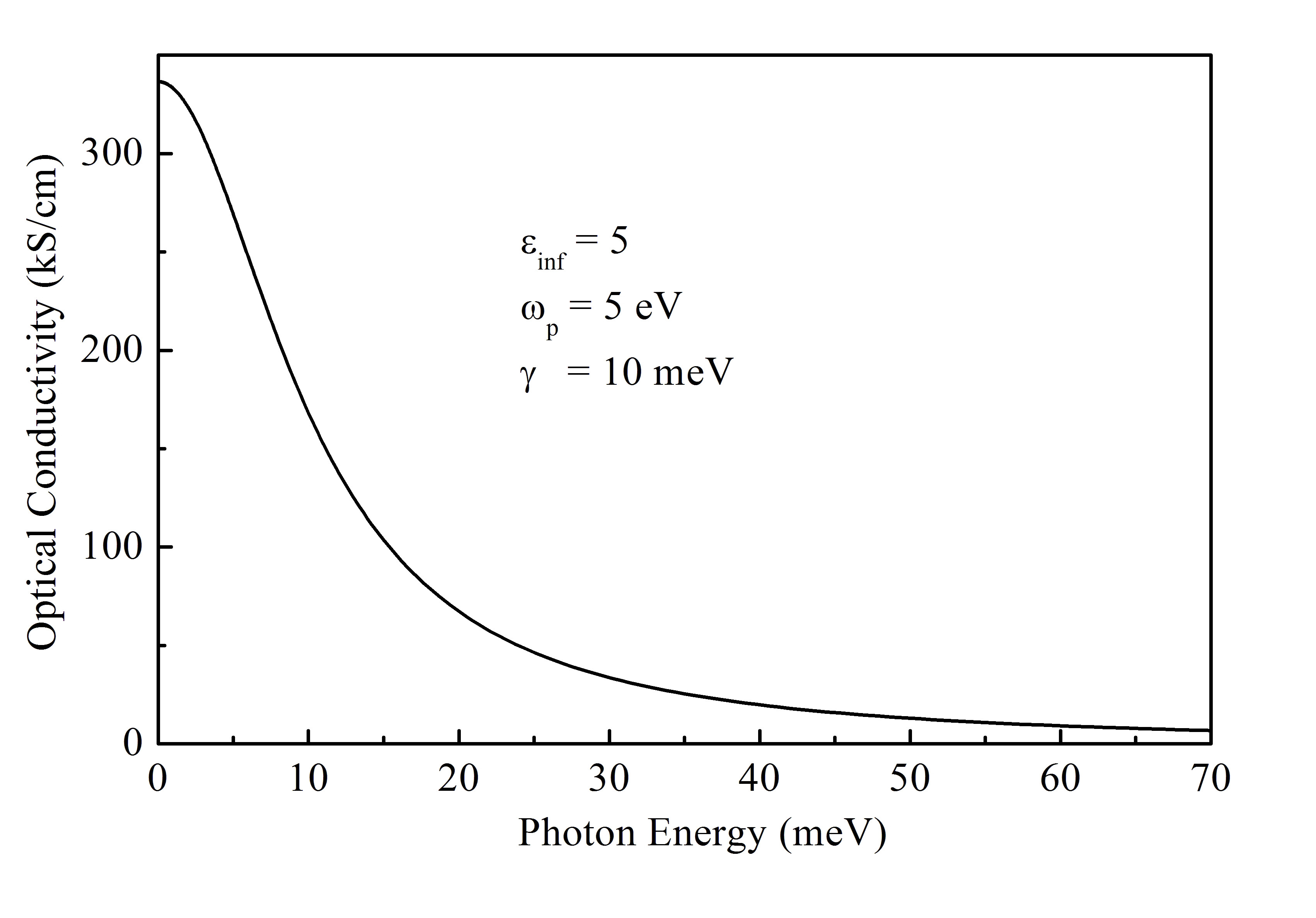}
\caption{\label{drudecond} Real part of the optical conductivity
for parameter values indicated in the graph. The curve is
calculated using equation (\ref{drude}).}
\end{minipage}\hspace{2pc}%
\begin{minipage}{8cm}
\includegraphics[width=8cm]{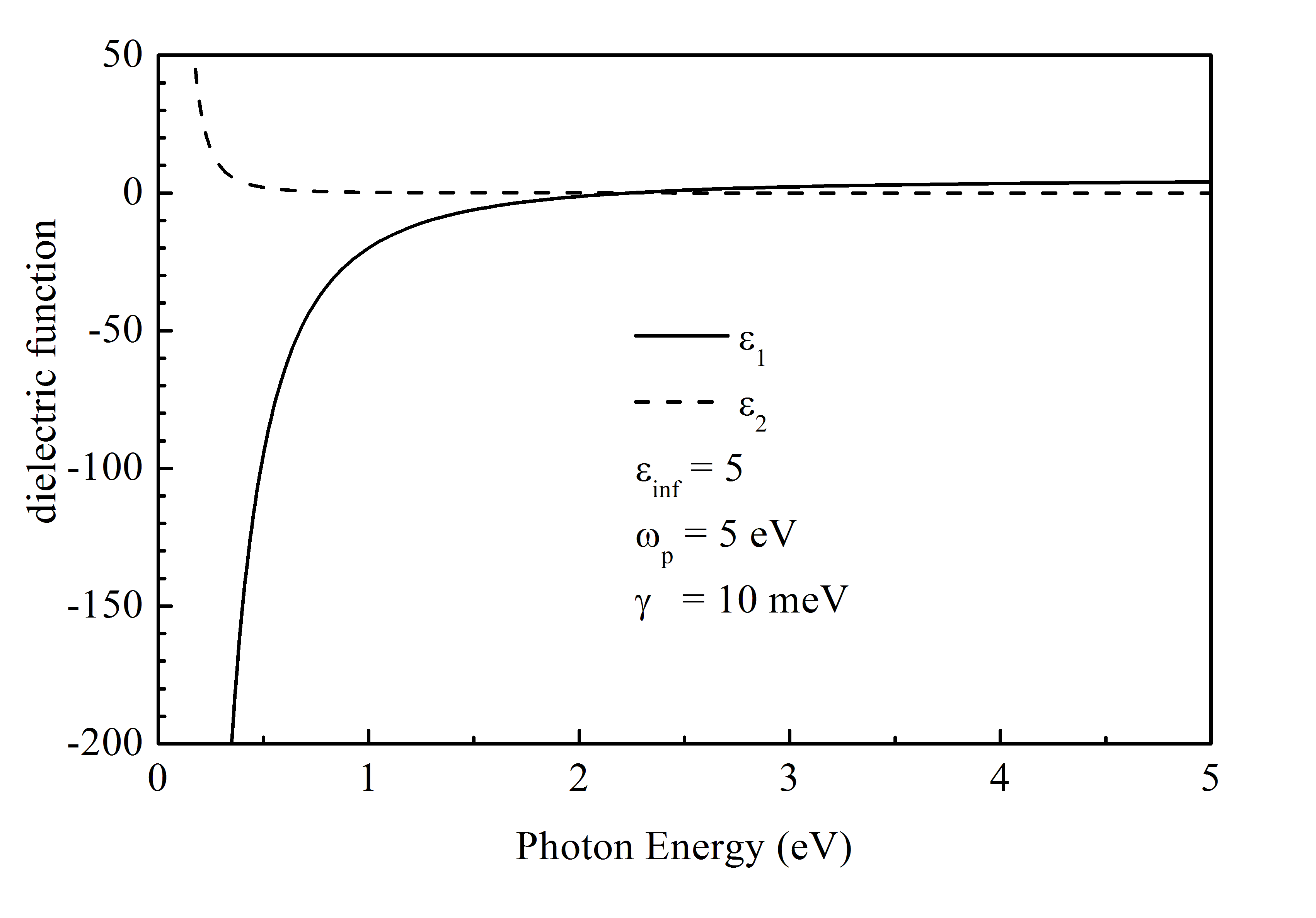}
\caption{\label{dielfunc} Dielectric function corresponding to
equation (\ref{drudediel}) with the same parameters as in figure
\ref{drudecond}.}
\end{minipage}
\end{figure}
Note that we can have $k>0$ if $Im(\varepsilon)=0$ and
$Re(\varepsilon)<0$ so that the wave gets attenuated even though
there is no absorption. In table \ref{skindepth} we indicate some
limits of the skin depth.
\begin{table}[tbh]
\begin{tabular}{lrlrl}
\hline\hline
Insulator &  & $\frac{4\pi\sigma_{1}}{\omega}\ll\varepsilon_{1}$ &
&
$\delta\approx\frac{c}{2\pi\sigma_{1}}\sqrt{\frac{\varepsilon_{1}}{\mu}}$\\
Metal & & $\frac{4\pi\sigma_{1}}{\omega}\gg\varepsilon_{1}$ &  &
$\delta\approx\frac{c}{\sqrt{2\pi\mu\sigma_{1}\omega}}$\\
Superconductor & &
$\frac{4\pi\sigma_{1}}{\omega}\ll\varepsilon_{1}=-\frac{c^{2}}{\lambda^{2}\omega^{2}}$
& & $\delta\approx\frac{\lambda}{\sqrt{\mu}}$\\
 \hline\hline
\end{tabular}
\caption{Some limiting cases of the general expression Eq.
(\ref{skin}). $\lambda$ in the last line is the London penetration
depth. }\label{skindepth}
\end{table}

To illustrate some of the previous results we now have a look at
the simplest model of a metal: the Drude-model. The optical
conductivity in the Drude model is,
\begin{equation}\label{drude}
\hat{\sigma}=\frac{ne^{2}}{m}\frac{1}{\tau^{-1}-i\omega}.
\end{equation}
Often $1/\tau$, the time in between scattering events, is
written as a scattering rate $\gamma$. The plasma frequency is
defined as $\omega^{2}_{p}\equiv 4\pi ne^{2}/2m$. The
dielectric function can now be written as,
\begin{equation}\label{drudediel}
\varepsilon(\omega)=1+4\pi\chi_{bound}-\frac{4\pi
ne^{2}}{m}\frac{1}{\omega(\gamma-i\omega)}=\varepsilon_{\infty}-\frac{4\pi
ne^{2}}{m}\frac{1}{\omega(\gamma-i\omega)},
\end{equation}
where for completeness we have included the contribution due to
the bound charges, represented by a high energy contribution
$\varepsilon_{\infty}$. Figure \ref{drudecond} shows the optical
conductivity given by equation (\ref{drude}) for parameter values
typical of a metal. Using the same parameters we can calculate the
dielectric function given by equation (\ref{drudediel}). The
results are shown in figure \ref{dielfunc}. We note that the real
part of the dielectric function is negative for
$\omega<\omega_{p}/\sqrt{\varepsilon_{\infty}}$ and positive for
$\omega>\omega_{p}/\sqrt{\varepsilon_{\infty}}$. The point where
it crosses zero is called the screened plasma frequency
$\omega^{*}_{p}$ (screened by interband transitions).

\begin{figure}[hbt]
\begin{minipage}{8cm}
\includegraphics[width=8cm]{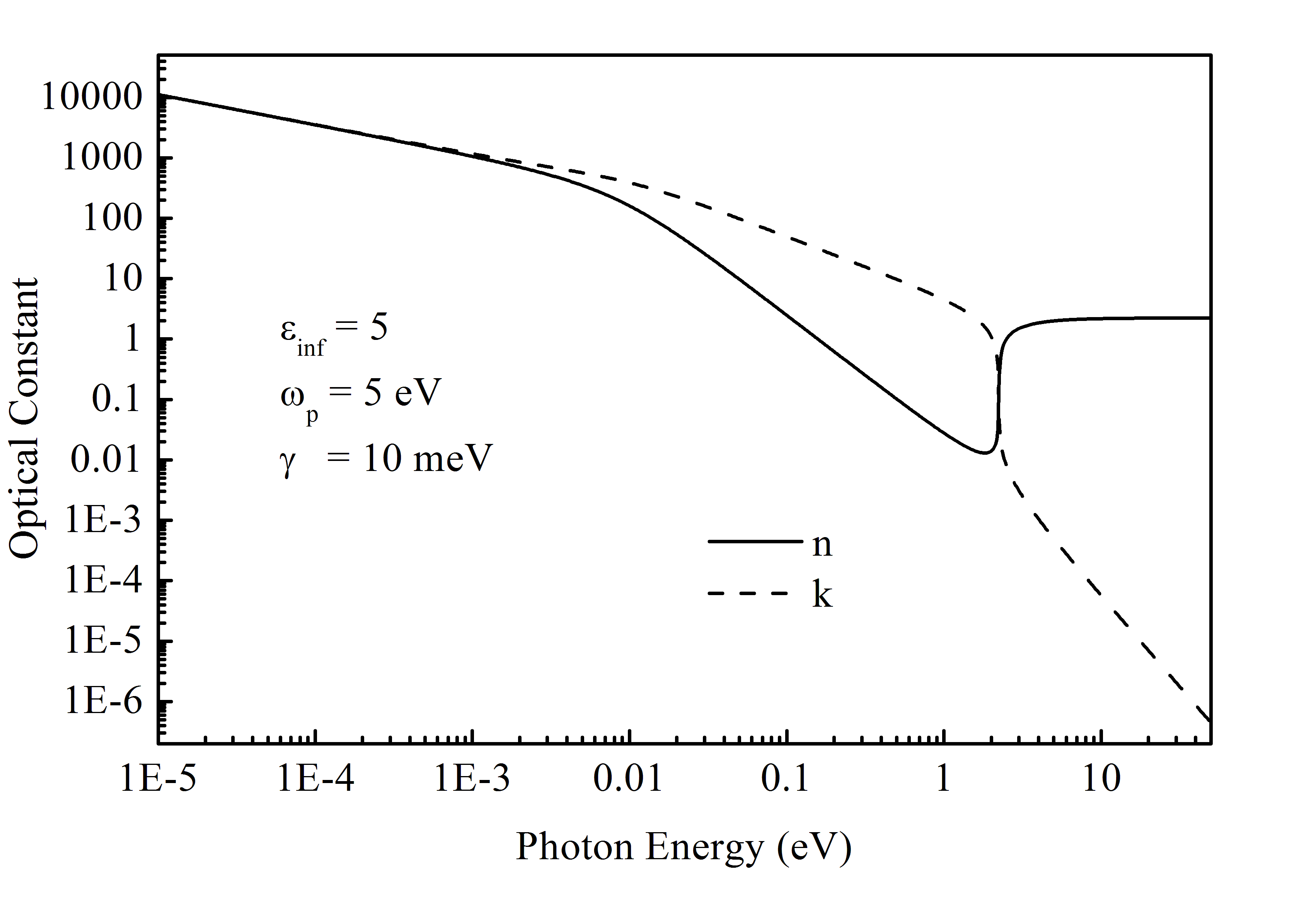}
\caption{\label{druden} Optical constants corresponding to
equation (\ref{drudediel}) with the same parameters as in figure
\ref{drudecond}.}
\end{minipage}\hspace{2pc}%
\begin{minipage}{8cm}
\includegraphics[width=8cm]{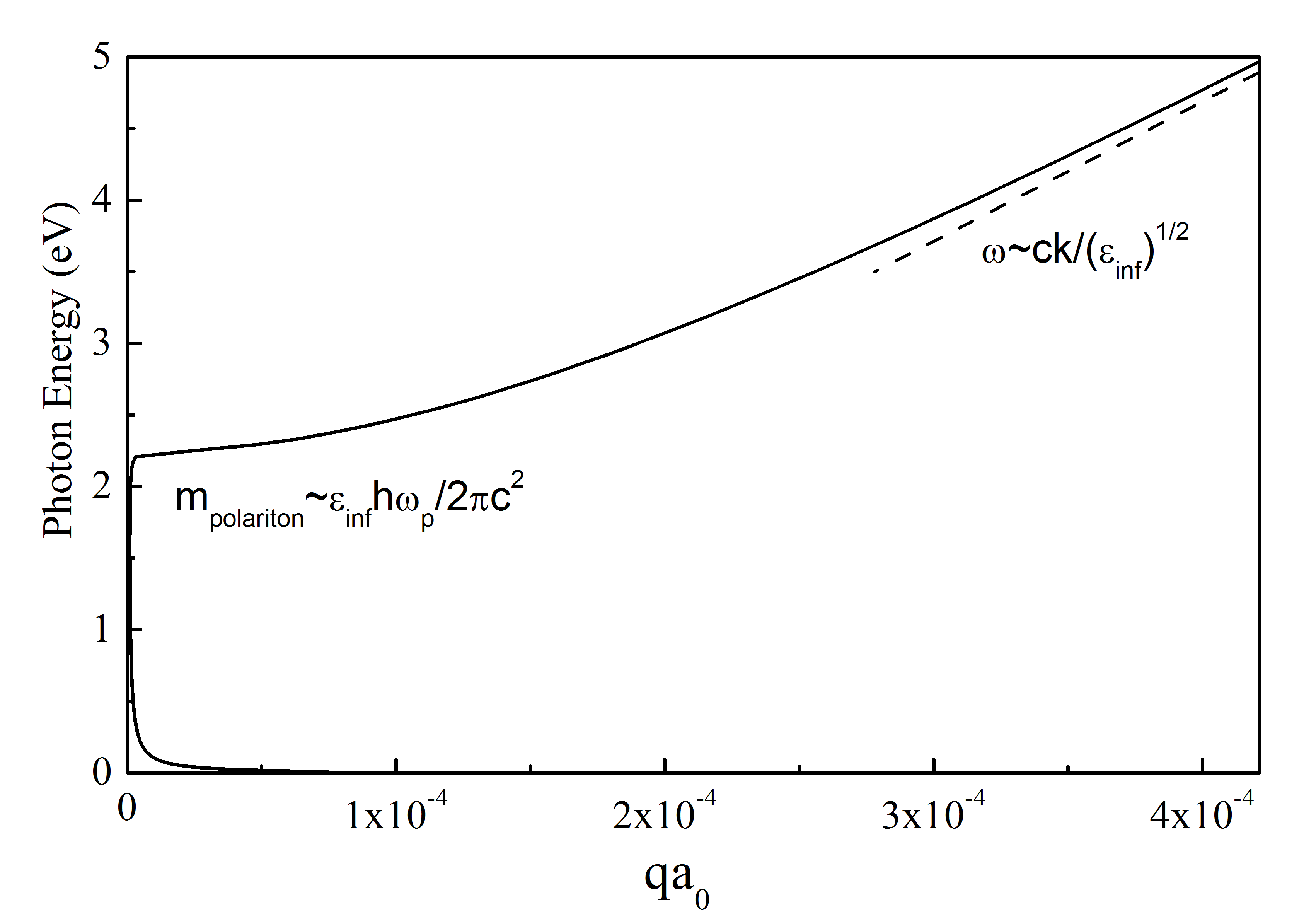}
\caption{\label{figpoldisp} Polariton dispersion calculated with
the same parameters as in figure \ref{drudecond}.}
\end{minipage}
\end{figure}

We can also easily calculate the optical constants,
\begin{equation}\label{Druden}
\hat n = \sqrt {\varepsilon _\infty   - \frac{{\omega _p^2 }}
{{\omega \left( {\omega  + i\tau ^{ - 1} } \right)}}}.
\end{equation}
The real and imaginary part are displayed in figure
\ref{druden}. We see that at the screened plasma frequency both
$n$ and $k$ show a discontinuity.

The polariton dispersion follows from equation (\ref{poldisp}).
Here we assume that $\mu$ = 1 and frequency independent and use
Eq. (\ref{drudediel}) to solve (\ref{poldisp}) for $\omega(q)$.
The polariton dispersion consists of two branches the lowest
one for 0 $\le$ $\omega$ $\le$ $1/\tau$ and one for $\omega$
$\ge$ $\omega_{p}/\sqrt{\varepsilon_{\infty}}$.

\begin{figure}[tbh]
\includegraphics[width=8.5 cm]{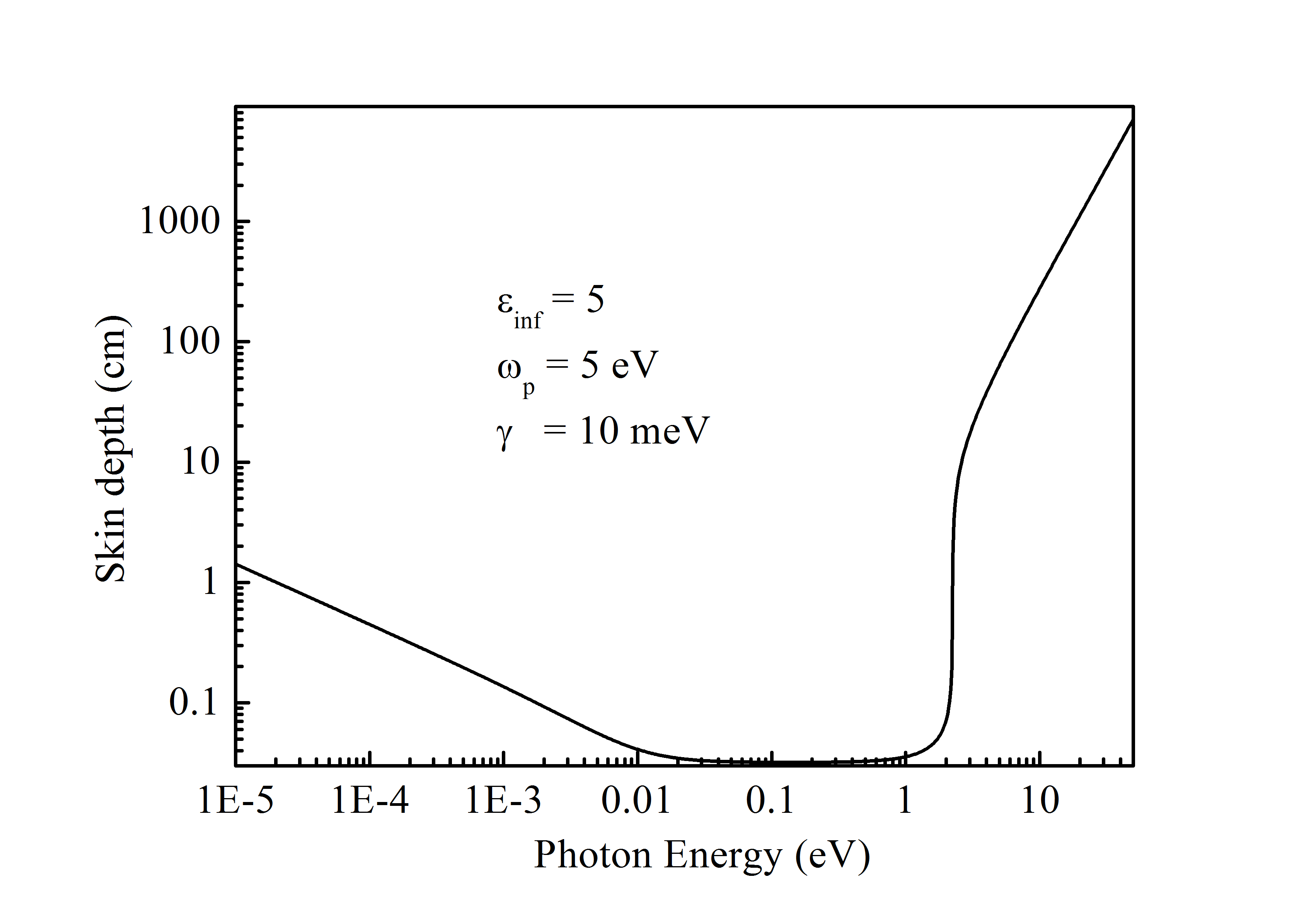}
\caption{\label{figskindepth} Skin depth calculated with the same
parameters as in figure \ref{drudecond}.}
\end{figure}

Finally we show the skin depth in figure \ref{figskindepth}.
We see that for frequencies smaller than the scattering rate,
$\gamma$,  light waves can enter the material. This is called the
classical skin effect. For frequencies larger than the screened
plasma frequency the material becomes transparent again.

\section{Experimental Techniques}
The goal of optical spectroscopy is to determine the complex
dielectric function or equivalently the complex optical
conductivity. Since electromagnetic waves have small momenta
compared to the typical momenta of a solid, i.e. $\textbf{q}\ll
1/a_{0}$, we usually only probe the $q\to 0$ limit of the optical
constants. In this limit,
\begin{eqnarray}
\lim_{q\to
0}(\varepsilon^{T}(\mathbf{q},\omega)-\varepsilon^{L}(\mathbf{q},\omega))=0,\\
\varepsilon(\mathbf{q\to
0},\omega)=\varepsilon_{1}(\omega)+i\frac{4\pi}{\omega}\sigma_{1}(\omega).
\end{eqnarray}
In some cases we can directly obtain information on both real and
imaginary components seperately, but more often we obtain
information where the contributions are mixed. We then make use of
some form the KK-relations to disentangle the two.

\subsection{Reflection and Transmission at an interface}
\begin{figure}[tbh]
\includegraphics[width=8.5 cm]{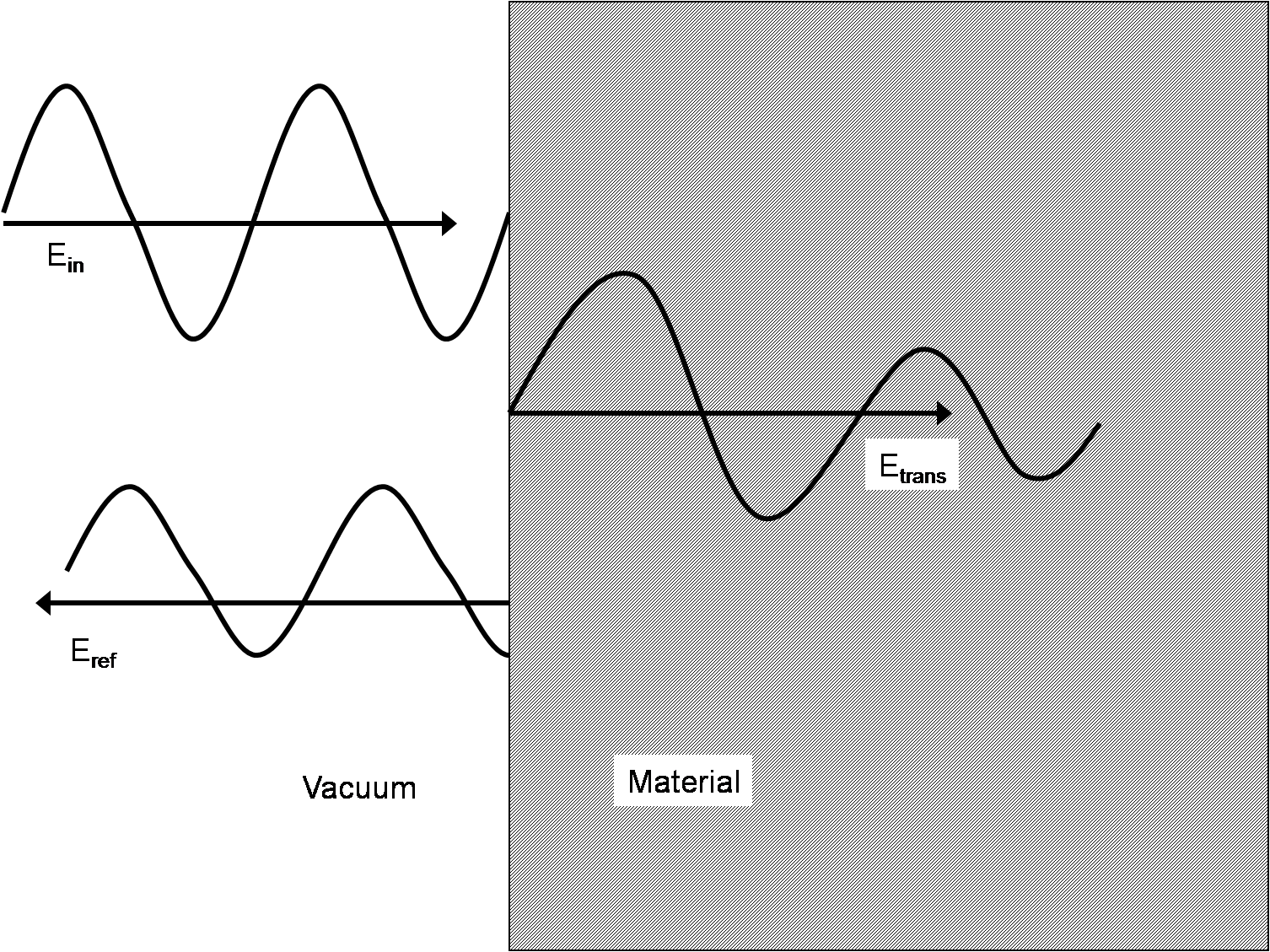}
\caption{\label{reflection}Electromagnetic waves reflecting from a
material. The reflected wave has a smaller amplitude and is phase
shifted with respect to the incoming wave. The transmitted wave is
continuously attenuated inside the material.}
\end{figure}

When we shine light on an interface between vacuum and a material,
part of the light is reflected and another part is transmitted as
in figure \ref{reflection}.
At the boundary the electromagnetic waves have to obey the
following boundary conditions,
\begin{eqnarray}
\mathbf{E}_{i}+\mathbf{E}_{r}=\mathbf{E}_{t},\label{boundcond1}\\
\mathbf{E}\times\mathbf{H}\;//\;\mathbf{k}.
\end{eqnarray}
From these two equations it follows that the reflected magnetic
field suffers a phase shift at the boundary,
\begin{equation}\label{boundcond2}
\mathbf{H}_{i}-\mathbf{H}_{r}=\mathbf{H}_{t}.
\end{equation}
Using equation (\ref{E}) in equation (\ref{linresmax2}) we obtain,
\begin{equation}
iqc\mathbf{E}^{T}=i\omega\mu\mathbf{H}.
\end{equation}
so that, using the dispersion relation (\ref{poldisp}),
\begin{equation}
\frac{\mathbf{H}}{\mathbf{E}^{T}}=\sqrt{\frac{\varepsilon}{\mu}}.
\end{equation}
From now on we set $\mu=1$ unless otherwise indicated. In that
case $\mathbf{H}/\mathbf{E}^{T}=\hat{n}$. Combining this result
with Eq. (\ref{boundcond2}) we get,
\begin{equation}
\mathbf{E}_{i}-\mathbf{E}_{r}=\hat{n}.
\end{equation}
Together with Eq. (\ref{boundcond1}) we can now solve for
$\textbf{E}_{r}/\textbf{E}_{i}$ and
$\textbf{E}_{t}/\textbf{E}_{i}$,
\begin{eqnarray}
\hat{r}\equiv\mathbf{E}_{r}/\mathbf{E}_{i}=\frac{1-\hat{n}}{1+\hat{n}},\label{rhat}\\
\hat{t}\equiv\mathbf{E}_{t}/\mathbf{E}_{i}=\frac{2}{1+\hat{n}}.\label{that}
\end{eqnarray}
The two quantities $\hat{r}$ and $\hat{t}$ are the complex
reflectance and transmittance.

\subsection{Reflectivity experiments}
The real reflection coefficient $R(\omega)$ which is measured in a
reflection experiment is related to $\hat{r}$ via
\begin{equation}
R=|\hat{r}|^{2}=|\frac{(n-1)^{2}+k^{2}}{(n+1)^{2}+k^{2}}|.
\end{equation}
Note that in this experiment we obtain no information on the phase
of $\hat{r}$. In these experiments the angle of incidence is as
close to normal incidence as possible. To measure $R(\omega)$ one
first measures the reflected intensity $I_{s}$ from the sample
under study. To normalize this intensity one then has to take a
reference measurement. This can be done by replacing the sample
with a mirror (i.e a piece of aluminum or gold) and again measure
the reflected intensity, $I_{ref}$. The reflection coefficient is
then $R(\omega)\equiv I_{s}(\omega)/I_{ref}(\omega)$. A better way
is to evaporate a layer of gold or aluminum \textit{in-situ} and
measure the reflected intensity as a reference. This way one
automatically corrects for surface imperfections and, if done
properly, there are no errors due to different size and shape of
the reflecting surface. To obtain the optical constants from such
an experiment we have to make use of KK-relations. If we define,
$\hat{r}(\omega)\equiv\sqrt{R(\omega)}e^{i\theta}$, then the
logarithm of $\hat{r}(\omega)$ is
\begin{equation}
\ln\hat{r}(\omega)=\ln\sqrt{R(\omega)}+i\theta.
\end{equation}
The phase $\theta$ in this expression is the unknown we want to
determine. If we interpret $\hat{r}$ as a response function we can
use the same arguments as in the section on KK relations and
calculate $\theta({\omega})$ from,
\begin{equation}
\theta(\omega)=-\frac{\omega}{\pi}P\int_{0}^{\infty}\frac{\ln
R(\omega')}{\omega'^{2}-\omega^{2}}d\omega',
\end{equation}
which is just the same as the KK-relation for $\hat{\epsilon}$.
The complex dielectric function is calculated from $R(\omega)$ and
$\theta(\omega)$ using,
\begin{equation}\label{KKofR}
\hat{\varepsilon}(\omega ) = \left( {\frac{{1 - \sqrt {R\left(
\omega \right)} e^{i\theta \left( \omega  \right)} }}{{1 + \sqrt
{R\left( \omega \right)} e^{i\theta \left( \omega  \right)} }}}
\right)^{\rm 2} {\rm  }.
\end{equation}
Although in principle exact, this technique is in practice only
approximate. The reason is that we cannot measure $R(\omega)$ from
zero to infinite frequency. Most experiments probe a frequency
range between a few meV and a few eV. To do the integral in Eq.
(\ref{KKofR}) one then has to use extrapolations in the frequency
ranges where no data is available. For metals the low frequency
extrapolation which is most often used is the so-called
Hagen-Rubens approximation,
\begin{equation}
R(\omega)=1-\alpha\sqrt{\omega}.
\end{equation}
For frequencies above the interband transitions one often uses an
extrapolation that is proportional to $\omega^{-4}$.
\begin{figure}[tbh]
\includegraphics[width=8.5 cm]{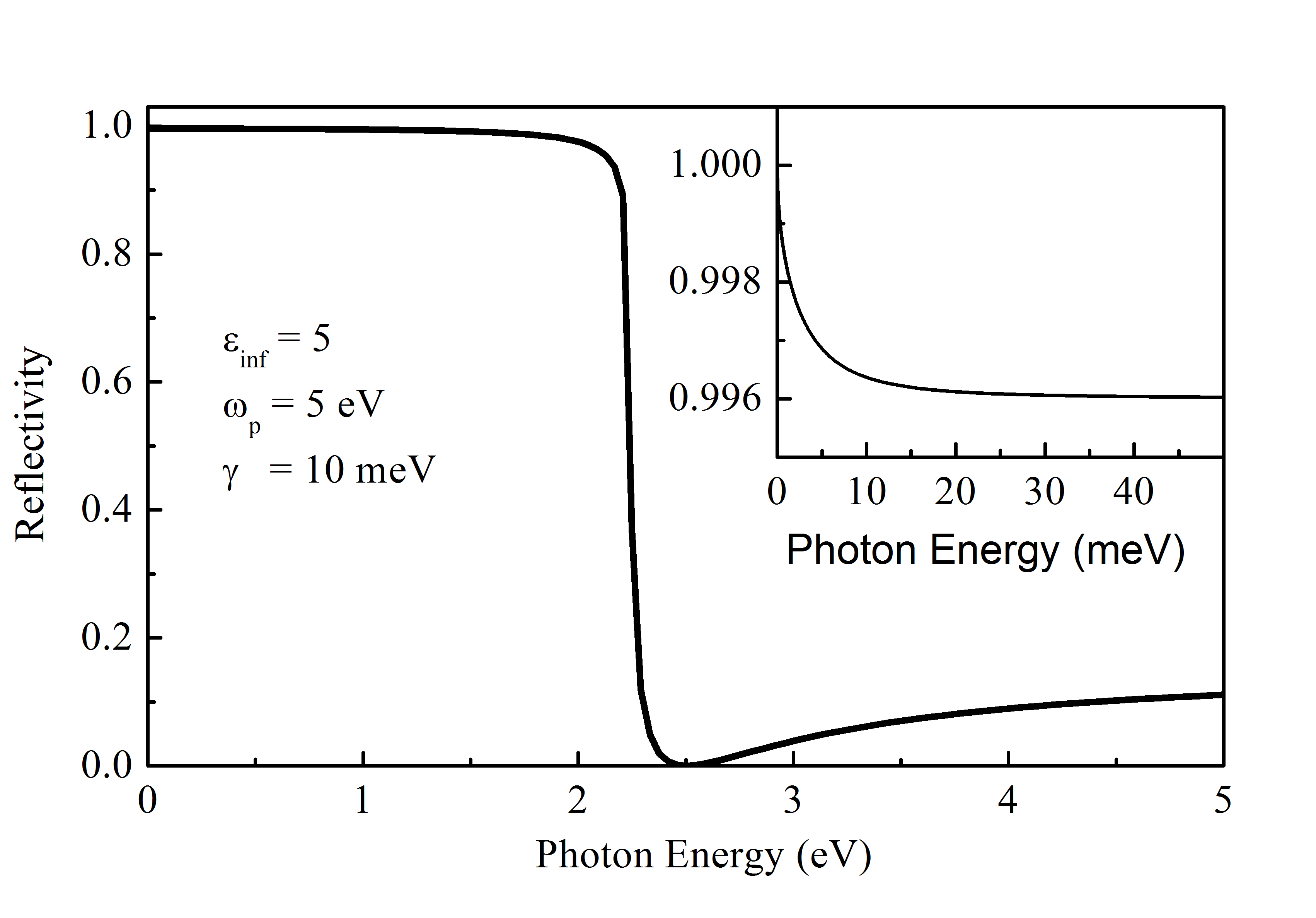}
\caption{\label{druderefl} Reflectivity calculated using
parameters typical for a metal. The inset shows the low energy
reflectivity on an enhanced scale.}
\end{figure}
As an example of a possible experimental result we show in figure
\ref{druderefl} the reflectivity calculated from the Drude model
for the same parameters as in section on polaritons.

The reflectivity is close to one until just below the plasma
frequency. At the zero crossing of $\varepsilon_{1}$ the
reflectivity has a minimum. The inset shows a blow up of the
"flat" region below 50 meV. Here one can clearly see the
Hagen-Rubens behavior mentioned above. If the sample under
investigation is anisotropic one has to use polarized light along
one of the principle crystal axes to perform the experiment.

\subsection{Grazing Incidence Experiments}
\begin{figure}[tbh]
\includegraphics[width=8.5 cm]{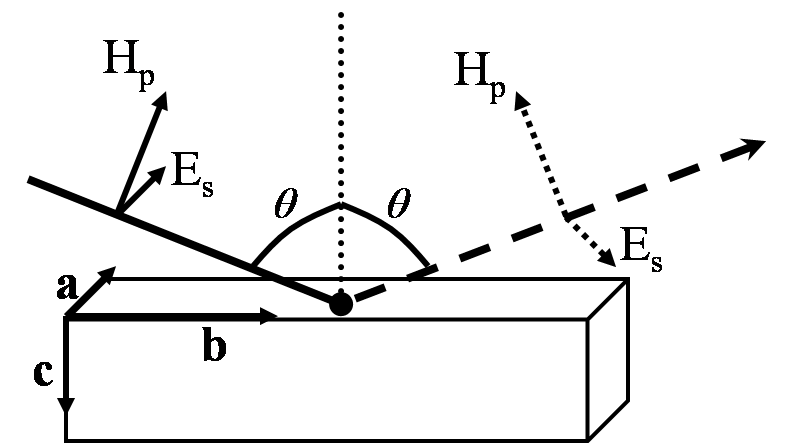}
\caption{\label{grazincid} Grazing incidence experiment. The
result of the experiment is extremely sensitive to the precise
orientation of the crystal axes with respect to the incoming
light.}
\end{figure}

\begin{figure}[tbh]
\includegraphics[width = 6 cm]{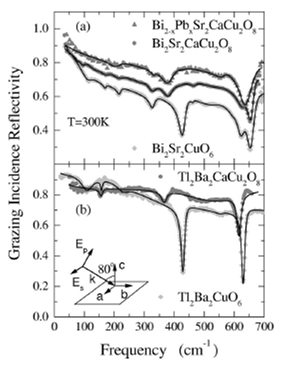}
\caption{\label{grazincidbi2212} Grazing incidence reflectivity of
Bi-2201, Bi-2212, Tl-2201 and Tl-2212. The inset in panel (b)
indicates the measurement geometry. The figure is adapted from
ref. \citep{tsvetkov}.}
\end{figure}

A closely related technique is to measure reflectance under a
grazing angle of incidence. Here one has to distinguish between
experiments performed with different incoming polarizations as
shown in figure \ref{grazincid}.
We distinguish between p-polarized light and s-polarized light.
For p-polarization the electric field is parallel to the plane of
incidence, whereas for s-polarization it is perpendicular to it (s
stands for senkrecht). Since in principal the optical constants
along the three crystal axes can be different, we use the labels
$a$, $b$ and $c$ for the optical constants as indicated in figure
\ref{grazincid}. For p-polarized light the complex reflectance is,
\begin{equation}\label{rp}
{\rm   }r_{\rm p}  = \frac{{{ \hat n}_c { \hat n}_b \cos \theta -
\sqrt {{ \hat n}_c ^2  - \sin ^2 \theta } }}{{{ \hat n}_c { \hat
n}_b \cos \theta  + \sqrt {{ \hat n}_c ^2 - \sin ^2 \theta } }}.
\end{equation}
The angle $\theta$ in this equation is the angle relative to the
surface normal under which the experiment is performed. For
s-polarized light the complex reflectance is,
\begin{equation}\label{rs}
{\rm  }r_{\rm s}  = \frac{{\cos \theta  - \sqrt {{ \hat n}_a ^2 -
\sin ^2 \theta } }}{{\cos \theta  + \sqrt {{ \hat n}_a ^2  - \sin
^2 \theta } }}.
\end{equation}
An example of such an experiment is shown in figure
\ref{grazincidbi2212}.
In this example the samples are from the bismuth based family of
cuprates \cite{tsvetkov}. They have a layered structure consisting
of conducting copper-oxygen sheets, interspersed with insulating
bismuth-oxygen layers. Since the bonding between layers is not
very strong it is very difficult to obtain samples that are
sufficiently thick along the insulating c-direction. The grazing
incidence technique is used here to probe the optical constants of
the c-axis without the need of a large ac-face surface area. A
disadvantage in this particular experiment is that it is not
possible to determine accurately the absolute value of the optical
constants. It is possible however to determine the so-called loss
function $Im(-1/\hat{\varepsilon}_{c})$. The experiment is
performed on the ab-plane of the sample using p-polarized light
and we can simplify the expression for $\hat{r}_{p}$ by using the
fact that the $a$ and $b$ direction are almost isotropic. The
resulting expression for $\hat{r}_{p}$ is,
\begin{equation}
\hat{r}_{\rm p}  = \frac{{\sqrt {\hat \varepsilon _b } \cos \theta
- \sqrt {1 - {{\sin ^2 \theta } \mathord{\left/
 {\vphantom {{\sin ^2 \theta } {\hat \varepsilon _c }}} \right.
 \kern-\nulldelimiterspace} {\hat \varepsilon _c }}} }}{{\sqrt {\hat \varepsilon _b } \cos \theta  + \sqrt {1 - {{\sin ^2 \theta } \mathord{\left/
 {\vphantom {{\sin ^2 \theta } {\hat \varepsilon _c }}} \right.
 \kern-\nulldelimiterspace} {\hat \varepsilon _c }}} }}.
\end{equation}
From this equation we can derive the following relation between
the grazing incidence reflectivity and a pseudo loss-function
$L(\omega)$ \cite{dvdm},
\begin{equation}
L(\omega)\equiv\frac{{\left( {1 - R_p } \right)}} {{\left( {1 +
R_p } \right)}} \approx Im \frac{{2e^{i\phi _p } }} {{\left| {n_b
} \right|\cos \theta }}\sqrt {1 - \frac{{\sin ^2 \theta }} {{\hat
\varepsilon _c }}}.
\end{equation}
The function
$\sqrt{1-\frac{\sin^{2}\theta}{\hat{\varepsilon}_{c}}}$ has maxima
at the same position as the true loss-function. In this way
information was gained on the phonon structure of the c-axis of
this material.

\subsection{Spectroscopic Ellipsometry}
The third technique we introduce here is spectroscopic
ellipsometry. This relatively new technique has two major
advantages over the previous techniques. Firstly, the technique is
self-normalizing meaning that no reference measurement has to be
done and secondly, it provides directly both the real and
imaginary parts of the dielectric function.

As with the grazing incidence technique we have to distinguish
between s- and p-polarized light and label the crystal axes.
Instead of measuring $R_{p}$ or $R_{s}$ independently, we now
measure directly the amplitude and phase of the ratio
$\hat{r}_{p}/\hat{r}_{s}=|\hat{r}_{p}/\hat{r}_{s}|e^{i(\eta_{p}-\eta_{s})}$.
To see how this can be done we first describe the experimental
setup. There are a number of different setups one can use and here
we describe the simplest. This setup consists of a source followed
by a polarizer. With this polarizer we can change the orientation
of the polarization impinging on the sample.
\begin{figure}[tbh]
\includegraphics[width=8.5 cm]{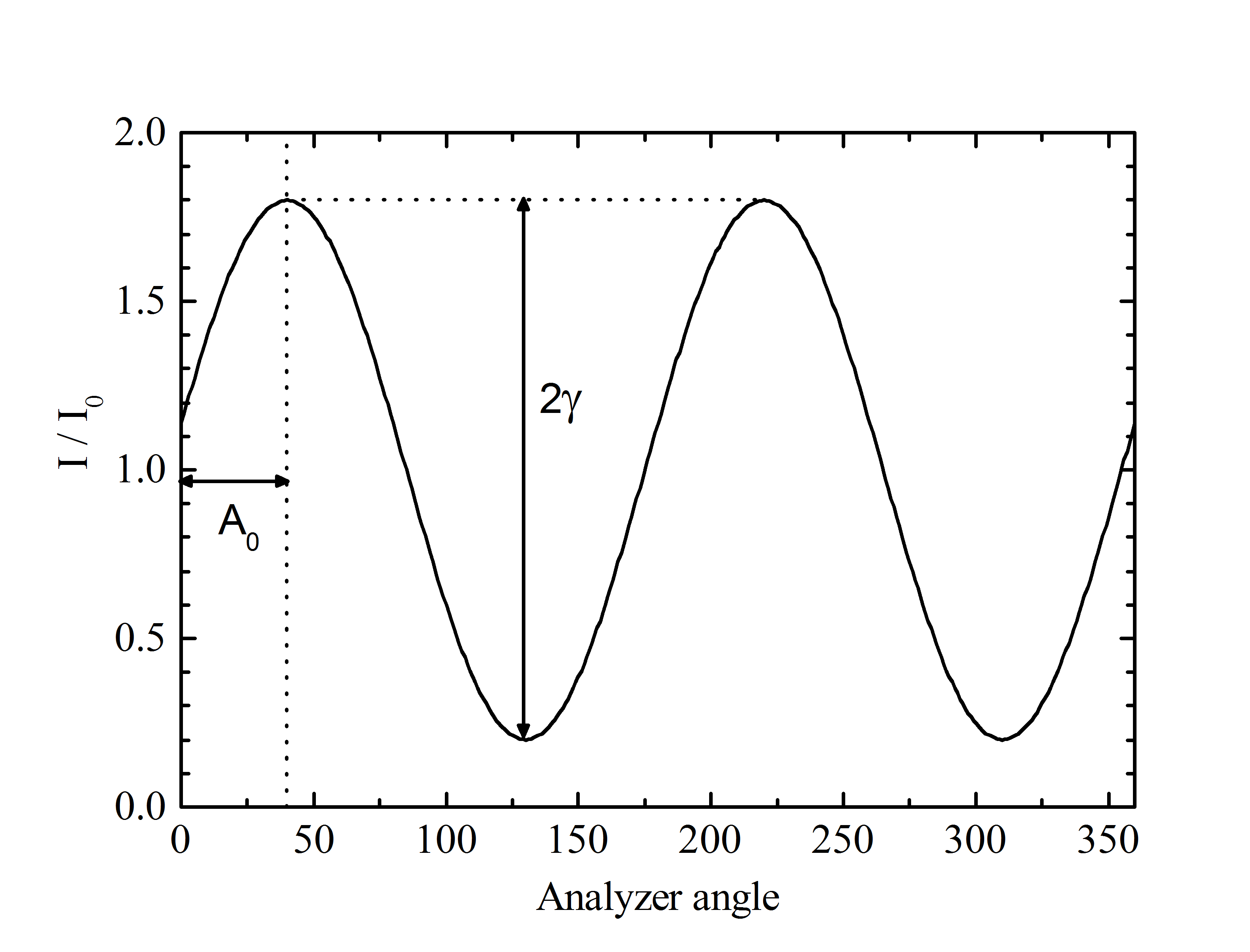}
\caption{\label{ellipsometry} Result of an ellipsometric
measurement. The phase shift $A_{0}$ and amplitude $2\gamma$ are
the two quantities that we are interested in.}
\end{figure}
The light reflected from the sample passes through another
polarizer (called analyzer) and then hits the detector. Depending
on the orientation of the first polarizer we can change the
electric field strength of s- and p-polarized light according to,
\begin{eqnarray}
 E_p  = \left| {E_i } \right|\cos \left( P \right), \\
 E_s  = \left| {E_i } \right|\sin \left( P \right). \\
\end{eqnarray}
From the expressions for $\hat{r}_{p}$ and $\hat{r}_{s}$,
(\ref{rp}) and (\ref{rs}), in the previous section it follows
that,
\begin{equation}\label{rp/rs}
{\rm  }\hat \rho  \equiv {\rm }\frac{{r_{\rm p} }}{{r_{\rm s} }} =
\frac{{\sqrt {{ \hat n}_c ^2  - \sin ^2 \theta }  - { \hat n}_c {
\hat n}_b \cos \theta }}{{\sqrt {{ \hat n}_c ^2  - \sin ^2 \theta
}  + { \hat n}_c { \hat n}_b \cos \theta }} \cdot \frac{{\sqrt {{
\hat n}_a ^2  - \sin ^2 \theta }  + \cos \theta }}{{\sqrt {{ \hat
n}_a ^2  - \sin ^2 \theta }  - \cos \theta }}.
\end{equation}
Our task is now to invert this equation and express the optical
constants in terms of measured quantities and instrument
parameters. For an isotropic sample this can be done quite easily.
We define the pseudodielectric function $\hat{\varepsilon}$ such
that:
\begin{equation}
\hat \rho  \equiv \frac{{\sin \theta \tan \theta  - \sqrt {{
\tilde \varepsilon } - \sin ^2 \theta } }}{{\sin \theta \tan
\theta + \sqrt {{ \tilde \varepsilon } - \sin ^2 \theta } }},
\end{equation}
where we note that
$\hat{\varepsilon}=\varepsilon_{a}=\varepsilon_{b}=\varepsilon_{c}$
in an optically isotropic medium. This equation can be inverted to
obtain $\tilde \varepsilon$,
\begin{equation}\label{ellipseps}
\tilde \varepsilon (\omega ) = \sin ^2 \theta \left[ {1 + \tan ^2
\theta \left( {\frac{{1 - \rho }}{{1 + \rho }}} \right)^2 }
\right].
\end{equation}
So all that is left to do is to express $\hat{\rho}$ in terms of
experimental parameters. The experiment is done in the following
way: we fix the polarizer at some angle $0<P<90$ and then we
record the intensity while rotating the analyzer 360 degrees. The
result is shown in figure \ref{ellipsometry}. We then measure the
amplitude of the resulting sine wave, $\gamma$ and the phase
offset with respect to zero, $A_{0}$ (we assume here that for
$P=0$ the polarizer and analyzer are aligned parallel to each
other). With some goniometry and figure \ref{ellipsometry} we can
show that,
\begin{equation}
\tan A_0  = \frac{{2\tan \left( P \right)}}{{\left| \rho \right|^2
+ \tan ^2 \left( P \right)}}\rho _1,
\end{equation}
and
\begin{equation}
\sqrt {1 - \gamma ^2 }  = \frac{{2\tan \left( P \right)}}{{\left|
\rho  \right|^2  + \tan ^2 \left( P \right)}}\rho _2.
\end{equation}
Combining these two equations leads to,
\begin{equation}\label{ellipsrho}
\rho  = \frac{{1 \pm \sqrt {\gamma ^2  - \tan ^2 A_0 } }}{{\tan
A_0  - i\sqrt {1 - \gamma ^2 } }}\tan \left( P \right).
\end{equation}
\begin{figure}[bth]
\includegraphics[width=8.5 cm]{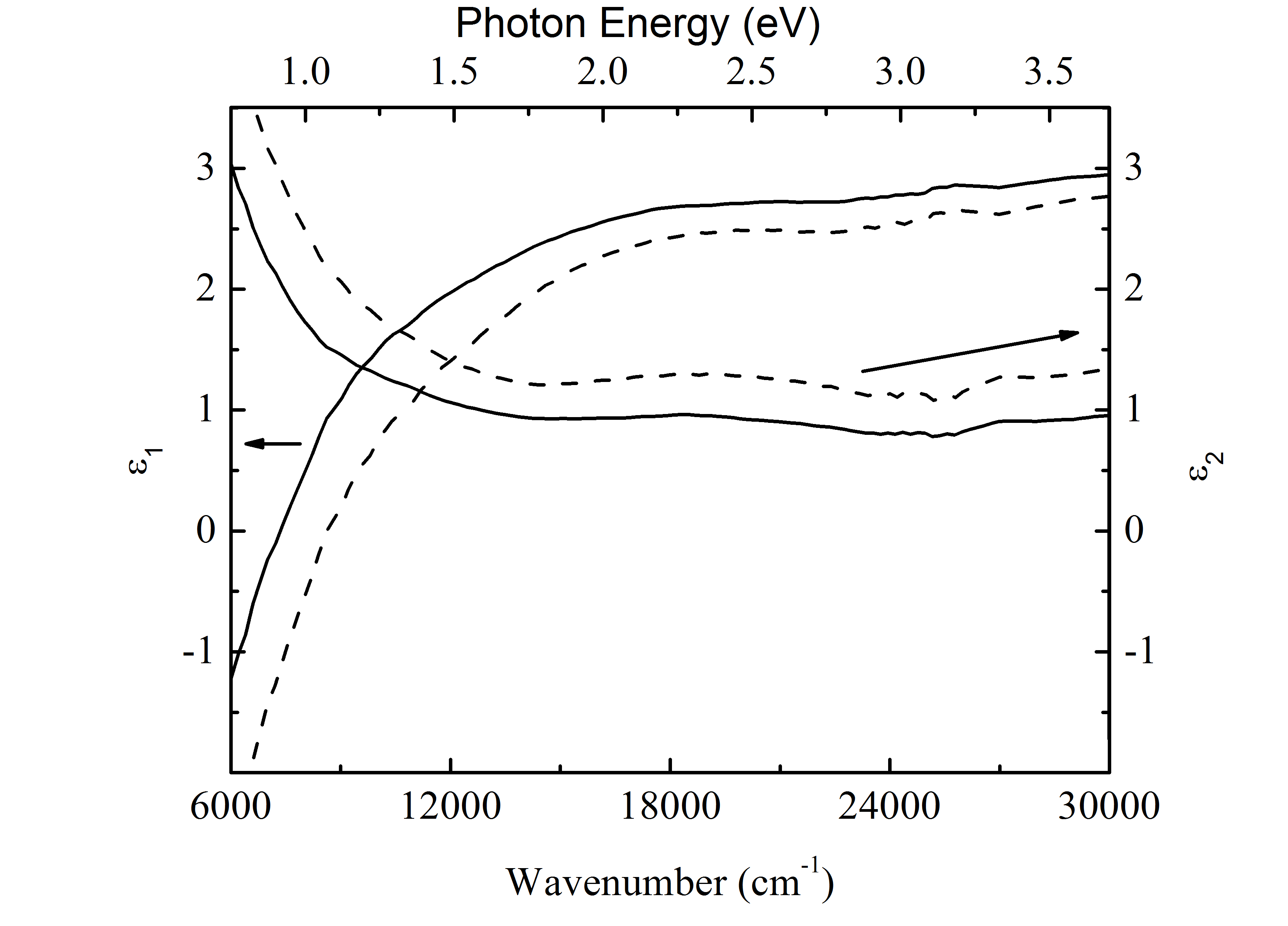}
\caption{\label{ellipshg1201}Dielectric function measured
ellipsometrically on a HgBa$_{2}$CuO$_{4}$ sample. The true
dielectric function is shown in solid lines. The pseudo
dielectric function (i.e. actually measured) is shown as a
dashed line. Data taken from ref. \citep{heumen}.}
\end{figure}
The combination of Eq. (\ref{ellipsrho}) with Eq.
(\ref{ellipseps}) is all we need to describe an ellipsometric
experiment on an isotropic sample. For an anisotropic sample
the problem is slightly more difficult. However, there exists a
theorem due to Aspnes which states that the inversion of Eq.
(\ref{rp/rs}) results in Eq. (\ref{ellipseps}) but now the
dielectric function on the left-hand side is a so-called
pseudo-dielectric function. This pseudo-dielectric function is
mainly determined by the component parallel to the intersection
of sample surface and plane of incidence (component along $b$
in figure \ref{grazincid}), but still contains a small
contribution of the two other components. If we perform three
measurements, each along a different crystal axis, we can
correct the pseudo dielectric functions and obtain the true
dielectric functions. If the sample is isotropic along two
directions, as is the case in high temperature superconductors
for example, only two measurements are required. Figure
\ref{ellipshg1201} shows in dashed lines the pseudo dielectric
function of HgBa$_{2}$CuO$_{4}$. In this case the $a$ and $b$
axes have the same optical constants. The c-axis dielectric
function was determined from reflectivity measurements and
subsequently used to correct the pseudo dielectric function
measured by ellipsometry on the ab-plane. The true dielectric
function after this correction is shown as the solid line.

\subsection{Transmission Experiments}
\begin{figure}[tbh]
\includegraphics[width=8.5 cm]{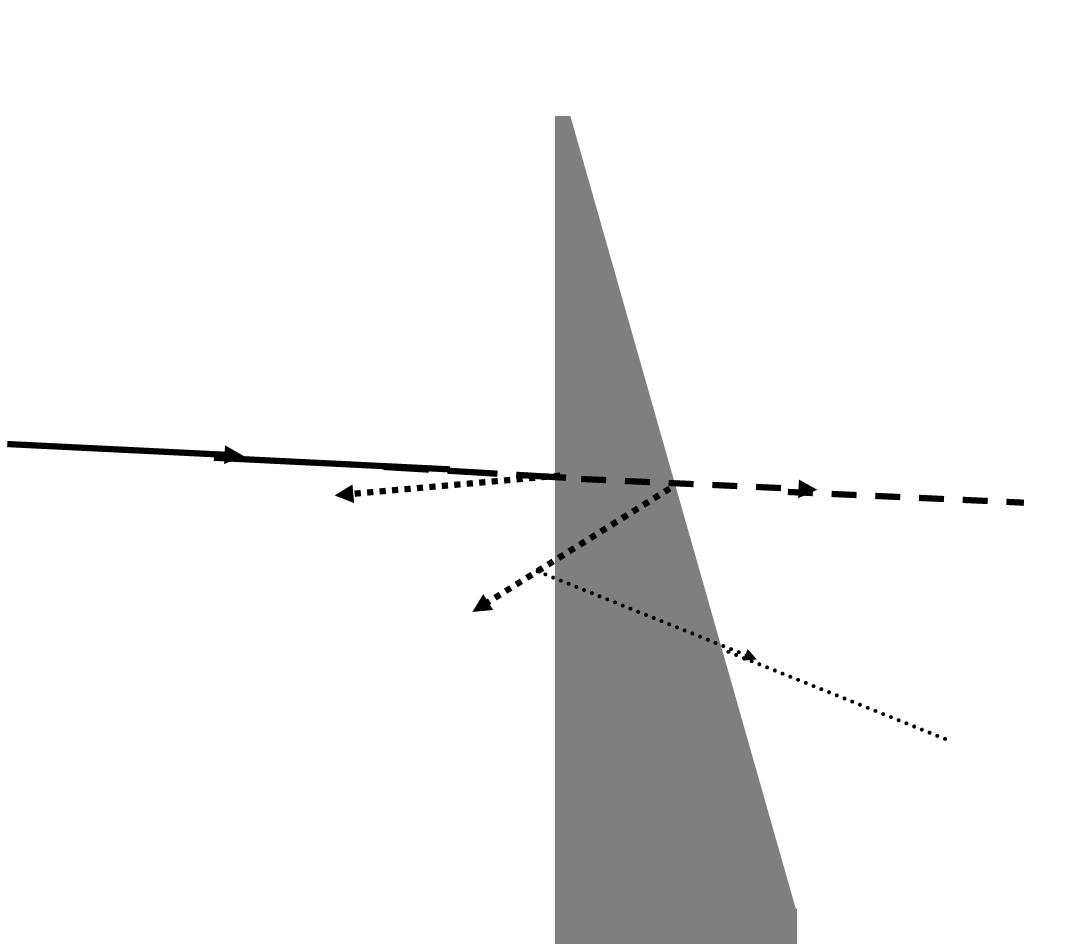}
\caption{\label{wedgedsample}Transmission experiment on a wedged
sample. After the initial ray is partially reflected back from the
front surface all following rays are no longer parallel to the
first transmitted ray.}
\end{figure}

A technique complementary to the reflection techniques is
transmission spectroscopy. This technique is, obviously, most
suitable for transparent samples. In principle the technique can
also be applied to metallic samples but this requires very thin
samples or films. The reflection experiments discussed above are
usually good methods to obtain accurate estimates of the real part
of the optical conductivity. In contrast the transmission
experiments discussed below are more sensitive to weak absorptions
or, in other words to the imaginary part of the optical
conductivity. Note that the simultaneous knowledge of reflection
and transmission spectra allows one to directly determine the full
complex dielectric function without any further approximations.
Examples of weak absorptions which are better probed in a
transmission experiment are multi-phonon or magnon absorptions.
The equations for transmission experiments are slightly more
difficult then those for the reflection experiments. These
equations simplify if we do the experiment on a wedged sample as
shown in figure \ref{wedgedsample}.
At the boundary between vacuum and the sample, part of the light
is reflected and part transmitted. The part that is transmitted is
given by,
\begin{equation}\label{tcoef1}
{ \hat t}_{v,s}  = {\rm  }\frac{2}{{1 + { \hat n}}}{\rm }.
\end{equation}
Inside the wave propagates according to $ e^{i\psi}$ where,
\begin{equation}
 \psi  \equiv \hat nd\omega /c.
\end{equation}
At the next boundary again part of the beam is reflected back into
the sample and part is transmitted. Now we can see the advantage
of the wedged sample: the part of the light that is reflected
propagates away at an angle and after another reflection the
second transmitted ray is no longer parallel to and spatially
separated from the first transmitted ray. This means that we only
have to care about the first transmitted ray. The transmission
coefficient at the boundary from sample to vacuum is given by,
\begin{equation}\label{tcoef2}
{ \hat t}_{s,v}  = \frac{{2{ \hat n}}}{{{ \hat n} + 1}}{\rm },
\end{equation}
so that the total transmission coefficient is,
\begin{equation}\label{tcoef3}
{ \hat t}_{v,s} e^{{\rm i}\psi } { \hat t}_{s,v}.
\end{equation}
Putting Eq.'s (\ref{tcoef1})-(\ref{tcoef3}) together and taking
the absolute value to calculate the transmission $T$ gives,
\begin{equation}\label{T}
T = \frac{{\left| {4{ \hat n}} \right|^2 }}{{\left| {1 + { \hat
n}} \right|^4 }}\exp \left\{ { - \frac{{2d}}{\delta }} \right\}.
\end{equation}
In most transmission experiments
$\varepsilon_{1}\gg\varepsilon_{2}$ and the classical skindepth
$\delta$ can be approximated by,
\begin{equation}
\delta \left( \omega  \right) \approx \frac{{2\pi \sigma _1 \left(
\omega  \right)}}{{{\rm c}\sqrt {\varepsilon _1(\omega)} }}.
\end{equation}
Moreover in these cases $\varepsilon_{1}$ is often dispersion-less
so that we can use the expression for $\delta$ to invert
(\ref{T}),
\begin{equation}
\sigma _1 \left( \omega  \right) = \left\{ { - \ln \left( T
\right) + 2\ln \left( {\frac{{4\left| {{ \hat n}}
\right|}}{{\left| {1 + { \hat n}} \right|^2 }}} \right)}
\right\}\frac{{c\sqrt {\varepsilon _1 } }}{{4\pi d}}.
\end{equation}
\begin{figure}[tbh]
\includegraphics[width=6 cm]{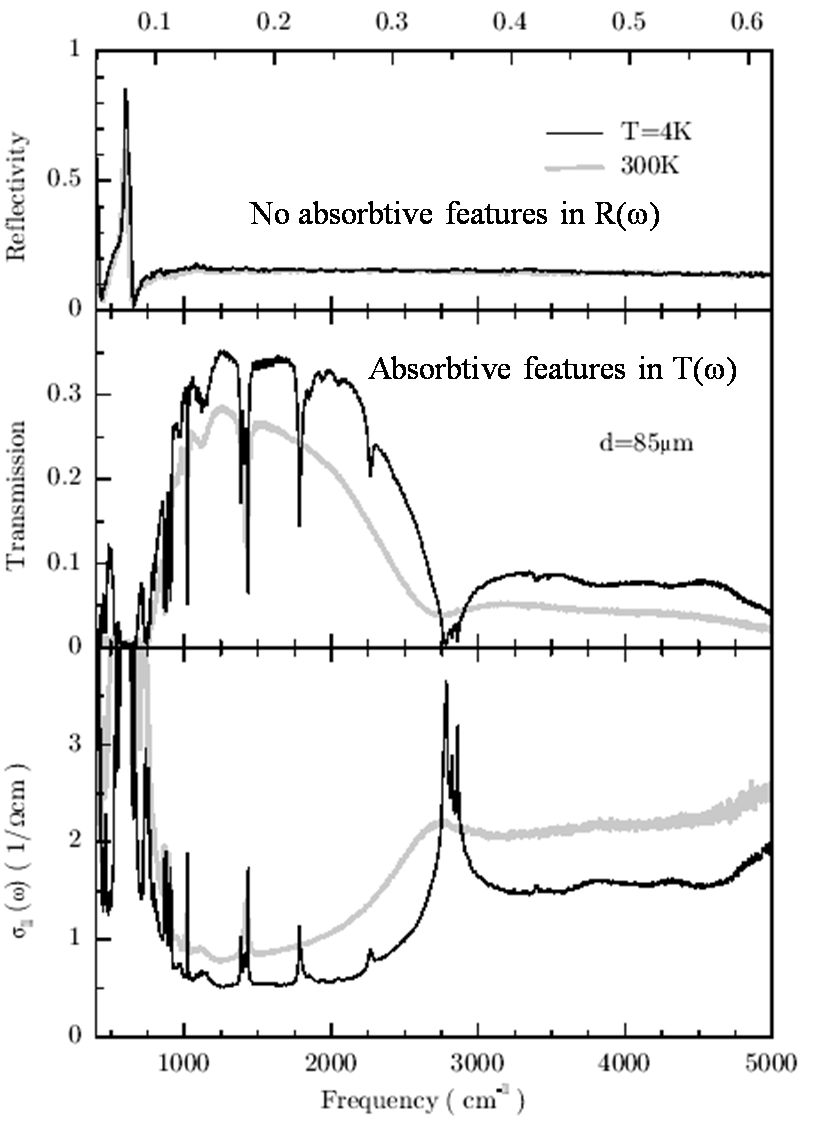}
\caption{\label{reftrans}Comparison of reflectivity and
transmission measured on the same sample. Note the strong
absorptive features present in the transmission spectrum that are
completely invisible in the reflectance spectra. Figure adapted
from \citep{Grueninger}}
\end{figure}
As an example of this technique we show in figure \ref{reftrans} a
comparison between the reflectivity and transmission spectra of
undoped YBa$_{2}$Cu$_{3}$O$_{7}$ \cite{Grueninger}. This material
is an (Mott) insulator which is clearly visible from the
reflectivity spectrum. The large structure at low energies is an
optical phonon. At higher energy the reflectivity spectrum appears
to be rather featureless. Focussing our attention on the
transmission spectrum we see that it is almost zero in the phonon
range but then above the phonon range a whole series of sharp dips
shows up. The optical conductivity consists of a set of smaller
peaks at energies between 100 meV and 300 meV which are due to
multi-phonon absorptions whereas the larger peak just above 300
meV is due to a two magnon plus one phonon absorption (see also
the section on spin interactions below).
\begin{figure}[tbh]
\includegraphics[width=6 cm]{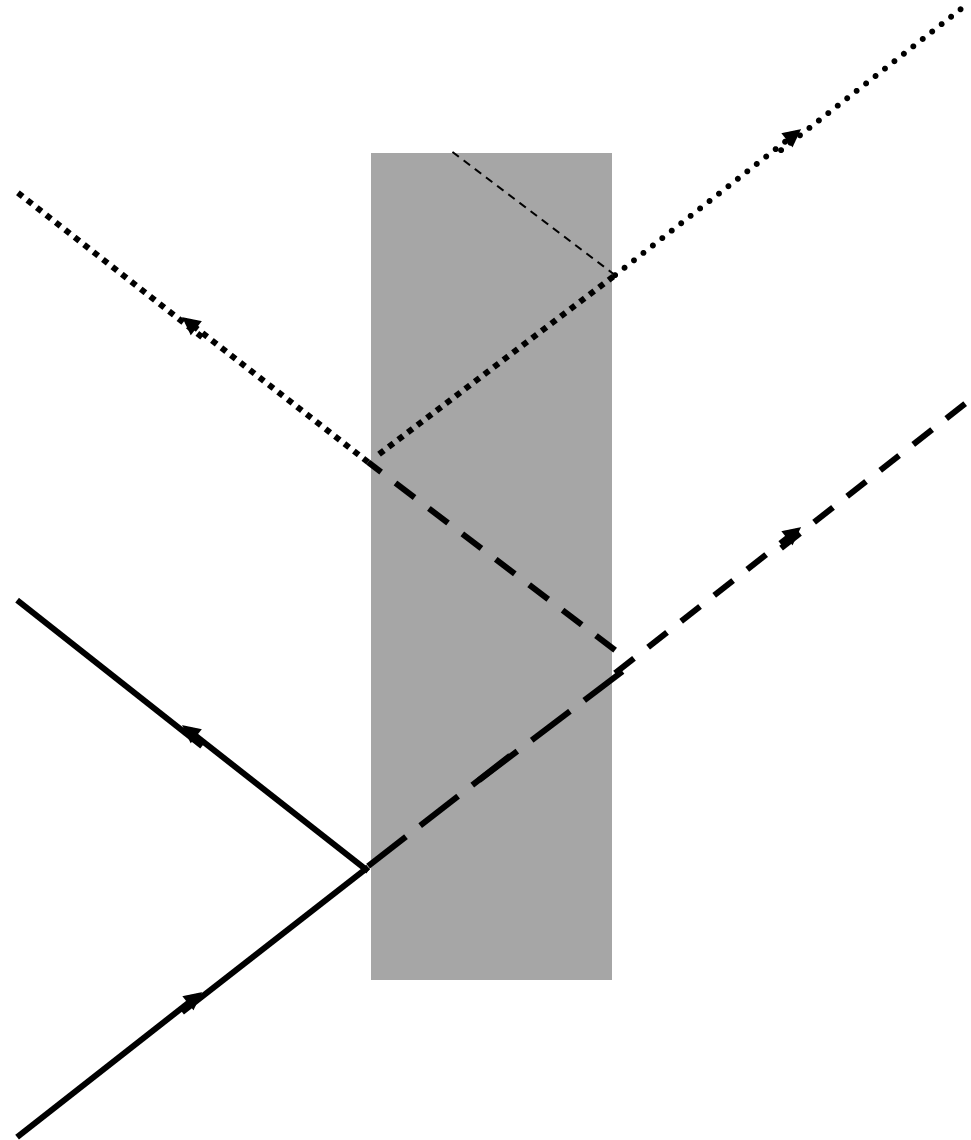}
\caption{\label{transmission}Pictorial of a transmission
experiment on a plan parallel sample.}
\end{figure}
We can also do the experiment on a sample with two
plan-parallel sides as depicted in figure \ref{transmission}.
We can immediately realize that for a given thickness of the
sample there will be interference effects between different
transmitted rays for certain frequencies. These will cause
oscillations in the transmission spectra which are called
Fabry-Perot resonances. We now analyze the transmission
coefficients for this experiment. The coefficient for the first
ray is off course the same as in Eq. (\ref{tcoef3}). The
coefficients for the higher order rays are formed by
multiplying $\hat{t}_{v,s}e^{i\psi}$ on the right with a factor
$f$,
\begin{equation}
f\equiv\hat{r}_{s,v}e^{i2\psi}\hat{r}_{s,v},
\end{equation}
followed by a factor $\hat{t}_{s,v}$. So the total transmission
coefficient for the second transmitted ray is given by,
\begin{equation}
{\hat t}_{v,s} e^{{\rm i}\psi } { \hat r}_{s,v} e^{{\rm i}\psi }
{\hat r}_{s,v} e^{{\rm i}\psi } { \hat t}_{s,v}={\hat t}_{v,s}
e^{{\rm i}\psi } { \hat t}_{s,v} f.
\end{equation}
The coefficients $\hat{t}_{v,s}$ and $\hat{t}_{s,v}$ are given
by Eq. (\ref{tcoef1}) and (\ref{tcoef2}). The coefficient for
reflection on a boundary from sample to vacuum is given by,
\begin{equation}
{\hat r}_{s,v}  = {\rm  }\frac{{{\hat n} - 1}}{{{\hat n} + 1}}.
\end{equation}
It is easy to see that if we sum over all transmitted rays the
total transmission coefficient is given by,
\begin{equation}\label{tpl}
\hat t  = {\hat t}_{v,s} e^{{\rm i}\psi } {\hat t}_{s,v} \left( {1
+ f + f^2  + ..} \right)= \frac{{2{\hat n}}}{{2{\hat n}\cos \psi -
i(1 + {\hat n}^2 )\sin \psi }}.
\end{equation}
For thin films the phase factor $\psi\ll1$ and we can simplify
this equation to,
\begin{equation}
\hat t \approx \frac{1}{{1 + \frac{{2\pi d}}{c}\sigma _1  -
i\frac{{\omega d}}{{2c}}(1 + \varepsilon ')}},
\end{equation}
and so,
\begin{equation}
T\left( \omega  \right) \approx \frac{1}{{1 + 4\pi dc^{ - 1}
\sigma _1 \left( \omega  \right)}}.
\end{equation}
\begin{figure}[htb]
\begin{minipage}{8.5 cm}
\includegraphics[width = 8.5cm]{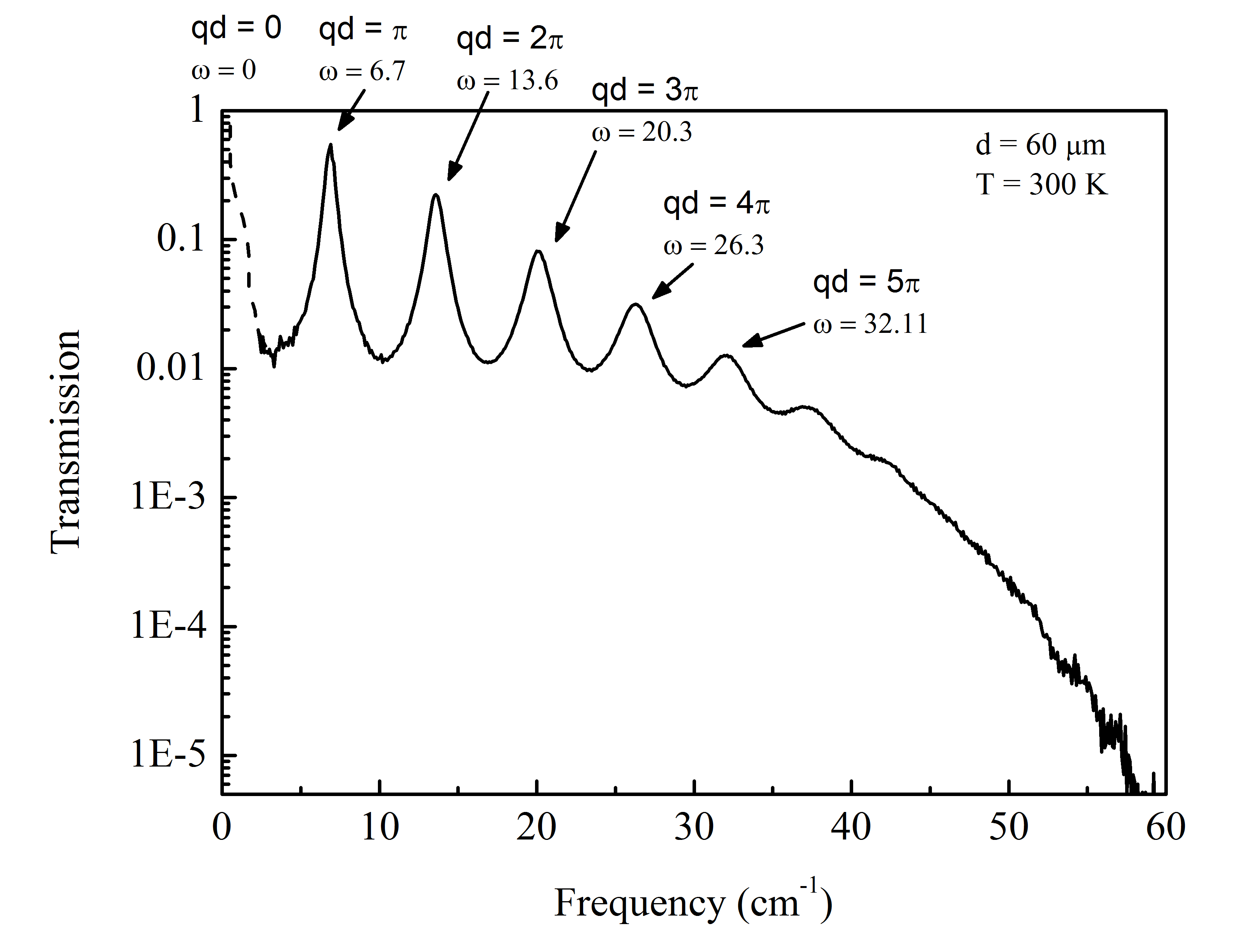}
\end{minipage}\hspace{2pc}%
\begin{minipage}{8.5 cm}
\includegraphics[width = 6 cm]{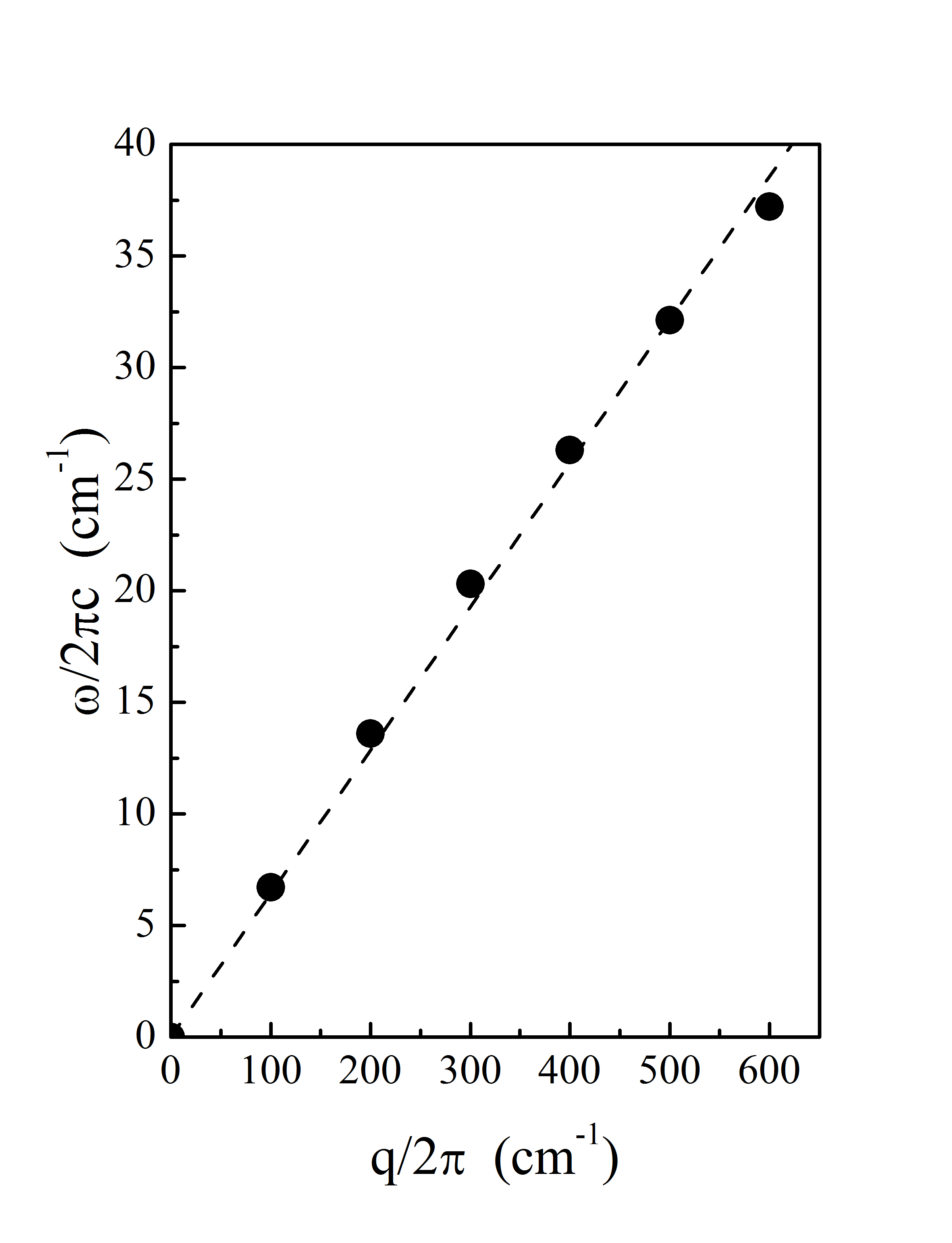}
\end{minipage}
\caption{\label{STO}Left: Far infrared transmission spectrum for
SrTiO$_{3}$. The positions of the peaks determine the polariton
dispersion. The dashed line at low frequency is an extrapolation
to zero frequency. Right: Dispersion relation of polaritons in STO
as derived from transmission spectrum in the left panel.}
\end{figure}
More generally from Eq. (\ref{tpl}) we obtain,
\begin{equation}
T_{LR}  = \frac{{4\left| \varepsilon  \right|}}{{\left|
{4\varepsilon \cos ^2 \psi } \right| + \left| {\left( {1 +
\varepsilon } \right)^2 \sin ^2 \psi } \right| + 2{\mathop{\rm
Im}\nolimits} \left\{ {(1 + \varepsilon )\sin 2\psi } \right\}}}.
\end{equation}
In the case that the sample under investigation has only weak
absorptions, i.e. $Im(\hat{n})\approx0$, this equation simplifies
to,
\begin{equation}\label{TLR}
T_{LR}  \approx \frac{{4n^2 }}{{4n^2  + \left( {1 - n^2 }
\right)^2 \sin ^2 \left( {nd\omega /c} \right)}}.
\end{equation}
This equation gives us some insight to the occurrence of
Fabry-Perot resonances: if $\omega=cm\pi/nd$ with m=0,1,2,... the
sinus is equal to zero and the transmission $T=1$. In between
these maxima the transmission has minima and
$T\approx4\hat{n}^{2}/(1+\hat{n}^{2})^{2}$. In reality the
transmission will never reach 1 due to the fact that
$Im(\hat{n})\neq0$ in which case our approximations are no longer
valid. As an example we display in the left panel of figure
\ref{STO} the transmission spectrum of SrTiO$_{3}$
\cite{mechelen}. This material is very close to being
ferroelectric and as a result it has a very large dielectric
constant. The non-sinusoidal shape of the peaks is due to this
large dielectric constant. One can use the Fabry-Perot resonances
to measure the polariton dispersion as we now show. Note that at
each maximum in the transmission spectrum we know precisely the
value of the argument of the sine function in Eq. (\ref{TLR}).
\begin{figure}[htb]
\begin{minipage}{8.5 cm}
\includegraphics[width = 8.5cm]{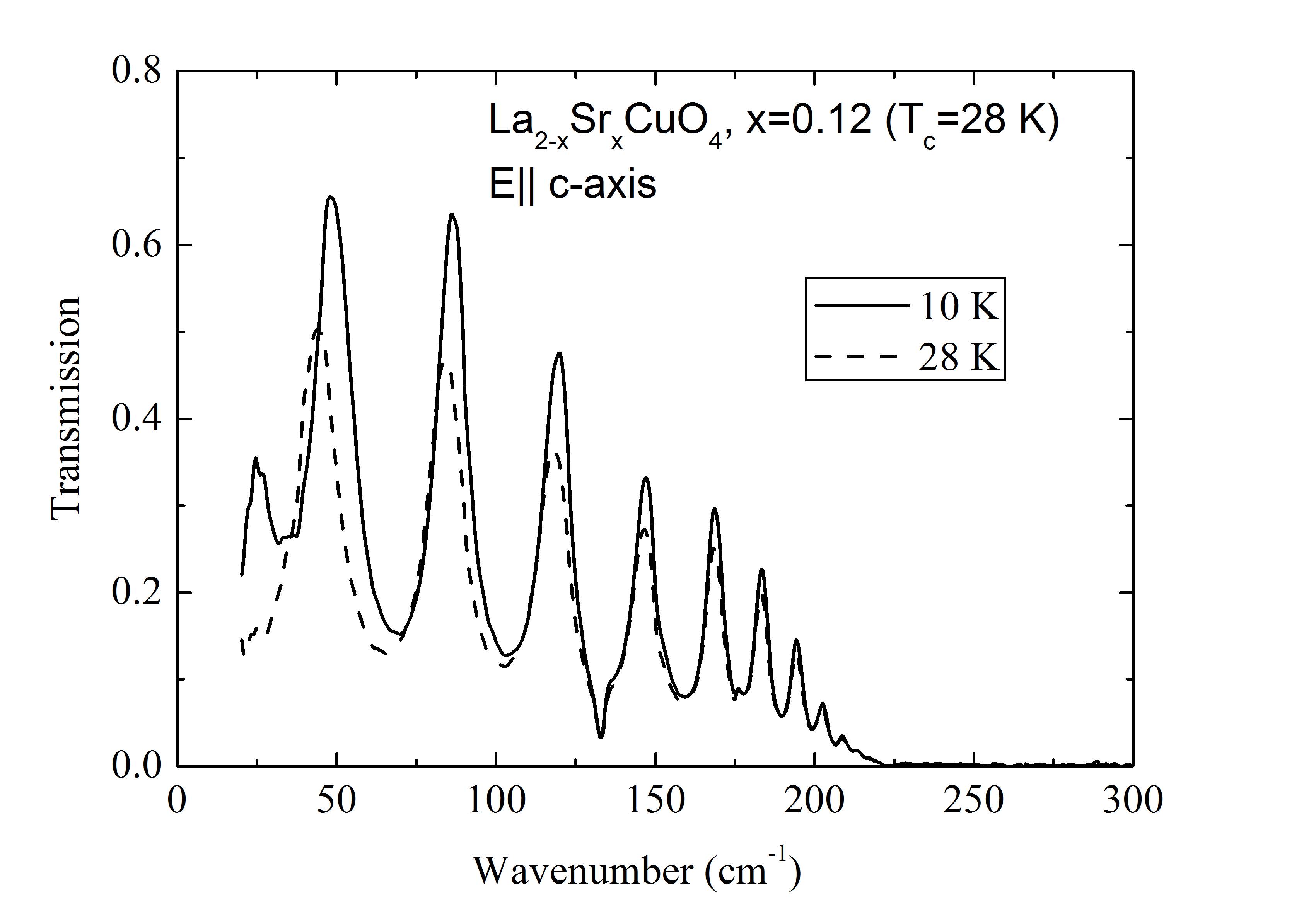}
\end{minipage}\hspace{2pc}%
\begin{minipage}{8.5 cm}
\includegraphics[width = 8.5 cm]{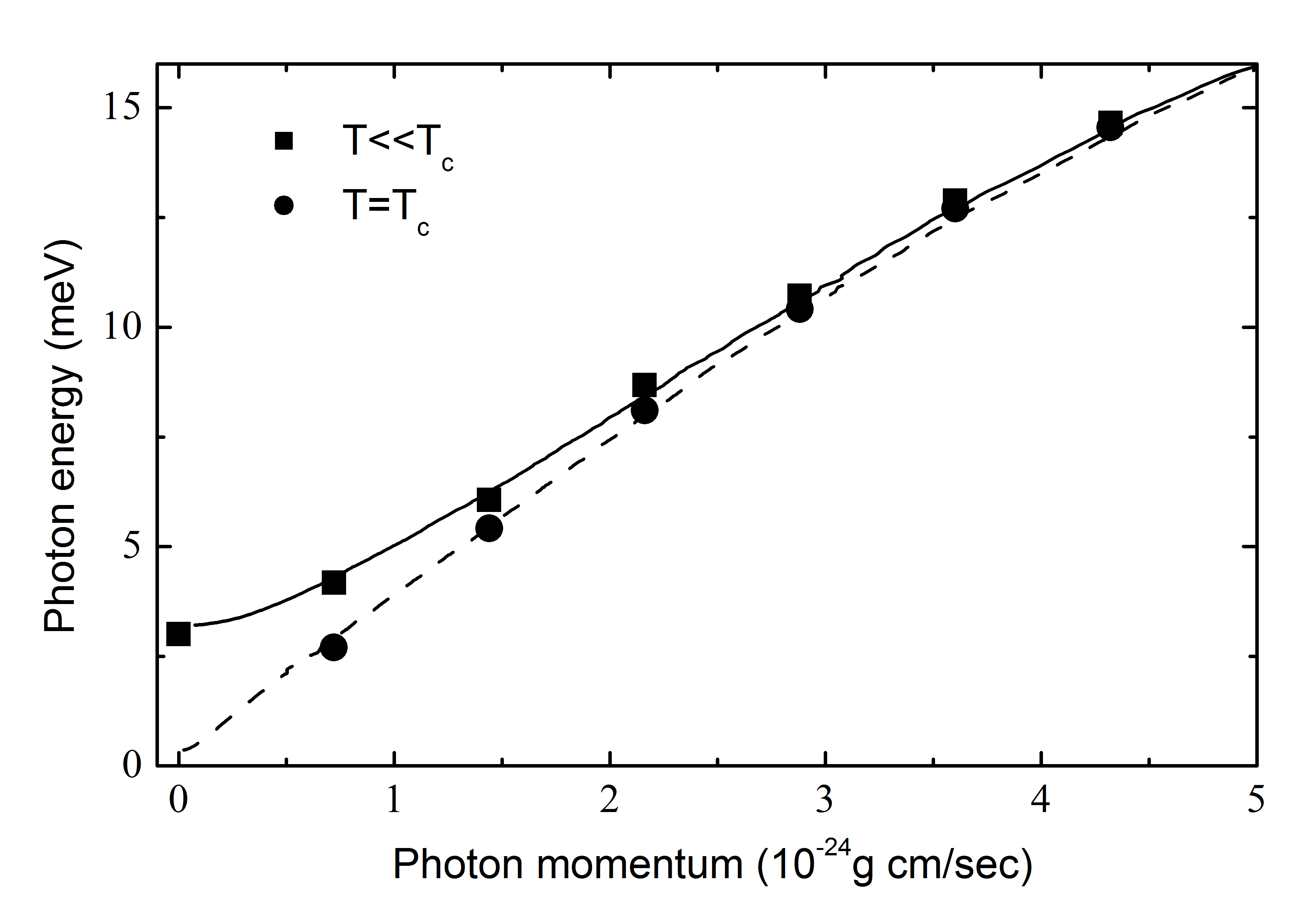}
\end{minipage}
\caption{\label{transLSCO}Left: Transmission spectrum of LSCO at a
temperature just above $T_{c}$ and one far below. Note the shift
in peak positions. Figure adapted from ref. \citep{kuzmenko}.
Right: Dispersion relation of polaritons in LSCO as derived from
left panel. The squares are derived from the spectrum in the
superconducting state whereas the circles are determined at a
temperature slightly above $T_{c}$.}
\end{figure}
We can read off the value of $\omega$ from the graph and using Eq.
(\ref{poldisp}) we can replace the argument in the sine function
by $n\omega d/c=qd$ so the momentum at a given maximum is,
\begin{equation}
q_{m}=\frac{m\pi}{d}\quad m=0,1,2,...
\end{equation}
So given the thickness of the sample we can make a plot of
$\omega(q)$. The result for STO is shown in the right panel of
figure \ref{STO}. We see that the dispersion is linear, indicating
that $n$ is dispersion-less in this range. The slope of the curve
directly gives us $n\approx20.5$. Another interesting application
of this is to superconductors. In figure \ref{transLSCO} we show
the transmission spectrum of LSCO at a temperature slightly above
T$_{c}$ and far below \cite{kuzmenko}. One can see that the
position of the maxima has changed and this shows up in the
polariton dispersion in an interesting way (see right panel figure
\ref{transLSCO}). As in STO we see that in the normal state the
dispersion is linear and extrapolates to zero. In the
superconducting state the dispersion has acquired a $q^{2}$
dependence and no longer extrapolates to zero for $q\to0$. This is
the result of the opening of the superconducting gap and it
implies that the polaritons in the superconducting state have
acquired a mass. This is an example of the Anderson-Higgs
mechanism\cite{anderson}, the same mechanism via which the
Higgs-field gives a mass to the W and Z bosons in elementary
particle physics. In the superconductor the order parameter plays
the role of the Higgs-field and the spontaneously broken symmetry
is that of the U(1) gauge symmetry.

\subsection{TeraHertz time-domain spectroscopy}
This relatively new technique is the last we will discuss here.
\begin{figure}[htb]
\includegraphics[width=8.5 cm]{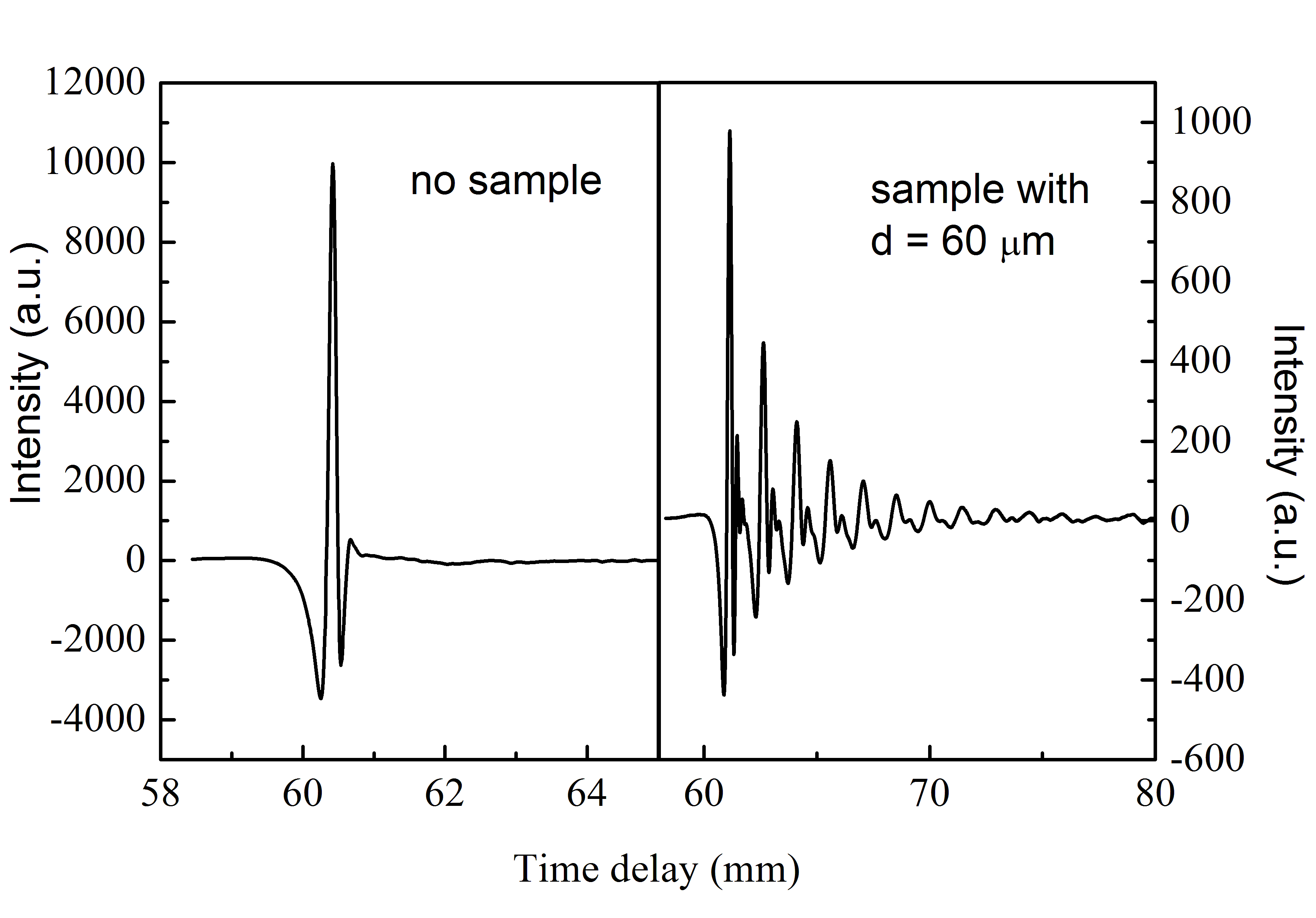}
\caption{\label{THz}Left: recorded signal v.s. delay distance
without sample. Right: recorded signal v.s. delay distance with
sample. Note the extra peaks in the signal on the right due to
multiple reflections in the sample.}
\end{figure}
This technique uses a powerful laser pulse and records the
detector output as a function of time, more often expressed in an
optical delay distance. The result for an experiment in vacuum is
shown in figure \ref{THz} on the left. If we now insert a sample
that is transparent to terahertz radiation in the path of the beam
we expect that due to the different optical path length in the
sample the pulse will arrive at a later time. In fact, if we use a
sample with to plane parallel surfaces we expect a series of peaks
due to multiple reflections in the sample (see right panel of
figure \ref{THz}. These peaks are just a different manifestation
of the Fabry-Perot oscillations observed in transmission
spectroscopy. By Fourier transforming this signal to the frequency
domain and doing the same for the signal without sample we can
again obtain the transmission spectrum. The frequency domain
spectrum corresponding to the time domain spectra of figure
\ref{THz} is shown in figure \ref{STO}.

\section{Quantum theory}
\begin{figure}[tbh]
\includegraphics[width=8.5 cm]{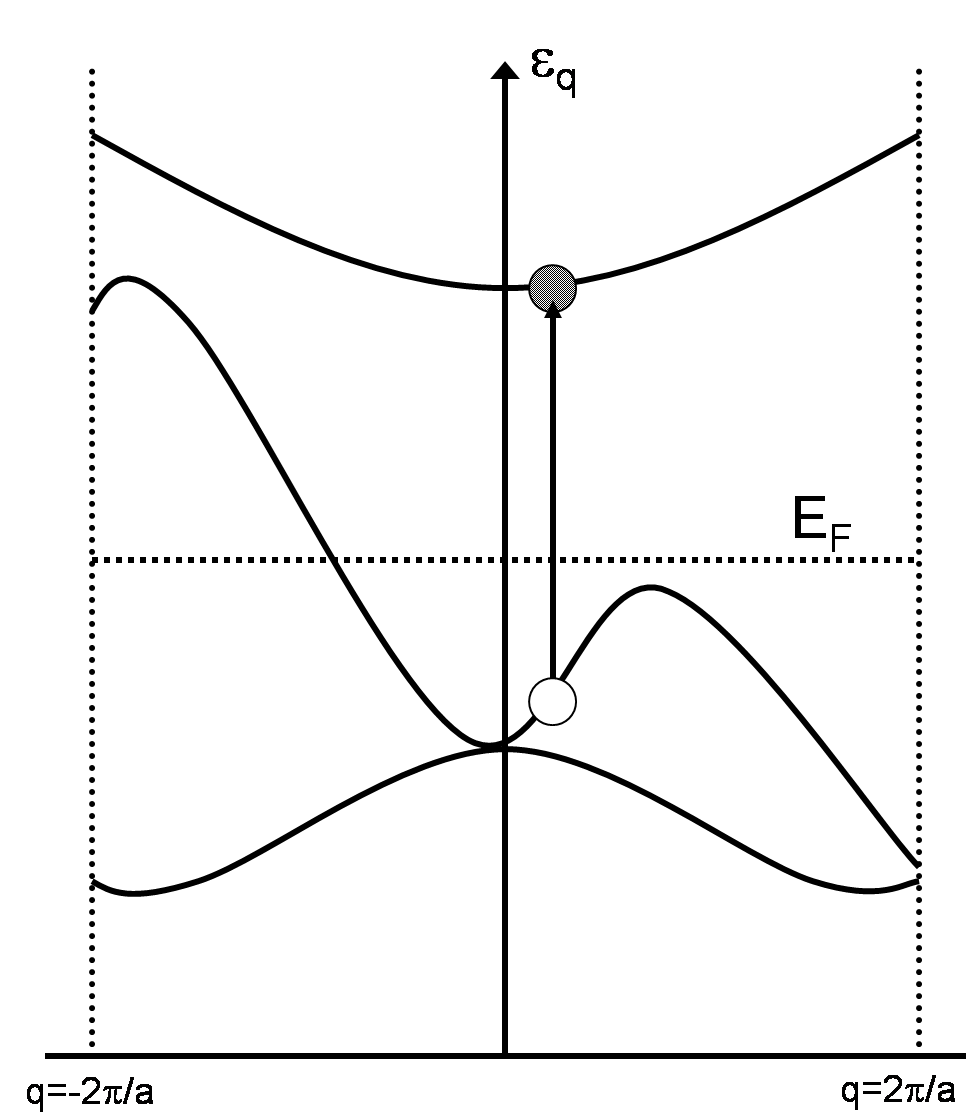}
\caption{\label{bandstruct}The indicated transition is an
interband transition. Al states below the dashed line indicated by
E$_{F}$ are occupied all states above are empty.}
\end{figure}

\begin{figure}[bth]
\includegraphics[width=6 cm]{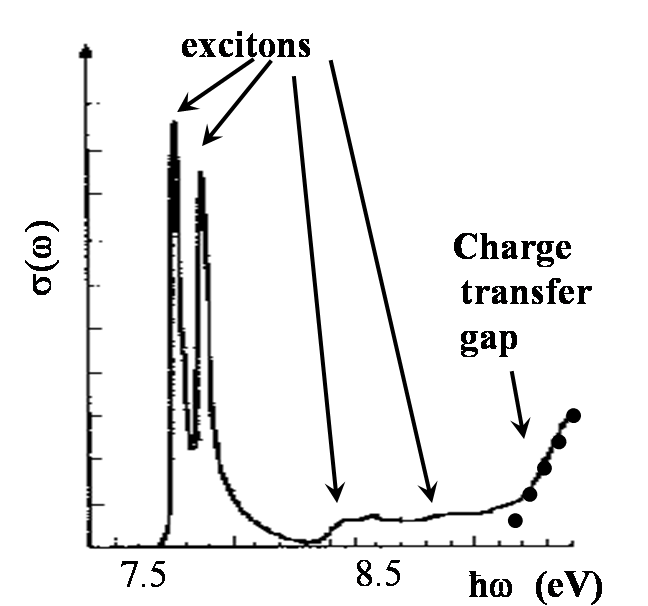}
\caption{\label{KCL}Optical conductivity of KCl. The series of
strong peaks are due to excitons. The onset in absorption
around 9 eV is the onset of charge transfer excitations.}
\end{figure}

We now move to the quantum theoretical description of the
interaction of light and matter using the Kubo-formalism. So far
we have been using a "geometrical" or macroscopic view of this
interaction, but in this section we will consider the effects of
the absorption of photons by electrons. Consider for simplicity a
metal. The electronic states of the system are described by a set
of bands, some of which are fully occupied, some partially and the
rest empty, figure \ref{bandstruct}.
When photons interact with these band electrons they can be
absorbed and in this process the electron is excited to a higher
lying band leaving behind a hole. In this way we create
electron-hole pairs and this (dipole) transition from a state
$|\Psi_{\nu}^{N}\rangle$ to a state $|\Psi_{\mu}^{N}\rangle$ is
characterized by an optical matrix element,
\begin{equation}
M_{\mu \nu } (\vec q) = \left\langle {\Psi _\mu ^N } \right.\left|
{{\bf \hat v}_q } \right|\left. {\Psi _\nu ^N } \right\rangle.
\end{equation}
If the transition is from one band to another band we call the
transition an \textit{interband} transition and if the
transition is within a band we call it a \textit{intraband}
transition. In figure \ref{KCL} we show the optical
conductivity of KCl. In this compound a strong onset is seen in
the optical conductivity around $\approx$8.7 eV. This onset is
due to the excitation of electrons from the occupied p-band
related to the Cl$^{-}$ ions to the unoccupied s-band of the
K$^{+}$ ions. Since this particular transition involves moving
charge from the chlorine atoms to the potassium atoms this type
of excitation is called a charge transfer (CT) excitation
\cite{zaanen}.
Another important feature in figure \ref{KCL} are the strong
peaks seen around 7.5 eV. Many theories often neglect so-called
vertex corrections because these corrections cancel if the
interactions between electrons are isotropic. However in real
materials interactions are more often than not anisotropic and
this means that these corrections have to be taken into
account. The peaks seen in figure \ref{KCL} are due to
transitions from bound states of electron-hole pairs, called
excitons, which arise due to the vertex corrections. Before we
start our display of the Kubo-formalism we first introduce some
notation. We introduce the field operators,
\begin{equation}
\psi^{\dagger}_{\sigma}(\mathbf{r})=\sum_{k}e^{-i\mathbf{k}\cdot\mathbf{r}}\hat{c}^{\dagger}_{k,\sigma}.
\end{equation}
The density operator is given by,
\begin{equation}
\hat{n}_{\sigma}(\mathbf{r})=\psi^{\dagger}_{\sigma}(\mathbf{r})\psi_{\sigma}(\mathbf{r}).
\end{equation}
The Fourier transform of $\hat{n}_{\sigma}(\mathbf{r})$ is,
\begin{equation}
\hat{n}_{\sigma}(\mathbf{r})=\frac{1}{V}\sum_{q}e^{-i\mathbf{q}\cdot\mathbf{r}}\rho_{q},
\end{equation}
with
\begin{equation}
\rho_{q}=\sum_{k,\sigma}\hat{c}^{\dagger}_{k-q/2,\sigma}\hat{c}_{k+q/2,\sigma}.
\end{equation}
The velocity operator is defined as,
\begin{equation}
\mathbf{\hat{v}}_{q}=\frac{\hbar}{m}\sum_{k,\sigma}\mathbf{k}\hat{c}^{\dagger}_{k-q/2,\sigma}\hat{c}_{k+q/2,\sigma}.
\end{equation}
Finally, we note that the operators $\hat{n}_{\sigma}(\mathbf{r})$
and $\mathbf{\hat{v}}_{q}$ satisfy,
\begin{equation}
\frac{i}{\hbar}\left[\hat{n}_{\sigma}(\mathbf{r}),\hat{H}\right]+\nabla\cdot\mathbf{\hat{v}}_{q}=0.
\end{equation}

\subsection{The Kubo-formalism}
To calculate the optical conductivity from a microscopic starting
point we have to add to the Hamiltonian of the system a term that
describes the interaction with the electromagnetic field described
by,
\begin{equation}\label{Eop}
{\bf E}^T ({\bf r},t) = \frac{{i\omega }}{c}\sum\limits_q {{\bf
A}_q e^{i({\bf q} \cdot {\bf r} - \omega t)} },
\end{equation}
Note that we have chosen the transverse gauge which we will use
throughout the rest of the chapter. The interaction Hamiltonian is
given by,
\begin{equation}
H' =  - \frac{{e\hbar }}{c}\sum\limits_q {e^{i({\bf q} \cdot {\bf
r} - \omega t)} {\bf A}_q  \cdot {\bf \hat v}_{ - q} },
\end{equation}
and in the presence of an electromagnetic field we use the minimal
coupling,
\begin{equation}
{\bf \hat v}_q \to{\bf \hat v}_q  - \frac{{e\hbar }}{{mc}}{\bf
A}_q e^{i({\bf q} \cdot {\bf r} - \omega t)} \hat \rho _q.
\end{equation}
We now start by examining the current operator
$\textbf{J}(\textbf{r},t)=\textbf{J}^{(1)}(\textbf{r},t)+\textbf{J}^{(2)}(\textbf{r},t)$.
It consists of two terms the first of which is called the
diamagnetic term,
\begin{equation}\label{J1}
{\bf J}^{(1)} (\mathbf{r},t) =  - \frac{{ne^2 }}{{mc}}{\bf
A}(r,t)=\frac{{ine^2 }}{{m\omega}}{\bf E}^{T}(r,t),
\end{equation}
where in the last equality we have used Eq. (\ref{Eop}). The
second term is more difficult. It is given by,
\begin{equation}
{\bf J}^{(2)} (\mathbf{r},t) = \frac{{e^2 }}{V}\int\limits_{ -
\infty }^{\rm t} {\left\langle {{\rm e}^{{\rm i}H'\tau } {\rm
e}^{{\rm  - i}H\tau } {\bf \hat v}(r,t){\rm e}^{{\rm i}H\tau }
{\rm e}^{{\rm - i}H'\tau } } \right\rangle {\rm e}^{{\rm i}\omega
\left( {{\rm t - }\tau } \right)} {\rm d}\tau } {\rm  }.
\end{equation}
We make here the approximation of using linear response theory: we
expand the exponentials $e^{iH'\tau}$ to first order in
$\textbf{A}(\textbf{r},t)$ and then stop the series expansion.
After some algebra we arrive at,
\begin{equation}\label{J2}
\frac{J^{(2)}}{\mathbf{E}(\mathbf{r},t)}=\frac{ie^2}{\omega
V}\sum_{n}\mathbf{v}_{-q}^{nm}\mathbf{v}_{q}^{mn}\left[\frac{1}{\omega-E_{n}+E_{m}+i0^{+}}-\frac{1}{\omega+E_{n}-E_{m}+i0^{+}}\right],
\end{equation}
where we have defined,
\begin{equation}
\mathbf{v}_{q}^{mn}\equiv\langle\Psi_{m}|\mathbf{\hat{v}}_{q}|\Psi_{n}\rangle.
\end{equation}
The result we have obtained is for zero temperature but is easily
generalized to finite $T$ if we use the grand canonical ensemble.
Combining Eq. (\ref{J1}) and Eq. (\ref{J2}) we find for the
optical conductivity,
\begin{equation}\label{conduc}
\sigma_{\alpha,\alpha}(\mathbf{q},\omega)=\frac{{iNe^2
}}{{mV\omega}}+\frac{{ie^2 }}{{V\omega}}\sum_{n,m\neq
n}e^{\beta(\Omega-E_{n})}\left[\frac{v_{\alpha,q}^{nm}
v_{\alpha,-q}^{mn}}{\omega-\omega_{mn}+i\eta}-\frac{v_{\alpha,-q}^{nm}
v_{\alpha,q}^{mn}}{\omega+\omega_{mn}+i\eta}\right],
\end{equation}
where we have defined $\omega_{mn}\equiv E_{m}-E_{n}$. The optical
conductivity consists now of three contributions: the diamagnetic
term followed by a contribution to positive frequencies and a
contribution to negative frequencies. We note that in general
$\sigma_{\alpha,\alpha}(\mathbf{q},\omega)$ is a tensor as indicated
by the $\alpha$ subscripts. We further note that the diamagnetic
term does not give a real contribution to the conductivity. This
term gives a $\delta$-function contribution at zero frequency and
this is exactly canceled by a delta function in the second part.
This can be seen by using the fact that for every $n$ we have the
following relationship,
\begin{equation}\label{velsum}
\sum_{n,m\neq n}\frac{v_{\alpha,q}^{nm}
v_{\alpha,-q}^{mn}}{\omega_{mn}}=\frac{N}{2m}.
\end{equation}
So we can rewrite Eq. (\ref{conduc}) as,
\begin{equation}\label{cond}
\sigma_{\alpha,\alpha}(\mathbf{q},\omega)=\frac{{ie^2}}{{V}}\sum_{n,m\neq
n}\frac{e^{\beta(\Omega-E_{n})}}{\omega_{mn}}\left[\frac{v_{\alpha,q}^{nm}
v_{\alpha,-q}^{mn}}{\omega-\omega_{mn}+i\eta}+\frac{v_{\alpha,-q}^{nm}
v_{\alpha,q}^{mn}}{\omega+\omega_{mn}+i\eta}\right],
\end{equation}
From here on we take the limit $q\to0$ and define a generalized
oscillator strength $\Omega_{mn}$ as,
\begin{equation}\label{Omeganm}
\Omega_{mn}^{2}\equiv\frac{8\pi
e^{2}e^{\beta(\Omega-E_{n})}|v^{nm}_{\alpha}|^{2}}{\omega_{mn}V}.
\end{equation}
With this definition we are lead to the Drude-Lorentz expansion of
the optical conductivity,
\begin{equation}\label{DL}
\sigma_{\alpha,\alpha}(\omega)=\frac{i\omega}{4\pi}\sum_{n,m\neq
n}\frac{\Omega_{mn}^{2}}{\omega(\omega+i\gamma_{mn})-\omega^{2}_{mn}}.
\end{equation}

\subsection{Sum Rules}
Sum rules play an important role in optics. Using the equations of
the previous section we derive the Thomas-Reich-Kuhn sum rule also
known as the f-sum rule. The f-sum rule states that, apart from
some constants, the area under $\sigma_{1}$ is proportional to the
number of electrons and inversely proportional to their mass. This
can be shown as follows: integrating Eq. (\ref{DL}) we have,
\begin{equation}
{\mathop{\rm Re}\nolimits} \int\limits_{ - \infty }^\infty {\sigma
\left( \omega  \right)d\omega  = \frac{1}{4}} \sum\limits_{n,m \ne
n} {\Omega _{_{mn} }^2 }.
\end{equation}
Using the expression for $\Omega_{mn}$, Eq. (\ref{Omeganm}), and
expression (\ref{velsum}) we can rewrite the sum on the right hand
side as,
\begin{equation}
\sum\limits_{n,m \ne n} {\Omega _{_{mn} }^2 }  = \frac{{4\pi e^2
N}}{{mV}}\sum\limits_n {e^{\beta \left( {\Omega  - E_n } \right)}
= } \frac{{4\pi e^2 N}}{{mV}}.
\end{equation}
So the f-sum rule states that,
\begin{equation}
\int_{-\infty}^{\infty}\sigma_{1}(\omega)d\omega=\frac{{\pi e^2
N}}{{mV}},
\end{equation}
\begin{figure}[tbh]
\includegraphics[width=8.5 cm]{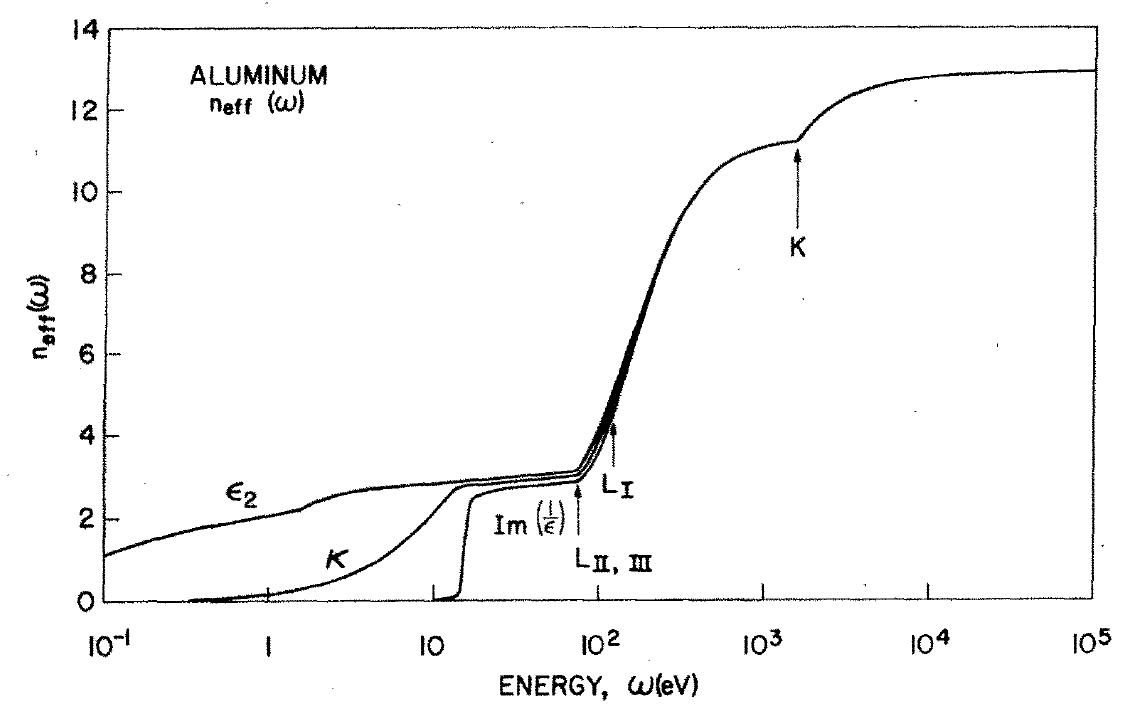}
\caption{\label{ALsumrule}Effective number of carriers
$n_{eff}(\Omega_{c})$ as a function of cutoff frequency
$\Omega_{c}$ for Al. Figure adapted from \citep{smith}.}
\end{figure}
as promised. This is the full universal sum rule. It is often
rewritten as an integral over positive frequencies only and
using the definition of the plasma frequency $\omega_{p}$,
\begin{equation}
\omega_{p}^{2}\equiv\frac{{4\pi e^2 N}}{{mV}},
\end{equation}
as,
\begin{equation}
\int_{0}^{\infty}\sigma_{1}(\omega)d\omega=\frac{\omega_{p}^{2}}{8}.
\end{equation}
We can also define \textit{partial} sum rules, i.e. sum rules
where we integrate up to a certain frequency cutoff $\Omega_{c}$.
In such a case the sum rule is not universal (this means for
instance that the value of this sum rule can depend on
temperature) and we can now define a plasma frequency that depends
on the chosen cutoff frequency,
\begin{equation}
\omega_{p,valence}^{2}\equiv\frac{4\pi
e^{2}}{m}n_{eff}(\Omega_{c}).
\end{equation}
A nice example of the application of the partial sum rule is shown
in figure \ref{ALsumrule}. Here the partial sum rule is applied to
the optical conductivity of aluminum \cite{smith}. Here the
effective number of carriers contributing to the sum rule is
plotted as a function of $\Omega_{c}$. $n_{eff}(\Omega_{c})$
slowly increases to a value of roughly three around 50 eV. This
means that as we increase the cutoff from zero to 50 eV we are
slowly integrating over the intraband transitions and when we
reach a value of 50 eV we have integrated over all transitions
involving the three valence electrons. For higher energies the
interband transitions start to contribute with a sharp onset near
80 eV. Finally at 10$^{4}$ eV the sum rule saturates at 13
electrons, the total number of electrons of aluminum.
\begin{figure}[tbh]
\includegraphics[width=8.5 cm]{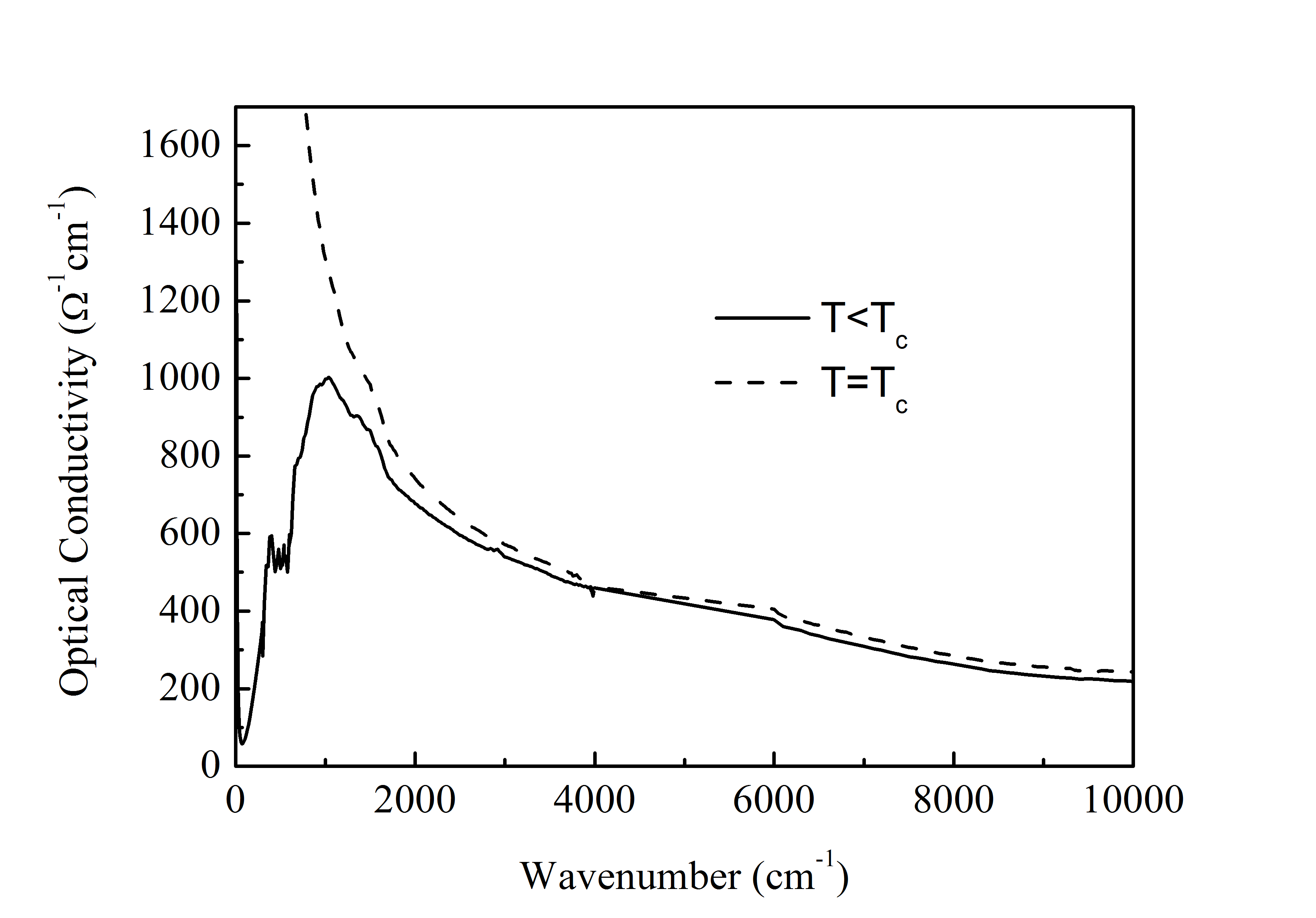}
\caption{\label{sigmaSC}Optical conductivity of Bi-2212 at $T_{c}$
and below. The difference in area between the two curves is an
estimate of the superfluid density.}
\end{figure}

Another application of sum rules can be found in superconductors.
In a superconductor the electrons form a superfluid condensate.
This condensate shows up in the optical conductivity as a delta
function at zero frequency (it contributes a diamagnetic term as
in Eq. (\ref{conduc})). At the same time a gap opens up in low
frequency part of the spectrum where the optical conductivity is
(close to) zero, see figure \ref{sigmaSC}. In the normal state the
system is usually metallic and characterized by a Drude peak. In
optical experiments we cannot measure the zero frequency response
and so we cannot directly measure the spectral weight
$\omega_{p,s}^{2}$ of the condensate. However, using sum rules we
can estimate its spectral weight because the total spectral weight
has to remain constant. This is summarized in the
Ferrel-Glover-Tinkham (FGT) sum rule \cite{ferrel}, which states
that the difference in spectral weight between the optical
conductivity in the superconducting and normal state is precisely
the spectral weight of the condensate,
\begin{equation}
\omega _{p,s} (T)^2  = 8\int\limits_{0^ +  }^\infty  {\left\{
{\sigma (\omega ,T_c ) - \sigma (\omega ,T)} \right\}d\omega }.
\end{equation}
Note that we integrate here from 0$^{+}$.

There also exist sum rules for mixtures of different types of
particles,
\begin{equation}\label{phonsum}
\int\limits_0^\infty  {\sigma _1 \left( {\omega '} \right)}
d\omega ' = \sum\limits_j {\frac{{\pi n_j q_j ^2 }}{{2m_j }}},
\end{equation}
\begin{figure}[tbh]
\includegraphics[width=8.5 cm]{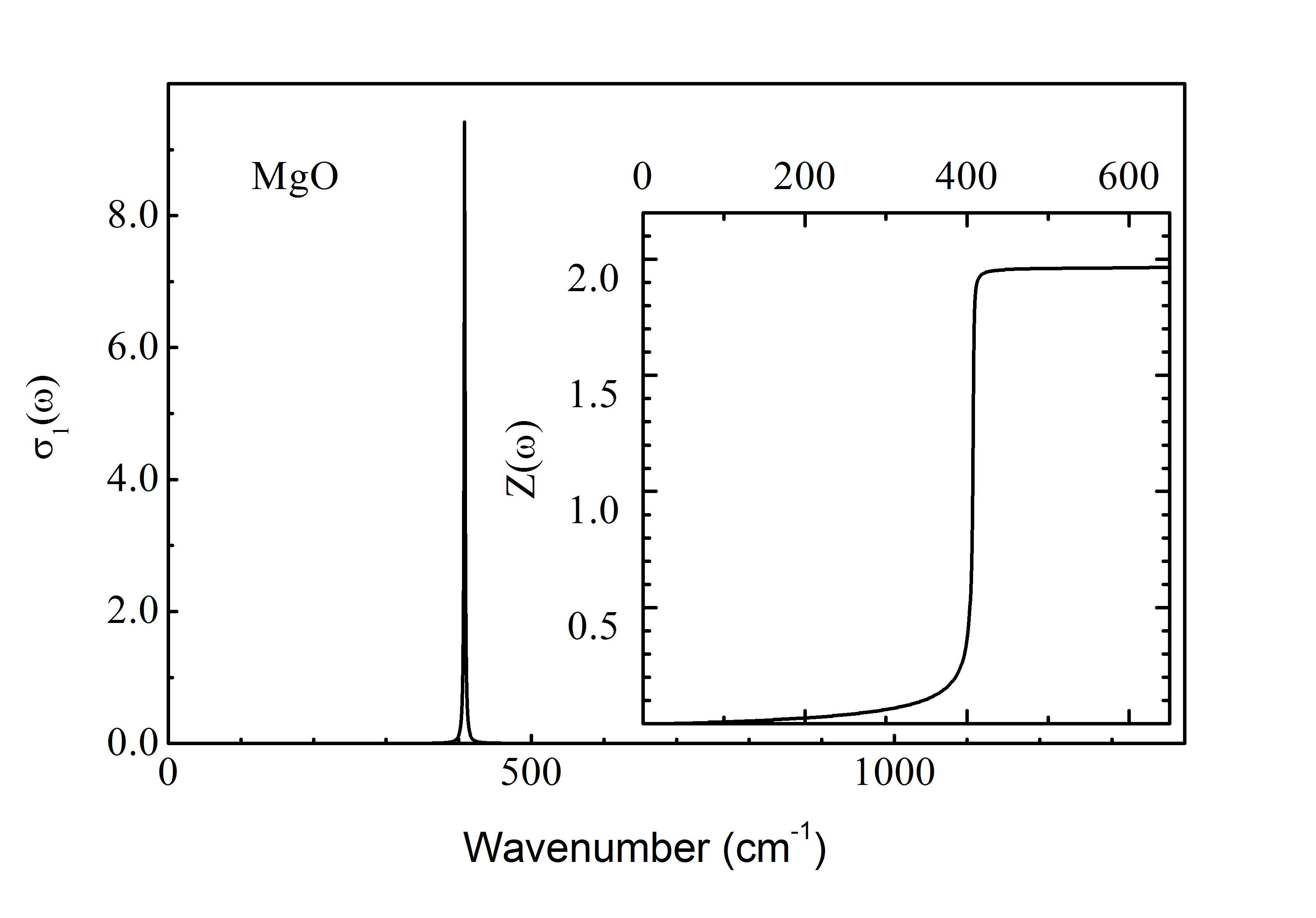}
\caption{\label{MgO}Optical conductivity due to phonon mode in
MgO. The area under the peak is proportional to the effective
charge of the mode. The inset shows the effective charge
calculated using (\ref{phonsum2}). Data from \citep{damascelli}.}
\end{figure}
here the index j labels the different species. This sum rule can
be applied to measure the charge of ions involved in vibrational
modes. If we can separate the contribution to the optical
conductivity due to the optical modes we can invert Eq.
(\ref{phonsum}) to calculate the effective charge related to the
mode. For example, in MgO (figure \ref{MgO}) both ions contribute
an equal charge $q_{Mg}=-q_{O}$. We define the effective mass
$\mu$ as $\mu^{-1}=m_{Mg}^{-1}+m_{O}^{-1}$ and assume that the
density of the two is equal. In that case we can rewrite Eq.
(\ref{phonsum}) as,
\begin{equation}\label{phonsum2}
Z(\omega )^2  \equiv \left( {\frac{{q_T^* (\omega )}}{e}}
\right)^2 \equiv \frac{2\mu}{{\pi ne^2
}}\int\limits_{\omega_{min}}^{\omega_{max}} {\sigma _{ph} \left(
{\omega '} \right)d\omega '},
\end{equation}
where the integral has to be taken in a frequency range such that
it includes the spectral weight of the optical phonon mode but
nothing else.

We will now derive expressions for the conductivity sum rule from
a more microscopic point of view. To do that we return to the Kubo
expression for the optical conductivity,
\begin{equation}
\sigma _1 \left( \omega  \right) = \frac{{\pi e^2
}}{V}Tr\left\langle {\Psi _n } \right|{\bf \hat v}\left\{
{\frac{{\delta \left( {\omega  - \hat H + E_n } \right)}}{{\hat H
- E_n }} + \frac{{\delta \left( {\omega  + \hat H - E_n }
\right)}}{{\hat H - E_n }}} \right\}{\bf \hat v}\left| {\Psi _n }
\right\rangle.
\end{equation}
The Hamiltonian in this expression is that of the system of
interacting electrons without the interaction of light. It
represents the optical conductivity for the system in an arbitrary
(ground or excited) many-body state $|\Psi\rangle$. A peculiar
point of this expression is that although the velocity operators
create a single electron-hole pair, due to the fact that the
hamiltonian in the denominator of this expression still contains
the interactions between all particles in the system, the optical
conductivity represents the response from the full collective
system of electrons. If we integrate this expression over
frequency we get,
\begin{equation}\label{microsum}
\int\limits_{ - \infty }^\infty  {\sigma _1 \left( \omega
\right)d\omega }  = \frac{{2\pi e^2 }}{V}Tr\left\langle {\Psi _n }
\right|{\bf \hat v}\frac{1}{{\hat H - E_n }}{\bf \hat v}\left|
{\Psi _n } \right\rangle.
\end{equation}
We now take a closer look at the right-hand side of this
expression. Remember that,
\begin{equation}\label{commutator}
{\bf \hat v} = \frac{i}{\hbar }\left[ {\hat H,{\bf \hat x}}
\right].
\end{equation}
Using the commutator we can rewrite,
\begin{equation}
 - 2i\hbar {\bf \hat v}\frac{1}{{\hat H - E_n }}{\bf \hat v} = \left( {\hat H{\bf \hat x} - {\bf \hat x}\hat H} \right)\frac{1}{{\hat H - E_n }}{\bf \hat v} + {\bf \hat v}\frac{1}{{\hat H - E_n }}\left( {\hat H{\bf \hat x} - {\bf \hat x}\hat H}
 \right).
\end{equation}
Inserting this back into Eq. (\ref{microsum}) we find after some
rearranging
\begin{equation}\label{sumrule}
\int\limits_{ - \infty }^\infty  {\sigma _1 \left( \omega
\right)d\omega }  = \frac{{i\pi e^2 }}{{\hbar V}}\left\langle
{[{\bf \hat v},{\bf \hat x}]} \right\rangle,
\end{equation}
where $\langle...\rangle$ stands for the trace over all many-body
states.
Here we have used that,
\begin{equation}
\mathbf{\hat{x}}\hat{H}\frac{1}{\hat{H}-E_{n}}\mathbf{\hat{v}}=\mathbf{\hat{x}}\left(\hat{H}-E_{n}\right)\frac{1}{\hat{H}-E_{n}}\mathbf{\hat{v}}+\mathbf{\hat{x}}E_{n}\frac{1}{\hat{H}-E_{n}}\mathbf{\hat{v}}=\mathbf{\hat{x}}\mathbf{\hat{v}}+\mathbf{\hat{x}}E_{n}\frac{1}{\hat{H}-E_{n}}\mathbf{\hat{v}},
\end{equation}
and the fact that $\hat{H}|\Psi_{n}\rangle=E_{n}|\Psi_{n}\rangle$.
We can now obtain different expressions for the sum rule by
working out the commutator on the right-hand side of Eq.
(\ref{sumrule}) based on different model assumptions. In table
\ref{modelexpr} we summarize some results.
\begin{table}[tbh]
\begin{tabular}{lrl}
\hline\hline Free electrons & \quad\quad & $[{\bf \hat v},{\bf
\hat x}]
=\frac{\hbar }{{im}}\sum\limits_{k\sigma } {\hat n_{k\sigma } }$\\
Band electrons & \quad\quad &  $[{\bf \hat v},{\bf \hat x}] =
\frac{\hbar }{{im}}\sum\limits_{k\sigma } {\hat n_{k\sigma } }
[{\bf \hat v},{\bf \hat x}] = \frac{\hbar }{i}\sum\limits_{k\sigma
} {\frac{{\partial ^2 \varepsilon _{k\sigma } }}{{\partial k^2
}}\hat n_{k\sigma } }$\\
N.N. & \quad\quad &  $[{\bf \hat v},{\bf \hat x}] =  -
\frac{{\hbar a^2 }}{i}\sum\limits_{k\sigma } {\varepsilon
_{k\sigma } \hat n_{k\sigma } }$\\
\hline\hline
\end{tabular}
\caption{Expressions for the commutator in Eq. (\ref{sumrule}) for
three different cases. N.N. stands for Nearest Neighbors tight
binding model} \label{modelexpr}
\end{table}
The sum rule for band electrons is in practice the most useful.
Suppose that we have a system with only a single reasonably well
isolated band around the Fermi level that can be approximated by a
tight binding dispersion $\varepsilon _k  =  - t\cos \left( {ka}
\right)$. In that case we find an interesting relation,
\begin{equation}
\int\limits_0^{\Omega _c } {\sigma _1 (\omega ,T)} d\omega  =  -
\frac{{\pi e^2 a^2 }}{{2\hbar ^2 V}}\sum\limits_{k,\sigma }
{\left\langle {\hat n_{k\sigma } \varepsilon _k } \right\rangle _T
}  =  - \frac{{\pi e^2 a^2 }}{{2\hbar ^2 V}}E_{kin} (T).
\end{equation}
This sum rule states that by measuring the optical spectral weight
we are in fact measuring the kinetic energy of the charge carriers
contributing to the optical conductivity. In real systems this
relation only holds approximately: usually there are other bands
lying nearby and the integral on the left contains contributions
from these as well. Often the bands are described by more
complicated dispersion relations in which case the relation
$\partial^{2}\varepsilon_{k}/\partial k^{2}=-\varepsilon_{k}$ does
not hold. We can make some other observations from the sum rule
for band electrons. Suppose again we have a single empty cosine
like band (it is only necessary that the band is symmetric but it
simplifies the discussion) at $T=0$. Since the band is empty, the
spectral weight is equal to zero. If we start adding electrons the
spectral weight starts to increase until we reach half-filling. If
we add more electrons the spectral weight will start to decrease
again because the second derivative becomes negative for
$k>\pi/2a$. If we completely fill the band the contributions from
$k>\pi/2a$ will precisely cancel the contributions from $k<\pi/2a$
and the spectral weight is again zero. Now consider what happens
if we have a half-filled band and start to increase the
temperature. Due to the smearing of the Fermi-Dirac distribution
higher energy states will get occupied leaving behind lower energy
empty states. The result of this is that the spectral weight
starts to decrease. One can show using the Summerfeld expansion
that the spectral weight follows a $T^{2}$ temperature dependence.
In the extreme limit of $T\to\infty$ something remarkable happens:
the Fermi-Dirac distribution is 1/2 everywhere and the electrons
are equally spread out over the band. The metal has become an
insulator!

\subsection{Applications of sum rules to superconductors}
Before we have a look at some applications of sum rules to
superconductors we first summarize some results from BCS theory.
We want to apply our ideas to cuprate superconductors so we use a
modified version from the original theory to include the
possibility of d-wave superconductivity. In other words we suppose
that there is some attractive interaction between the electrons
that has a momentum dependence. The energy difference between the
normal and superconducting state due to interactions can be
written as \cite{marel},
\begin{equation}
\langle\hat{H}_{s}^{int}\rangle-\langle\hat{H}_{s}^{int}\rangle=\int
d^{3}r g(r)V(r)=\sum_{k}g_{k}V_{k},
\end{equation}
where $g(r)$ and $g_{k}$ are the pair correlation function and its
fourier transform respectively.
\begin{figure}[tbh]
\includegraphics[width=6 cm]{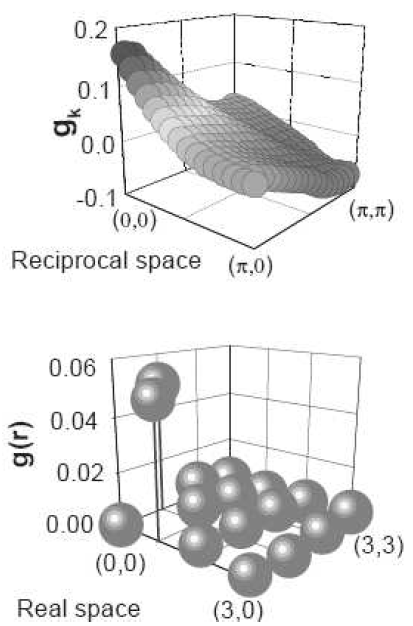}
\caption{\label{correlation} Real and momentum space picture of
the correlation functions $g(r)$ and $g_{k}$. Figure adapted from
\citep{marel}.}
\end{figure}
We can find an expression for $g_{k}$,
\begin{equation}
g_{k}=\sum_{q}\frac{\Delta_{q+k}\Delta^{*}_{q}}{4E_{q+k}E_{q}}.
\end{equation}
As usual,
\begin{equation}
E_{k}=\sqrt{(\varepsilon_{k}-\mu)^{2}+\Delta^{2}},
\end{equation}
and the temperature dependence of $\Delta_{k}$ is given by,
\begin{equation}
\Delta_{k}=\sum_{q}\frac{V_{q}\Delta_{q}}{2E_{q}}\tanh\left(\frac{E_{k}}{2k_{b}T}\right).
\end{equation}
We now use a set of parameters extracted from ARPES measurements
to do some numerical simulations. First of all we calculate
$g_{k}$ and fourier transform it to obtain $g(r)$. The results are
shown in figure \ref{correlation}.

\begin{figure}[tbh]
\includegraphics[width=8.5 cm]{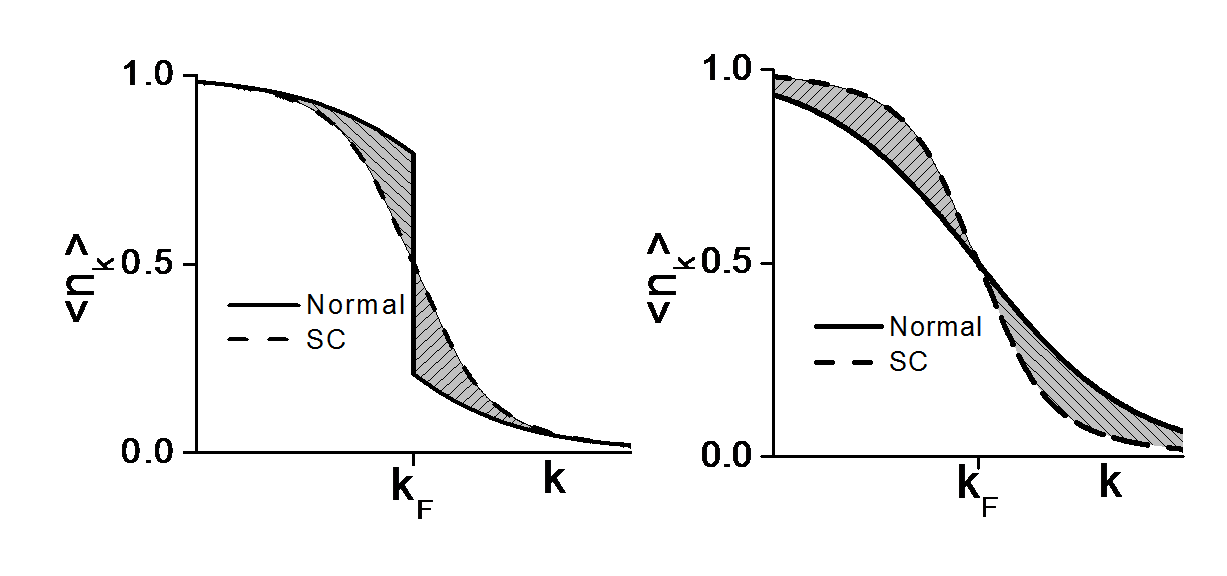}
\caption{\label{distribution} Left: Distribution function for the
normal (Fermi liquid like) state and the superconducting state.
Right: Distribution function for a non-Fermi Liquid like state and
the superconducting state.}
\end{figure}

\begin{figure}[tbh]
\includegraphics[width=6 cm]{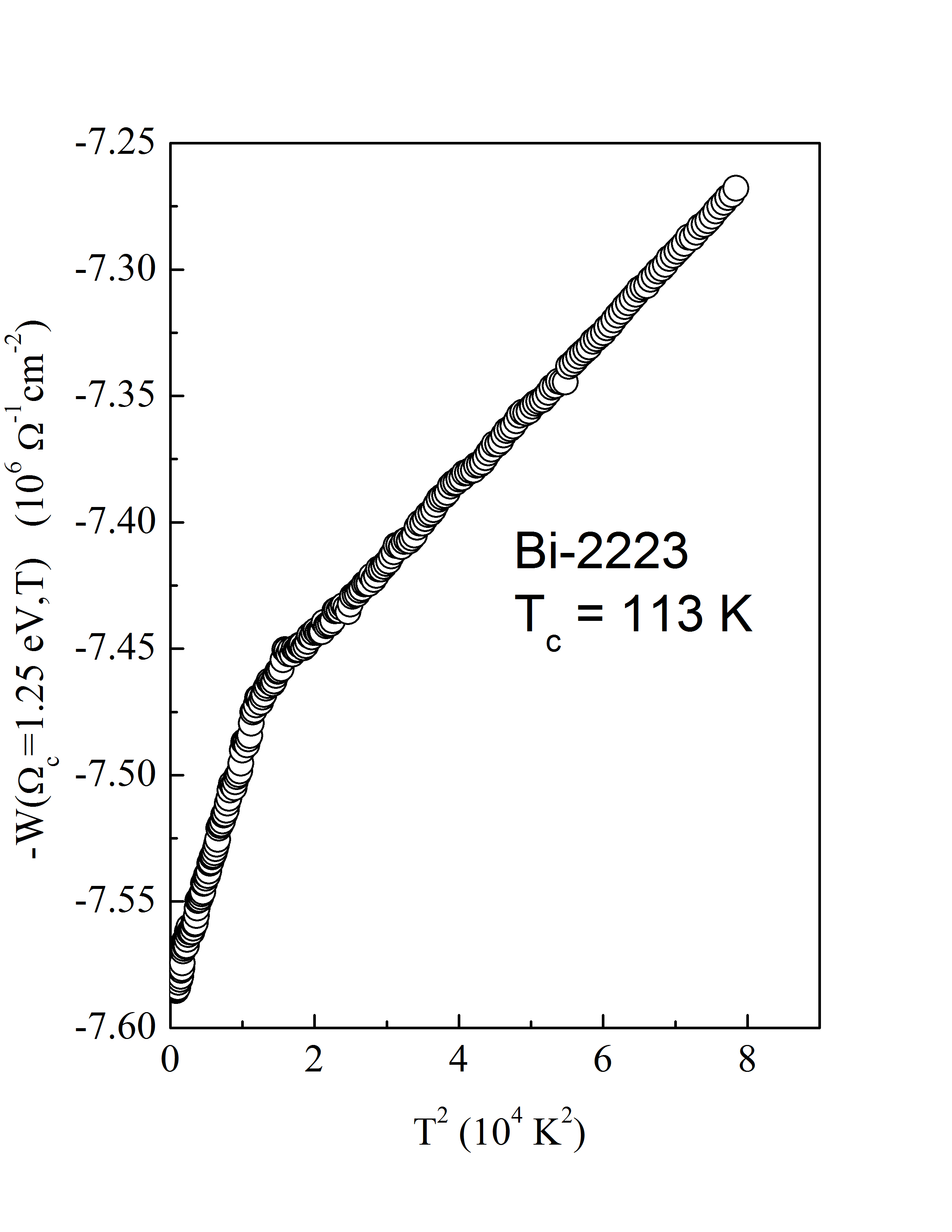}
\caption{\label{WBi2223} Temperature dependent spectral weight of
Bi-2223. Data taken from ref. \citep{carbone}.}
\end{figure}

Although $g_{k}$ is not so illuminating $g(r)$ is. This function
is zero at the origin and strongly peaked at the nearest neighbor
sites. This is a manifestation of the d-wave symmetry. We also
note that the correlation function drops off very fast for sites
removed further from the origin.
\begin{figure}[tbh]
\includegraphics[width=8.5 cm]{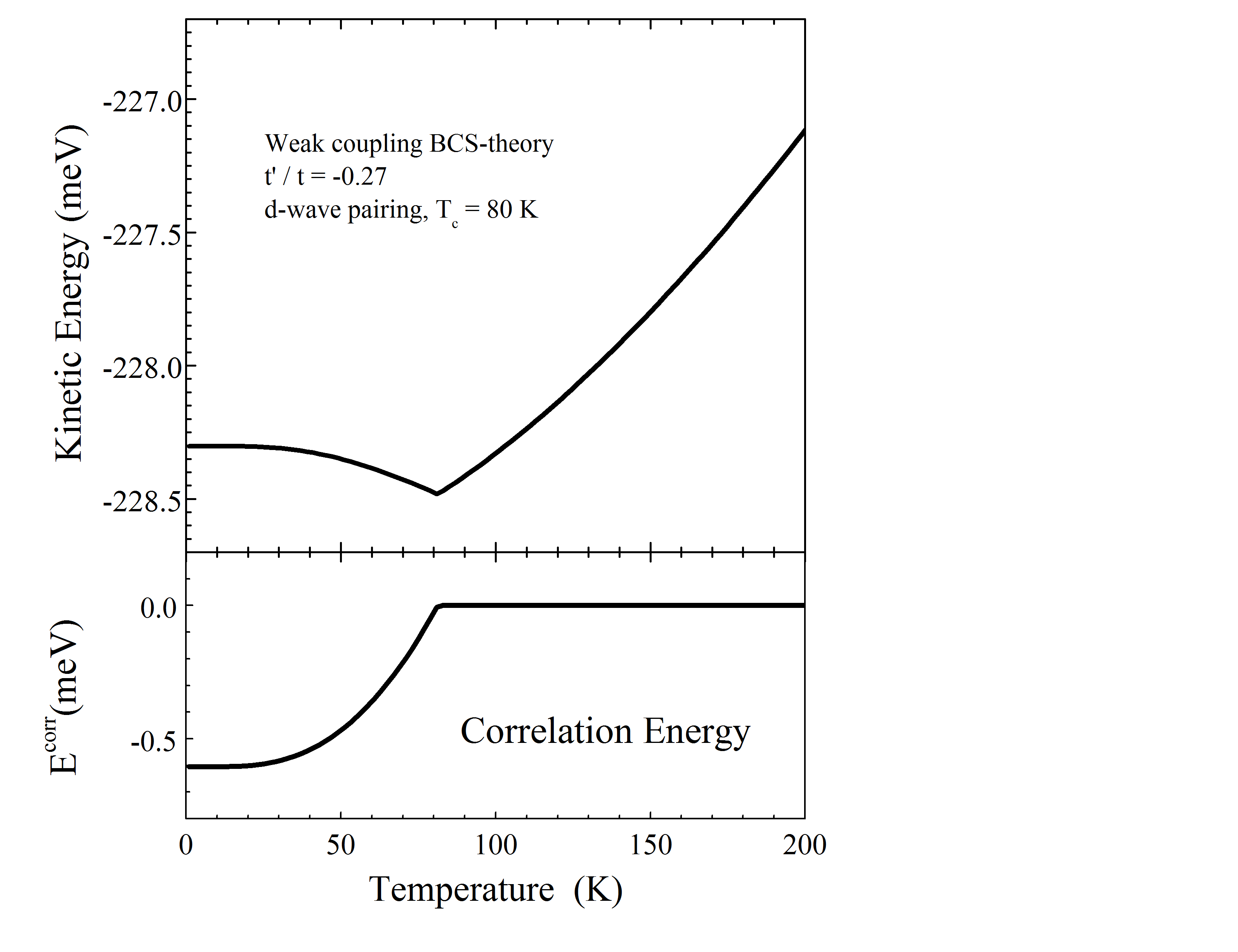}
\caption{\label{correnergy} Correlation energy and kinetic energy
as a function of temperature for a d-wave BCS superconductor.
Figure adapted from \citep{marel}.}
\end{figure}
In figure \ref{correnergy} we show the results for a calculation
of the correlation and kinetic energy using the parameters
extracted from ARPES measurements on Bi-2212. The kinetic energy
is calculated from,
\begin{equation}
\langle\hat{H}_{kin}\rangle=\sum_{k}\varepsilon_{k}\{1-\frac{\varepsilon_{k}-\mu}{E_{k}}\tanh\left(\frac{E_{k}}{2k_{b}T}\right)\}.
\end{equation}
We see that the kinetic energy \textit{increases} in the
superconducting state. This can be easily understood by looking at
what happens to the particle distribution function below $T_{c}$,
as indicated in the left panel of figure \ref{distribution}: when
the system enters the superconducting state the area below the
Fermi energy decreases and the area above the Fermi energy
increases thereby increasing the total kinetic energy of the
system.
Nevertheless the total internal energy, which is the sum of the
interaction energy and the kinetic energy, decreases and this is
of course why the system becomes superconducting. Now let us take
a look at what happens in the cuprates. In figure \ref{WBi2223} we
display the optical spectral weight $W(\Omega_{c},T)$ as a
function of $T^{2}$ for Bi-2223.
To compare this to the BCS kinetic energy we have plotted here
$-W(\Omega_{c},T)$. This result is contrary to the result from our
calculation: the kinetic energy decreases in the superconducting
state. This experimental result, observed first by Molegraaf
\textit{et al.} \cite{molegraaf}, has sparked a lot of interest
both experimentally \cite{heumen,basov,syro,kuz2,carbone} and
theoretically
\cite{hirsch,anderson2,eckl,wrobel,haule,toschi,marsiglio,mayer,norman}.
We note that DMFT calculations with the Hubbard model as starting
point have shown the same effect as observed here \cite{mayer}.
Roughly speaking the effect is believed to be due to the
"strangeness" of the normal state (right panel figure
\ref{distribution}). It is well known that the normal state of the
cuprates shows non Fermi-liquid behavior. So if the distribution
function in the normal state does not show the characteristic step
of the Fermi liquid at the Fermi energy but is rather a broadened
function of momentum it is very well possible that the argument we
made for the increase of the kinetic energy (see above) is
reversed.

\subsection{Applications of sum rules: the Heitler-London model}
Another interesting application of sum rules is that we can use
them in some cases to extract the hopping parameters of a system.
In order to see how this works we express the optical conductivity
at zero temperature,
\begin{equation}
\sigma _1 \left( \omega  \right) = \frac{{\pi e^2
}}{V}\left\langle {\Psi _g } \right|{\bf \hat v}\frac{{\delta
\left( {\omega  - \hat H + E_g } \right)}}{{\hat H - E_g }}{\bf
\hat v}\left| {\Psi _g } \right\rangle,
\end{equation}
in terms of the dipole operator. Here $|\Psi_{g}\rangle$ is the
groundstate of the system. To do this we make use of the
commutator Eq. (\ref{commutator}) and the insertion of a complete
set of states. After integrating over frequency we get
\begin{equation}\label{dipsumrule}
\int\limits_0^\infty  {\sigma _1 \left( \omega  \right)d\omega } =
\frac{{\pi e^2 }}{{\hbar ^2 V}}\sum\limits_n {\left( {E_n  - E_g }
\right)\left| {\left\langle n \right|{\bf \hat x}\left| g
\right\rangle } \right|^2 }.
\end{equation}
We note that this can be done only for finite system sizes. Now
consider the special case of a diatomic molecule with two energy
levels, one on each atom and a hopping parameter t and distance
$d$ between the two atoms. We also assume that there is a
splitting $\Delta$ between the two levels. The hamiltonian for
such a system is,
\begin{equation}
H = t{\rm  }\sum\limits_\sigma  {\left( {\psi _{L,\sigma }^t \psi
_{R,\sigma }  + \psi _{R,\sigma }^t \psi _{L,\sigma } } \right)} +
\frac{\Delta }{2}\left( {\hat n_R  - \hat n_L } \right) + U\left(
{\hat n_{L \uparrow } \hat n_{L \downarrow }  + \hat n_{R \uparrow
} \hat n_{R \downarrow } } \right).
\end{equation}
The indices $L$ and $R$ indicate the left and right atom
respectively. If we now put 1 electron in this system we have a
two-level problem that is easily diagonalized. As usual we make
bonding and anti-bonding states,
\begin{eqnarray}
\left| {\psi _{g,\sigma } } \right\rangle  = u\left| {\psi _{l,\sigma } } \right\rangle  + v\left| {\psi _{R,\sigma } } \right\rangle,  \\
\left| {\psi _{e,\sigma } } \right\rangle  = v\left| {\psi
_{l,\sigma } } \right\rangle  - u\left| {\psi _{R,\sigma } }
\right\rangle.
\end{eqnarray}
The coefficients $u$ and $v$ are given by,
\begin{equation}
u = \frac{1}{{\sqrt 2 }}\sqrt {1 + \frac{\Delta }{{E_{CT}
}}};\quad\quad v = \frac{1}{{\sqrt 2 }}\sqrt {1 - \frac{\Delta
}{{E_{CT} }}}.
\end{equation}
The bonding and anti-bonding states are split by an energy
$E_{CT}$,
\begin{equation}
E_{CT}  = \sqrt {\Delta ^2  + 4t^2 }.
\end{equation}
We are now in position to calculate the transition matrix element
appearing in Eq. (\ref{dipsumrule}). The position operator can be
represented by,
\begin{equation}
{\bf \hat x} = \frac{d}{2}{\rm  }\left( {\hat n_R  - \hat n_L }.
\right).
\end{equation}
So the matrix element is,
\begin{eqnarray}
\left\langle {\psi _{g,\sigma } } \right|{\bf \hat x}\left| {\psi
_{e,\sigma } } \right\rangle  =
\left(u\langle\Psi_{L}|+v\langle\Psi_{R}|\right)\frac{d}{2}{\rm
}\left( {\hat n_R  - \hat n_L }
\right)\left(u|\Psi_{L}\rangle-v|\Psi_{R}\rangle\right) \nonumber\\
=-\frac{d}{2}(uv)=- \frac{t}{{E_{CT} }}d
\end{eqnarray}
Using this in the sum rule Eq. (\ref{dipsumrule}) finally gives us
the spectral weight of this model,
\begin{equation}\label{tsum}
\int\limits_0^\infty  {\sigma _1 \left( \omega  \right)d\omega }
=\frac{{e^2 \pi d^2 }}{{\hbar ^2 V}}\frac{{t^2 }}{{\sqrt {\Delta
^2 + 4t^2 } }}.
\end{equation}
\begin{figure}[tbh]
\includegraphics[width=8.5 cm]{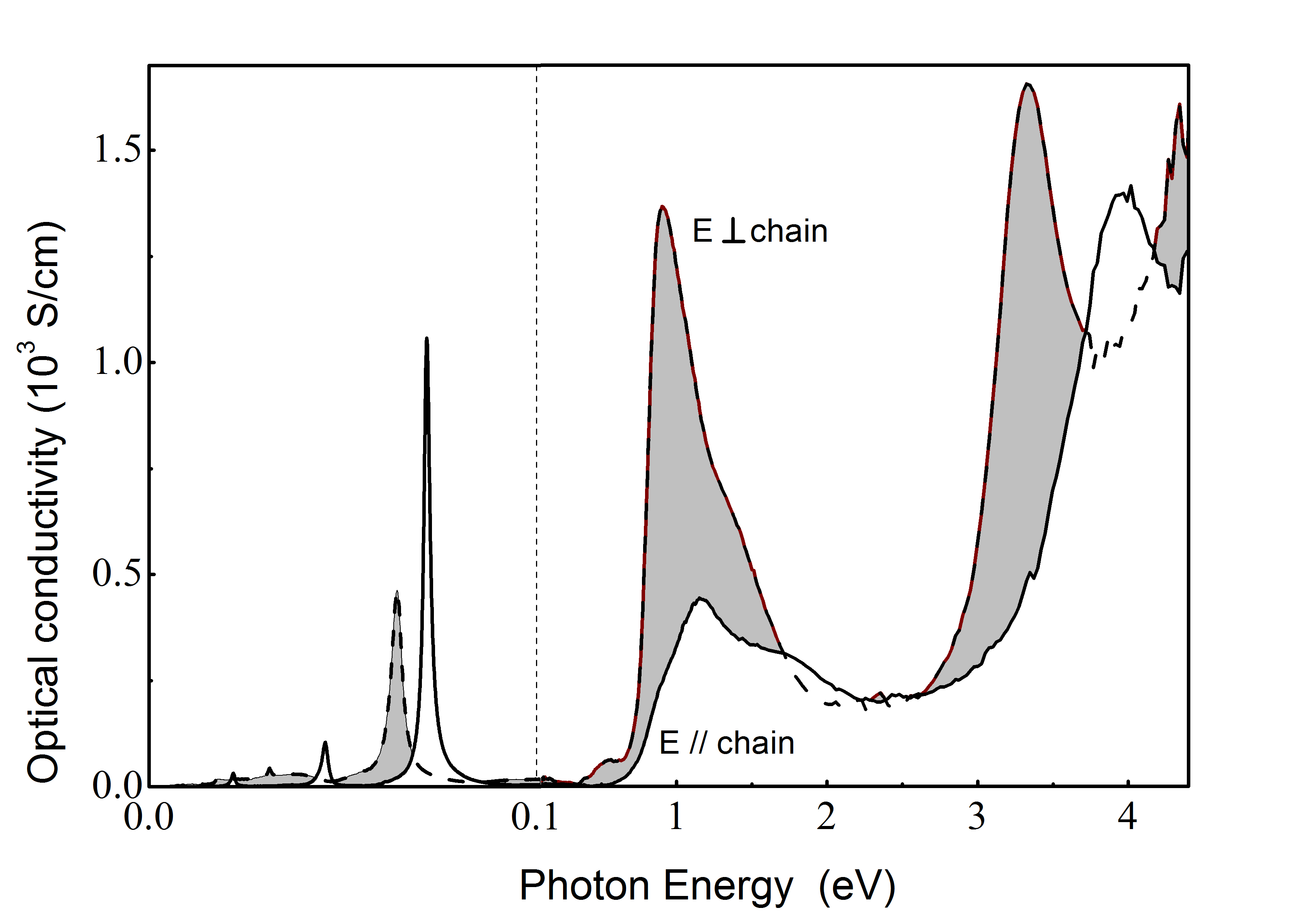}
\caption{\label{NaVO} Optical conductivity of
$\alpha$-NaV$_{2}$O$_{5}$ for two polarizations: one with the
field parallel to the chains and one perpendicular. Data taken
from ref. \citep{damascelli2}.}
\end{figure}
We see that there is a very simple relation between the spectral
weight of this model and the hopping parameter. This sumrule has
been applied to $\alpha$-NaV$_{2}$O$_{5}$ \cite{damascelli2}. This
compound is a so-called ladder compound. It consists of double
chains of vanadium atoms forming ladders which are weakly coupled
to each other. Each unit cell contains 4 V atoms and 2 valence
electrons. The vanadiums on the rungs of the ladder are more
strongly coupled than those along the legs, i.e. $t_{\perp}\gg
t_{\parallel}$. The Heitler-London model we have discussed above
can be applied to this system since each rung forms precisely a
two level system with different levels. The only difference that
we have to take into account is that this is a crystal consisting
of $N$ independent two level systems. Figure \ref{NaVO} shows the
optical conductivity of $\alpha$-NaV$_{2}$O$_{5}$.

There are two measurements shown: one with light polarized
parallel to the chains and one with light polarized perpendicular
to the chains. We can immediately read of $E_{CT}\approx 1$ eV.
Integrating the contribution to the optical conductivity of the
peaks we find that the spectral weight perpendicular to the chains
is roughly 4 times larger then the spectral weight parallel to the
chains, so $t_{\perp}\approx 4t_{\parallel}$. Inverting Eq.
(\ref{tsum}) we can calculate $t_{\perp}$ and we find
$t_{\perp}\approx0.3$ eV. The second strong peak at approximately
3.2 eV is a charge transfer peak from vanadium to oxygen. We will
come back to this example in the last section on spin
interactions.

\subsection{Generalized Drude formalism}
We have already encountered the Drude formula for the optical
conductivity of a metal (see the section on polaritons). Even
though this model is based on a classical gas of non-interacting
particles it describes amazingly well the optical properties of a
good metal. This is even more surprising if one realizes that in a
metal electrons reside in bands and that the transitions we are
making with photons are vertical due to the negligible photon
momentum. So from the band picture point of view, when we consider
a single band of electrons interacting with photons we should
expect a single delta function at the origin. The reason that this
is not what is observed is because we have neglected the other
interactions in the system. Electrons in solids interact with the
lattice vibrations, impurities and/or other collective modes. Due
to the electron-phonon interaction for instance we can have
processes where a photon creates an electron-hole pair in which
the electron "shakes off" a phonon. In this process the phonon can
carry away a much larger momentum then originally provided by the
photon. Due to this effect we can have phonon-assisted transitions
which give a width $1/\tau$ to the delta function. This width is
called the scattering rate. If the interactions are inelastic, as
in the interactions with impurities, this scattering rate is just
a constant. Otherwise, this scattering rate can depend on
frequency. However, if we define the scattering rate in Eq.
(\ref{drude}) to be frequency dependent,
$1/\tau\equiv1/\tau(\omega)$, the KK-relations force us to
introduce a frequency dependent effective mass as well. This is
what is done in the generalized Drude formalism \cite{Allen}. The
optical conductivity is written as,
\begin{equation}
\sigma \left( \omega  \right) = \frac{{ne^2 /m}}{{\tau ^{ - 1}
(\omega ){\rm  - i}\omega m*(\omega )/m}}.
\end{equation}
Having measured a conductivity spectrum we can invert these
equations to calculate 1/$\tau(\omega)$ or $m^{*}(\omega )/m$ via,
\begin{equation}\label{1/tau}
\tau ^{ - 1} (\omega ) \equiv {\mathop{\rm Re}\nolimits}
\frac{{ne^2 /m}}{{\sigma \left( \omega  \right)}} = \Sigma
''(\omega ),
\end{equation}
and
\begin{equation}\label{mstar/m}
\frac{{m^{*}(\omega )}}{m} \equiv {\mathop{\rm Im}\nolimits}
\frac{{ - ne^2 /m}}{{\omega \sigma \left( \omega  \right)}} = 1 +
\frac{{\Sigma '(\omega )}}{\omega }.
\end{equation}
In the last equality of these equations we have defined an optical
self-energy. Note that this quantity is \textit{not} equivalent to
the self-energy used in the context of Green's functions. We can
rewrite the optical conductivity in terms of $\Sigma(\omega)$ as,
\begin{equation}\label{optcondSigma}
\sigma \left( \omega  \right) = \frac{{ne^2 }}{m}\frac{i}{{\omega
+ \Sigma \left( \omega  \right)}}.
\end{equation}
In the case of impurity scattering $\Sigma(\omega)$ is simply
given by
\begin{equation}
\Sigma \left( \omega  \right) = i/\tau _0,
\end{equation}
\begin{figure}[tbh]
\includegraphics[width=8.5 cm]{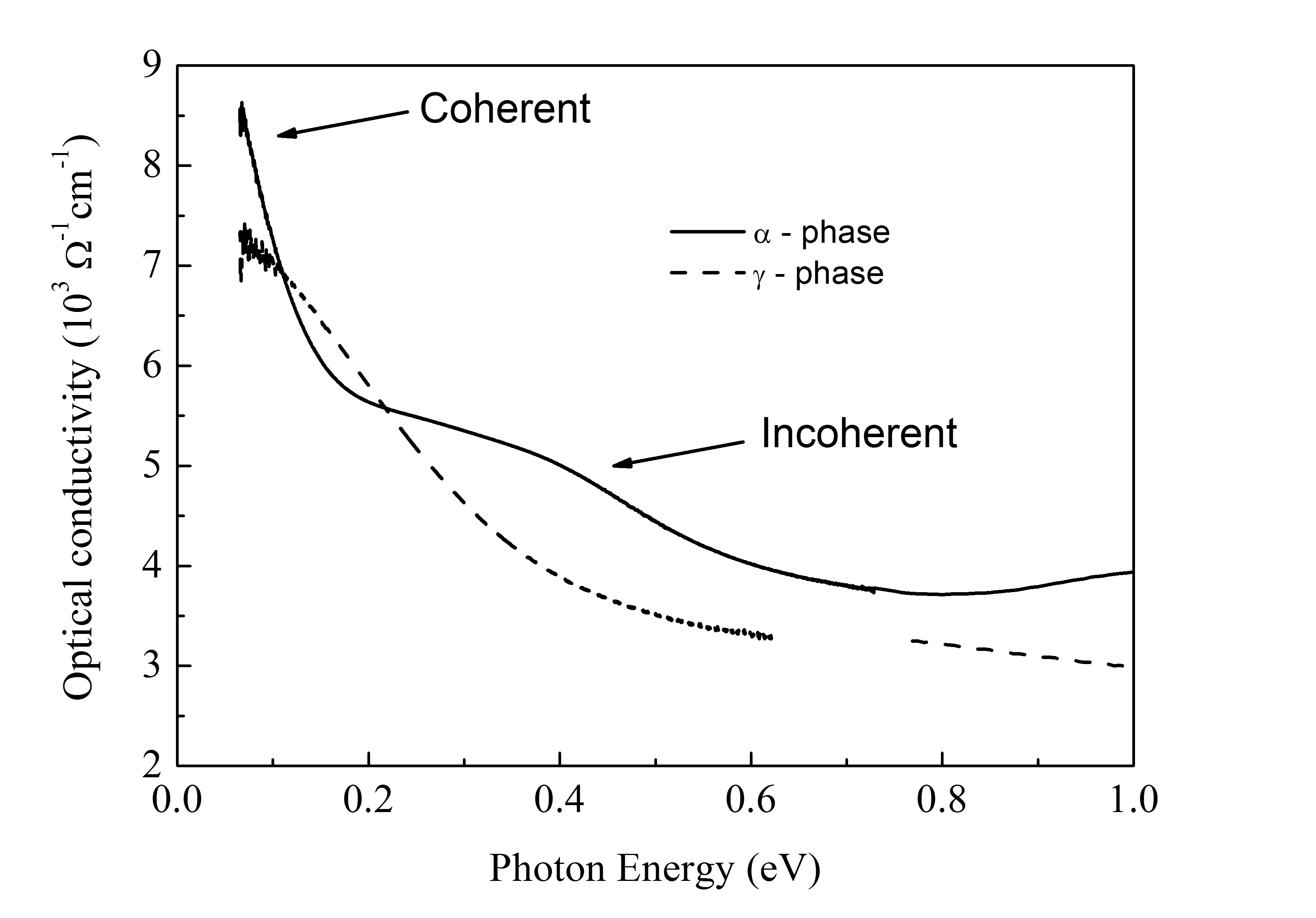}
\caption{\label{sigmaCe} Optical conductivity of Ce in the
$\alpha$ and $\gamma$ phases. Data taken from ref.
\citep{vandereb}.}
\end{figure}
so that $1/\tau(\omega)=1/\tau_{0}$ and $m*(\omega )/m=1$. We can
also capture the effect of the interaction of the electrons with
the static lattice potential in a self-energy,
\begin{equation}
\Sigma \left( \omega  \right) = \lambda \omega,
\end{equation}
which gives $\tau^{-1}(\omega )=0$ and
$m^{*}(\omega)/m=1+\lambda$. This is also called static mass
renormalization. Finally we consider dynamical mass
renormalization where the electrons couple to a spectrum of
bosons,
\begin{equation}\label{SEboson}
\Sigma \left( \omega  \right) = \frac{{\lambda \omega }}{{1 -
i\omega /T^* }},
\end{equation}
Here $\lambda$ is a coupling constant and $T^{*}$ is a
characteristic temperature scale related to the bosons. In this
case we find,
\begin{equation}
\tau ^{ - 1} (\omega ) = \lambda T^* \frac{{\omega ^2 }}{{T^{*2} +
\omega ^2 }},
\end{equation}
and
\begin{equation}
\frac{{m^{*}(\omega )}}{m} = 1 + \lambda \frac{{T^{*2} }}{{T^{*2}
+ \omega ^2 }}.
\end{equation}

As an example we will discuss the $\alpha$-phase to $\gamma$-phase
transition in pure Cerium. When Cerium is grown at elevated
temperatures it forms in the so called $\gamma$-phase. At low
temperatures a volume collapse occurs and the resulting phase is
called the $\alpha$-phase. This iso-structural transition is first
order. The reduction in volume can be as much as 20 to 30 $\%$. Ce
has 4 valence electrons and these can be distributed between
\begin{figure}[tbh]
\includegraphics[width=6 cm]{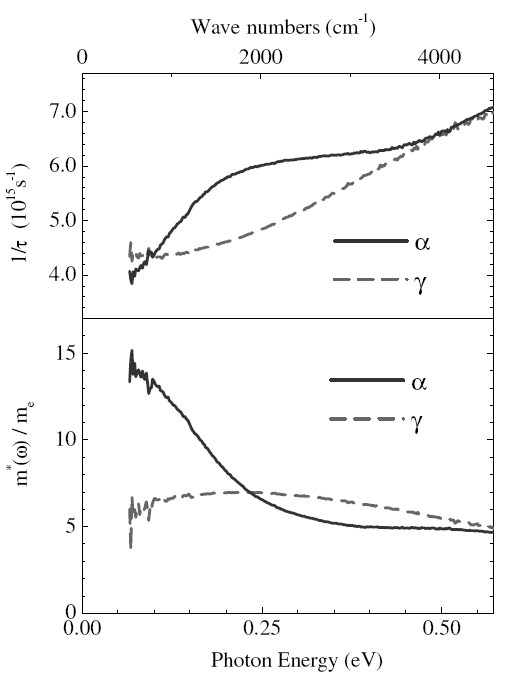}
\caption{\label{tauCe} Scattering rate and effective mass of Ce in
the $\alpha$ and $\gamma$ phases. Data taken from ref.
\citep{vandereb}.}
\end{figure}
localized $4f$ states and the $5d$ states that form the conduction
band. If occupied, the 4f states will act as paramagnetic
impurities. In the $\gamma$-phase the Kondo temperature
$T_{K}\approx$ 100 K whereas in the $\alpha$-phase $T_{K}\approx$
2000 K. This difference can be understood to be simply due to the
larger lattice spacing in the $\gamma$-phase: the hopping integral
$t$ is smaller and hence $T_{K}$ is smaller. Figure \ref{sigmaCe}
shows the optical conductivity of $\alpha$- and $\gamma$-Cerium.
These measurements were done by depositing Ce films on a substrate
at high and low temperature to form either the $\alpha$ or
$\gamma$ phase. We see that $\gamma$-Cerium is less metallic than
$\alpha$-Cerium. In the $\gamma$-phase there is only a weak Kondo
screening of the impurity magnetic moments and this gives rise to
spin flip scattering, which is the main source of scattering. In
the $\alpha$-phase the moments are screened and form renormalized
band electrons in a very narrow band. This suppresses the
scattering. The difference in scattering rates shows up in the
optical conductivity as a narrower Drude peak for the
$\alpha$-phase (see figure \ref{sigmaCe}). Figure \ref{tauCe}
displays the scattering rate and effective mass extracted from the
optical conductivity in figure \ref{sigmaCe} using Eq.
(\ref{1/tau}) and (\ref{mstar/m}). In the $\gamma$-phase
$1/\tau(\omega)$ extrapolates to a finite value due to local
moment or spin-flip scattering. We can rewrite the real part of
the optical conductivity in Eq. (\ref{optcondSigma}) with the
self-energy of Eq. (\ref{SEboson}) as follows,
\begin{equation}
\frac{{4\pi }} {{\omega _{_p }^{*2} }}\sigma \left( \omega \right)
= \frac{i} {\omega }\frac{1} {{1 + \frac{\lambda } {{1 - i\omega
/T^* }}}}.
\end{equation}
Note that we have defined a renormalized plasma frequency,
$\omega_{p}^{*}$, since the spectral weight is not conserved when
adding $\Sigma(\omega)$ to $\sigma(\omega)$. With some simple
algebra this can be rewritten as,
\begin{equation}
\frac{{4\pi }} {{\omega _{_p }^{*2} }}\sigma \left( \omega \right)
= \frac{i} {\omega } + \lambda \frac{{T^* }} {{\omega ^2  +
i\omega (1 + \lambda )T^* }}.
\end{equation}
It follows that the real part of this expression is then,
\begin{equation}
\frac{{4\pi }} {{\omega _{_p }^{*2} }}Re \sigma \left( \omega
\right) = \frac{\pi } {2}\delta (\omega ) + \lambda \frac{{T^* }}
{{\omega ^2  + (1 + \lambda )^2 T^{*2} }}.
\end{equation}
We see that the optical conductivity is split into two
contributions: a $\delta$-function which represents the
coherent part of the charge response and an incoherent
contribution. The $\delta$-function is usually broadened due to
other scattering channels present in the system. In this case
the $\delta$-function represents the contribution due to the
Kondo-peak whereas the incoherent contribution is due to the
side-bands. This splitting of the conductivity in a coherent
and incoherent contribution is nicely observed in the
$\alpha$-phase of Cerium as indicated in figure \ref{sigmaCe}.
\begin{figure}[thb]
\begin{minipage}{6cm}
\includegraphics[width=8cm]{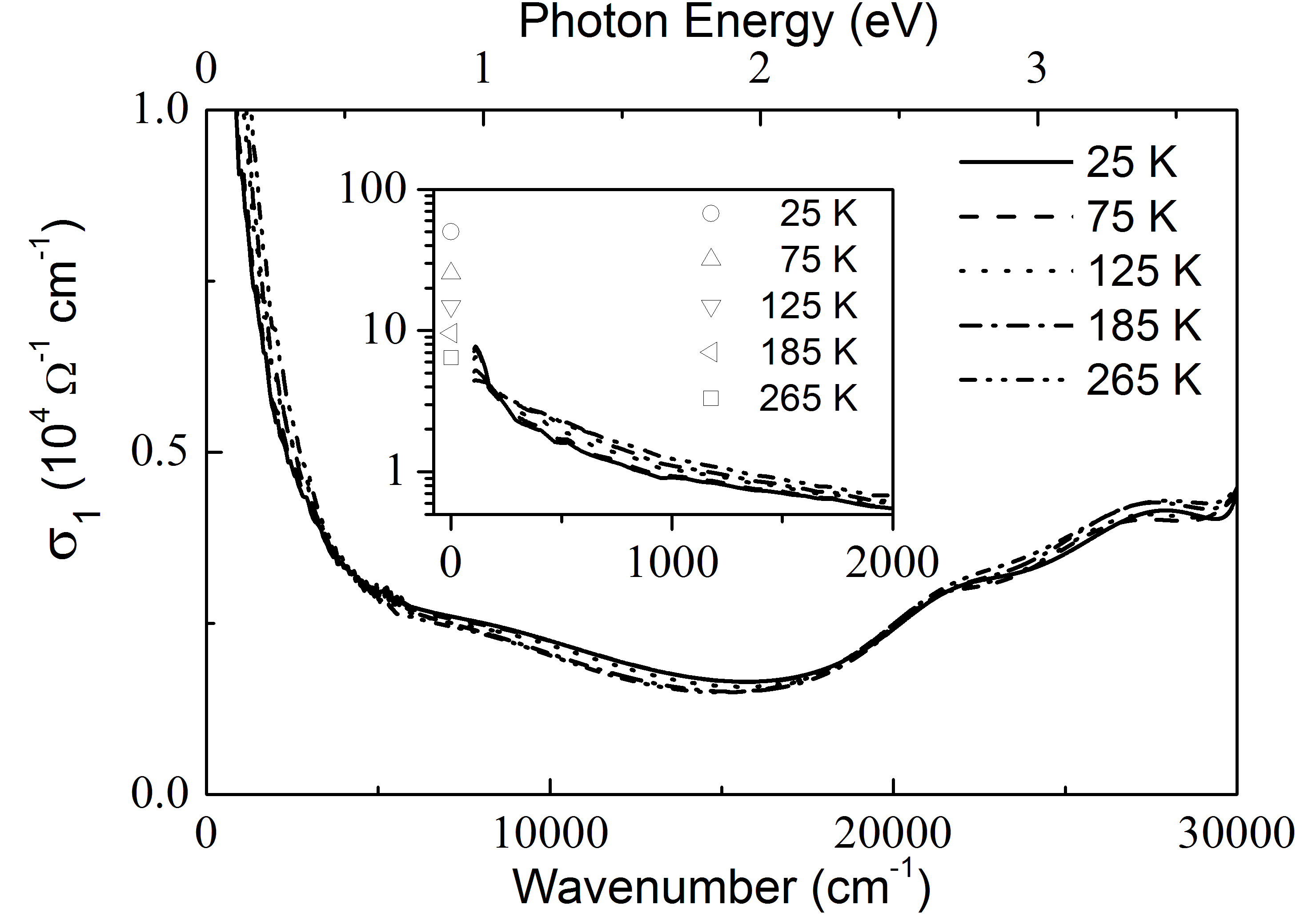}
\end{minipage}\hspace{2pc}%
\begin{minipage}{6cm}
\includegraphics[width=6cm]{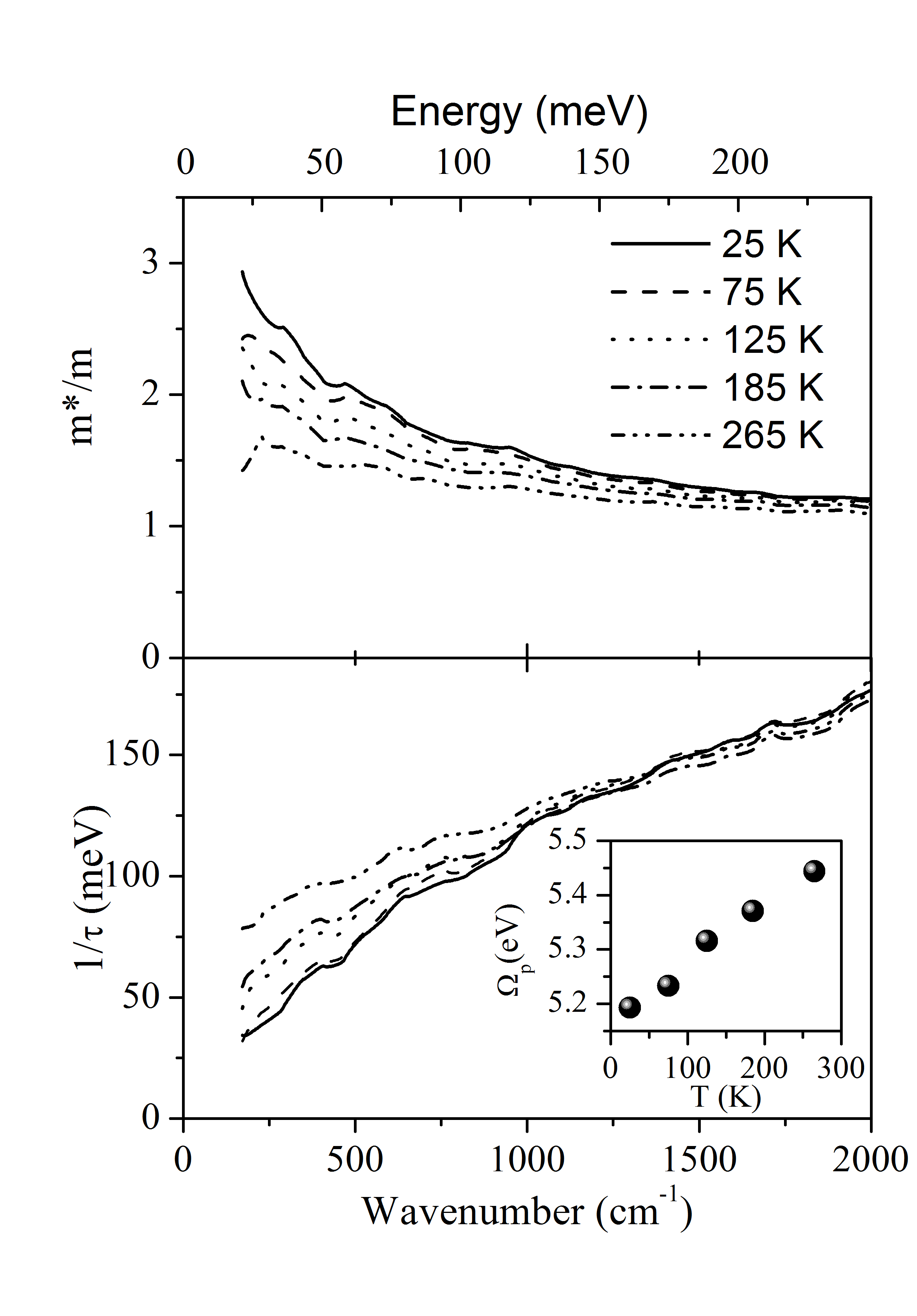}
\end{minipage}
\caption{\label{ZrB12sig}Left: Optical conductivity of ZrB$_{12}$
at selected temperatures. Right: $1/\tau$ and $m^{*}/m_{b}$ for
several temperatures. Data taken from ref. \citep{teyssier}.}
\end{figure}
This splitting of the optical conductivity
in coherent and incoherent contributions is much more general
however and is frequently observed in correlated electron
systems.
\section{Electron-phonon coupling}
\begin{figure}[tbh]
\includegraphics[width=8.5 cm]{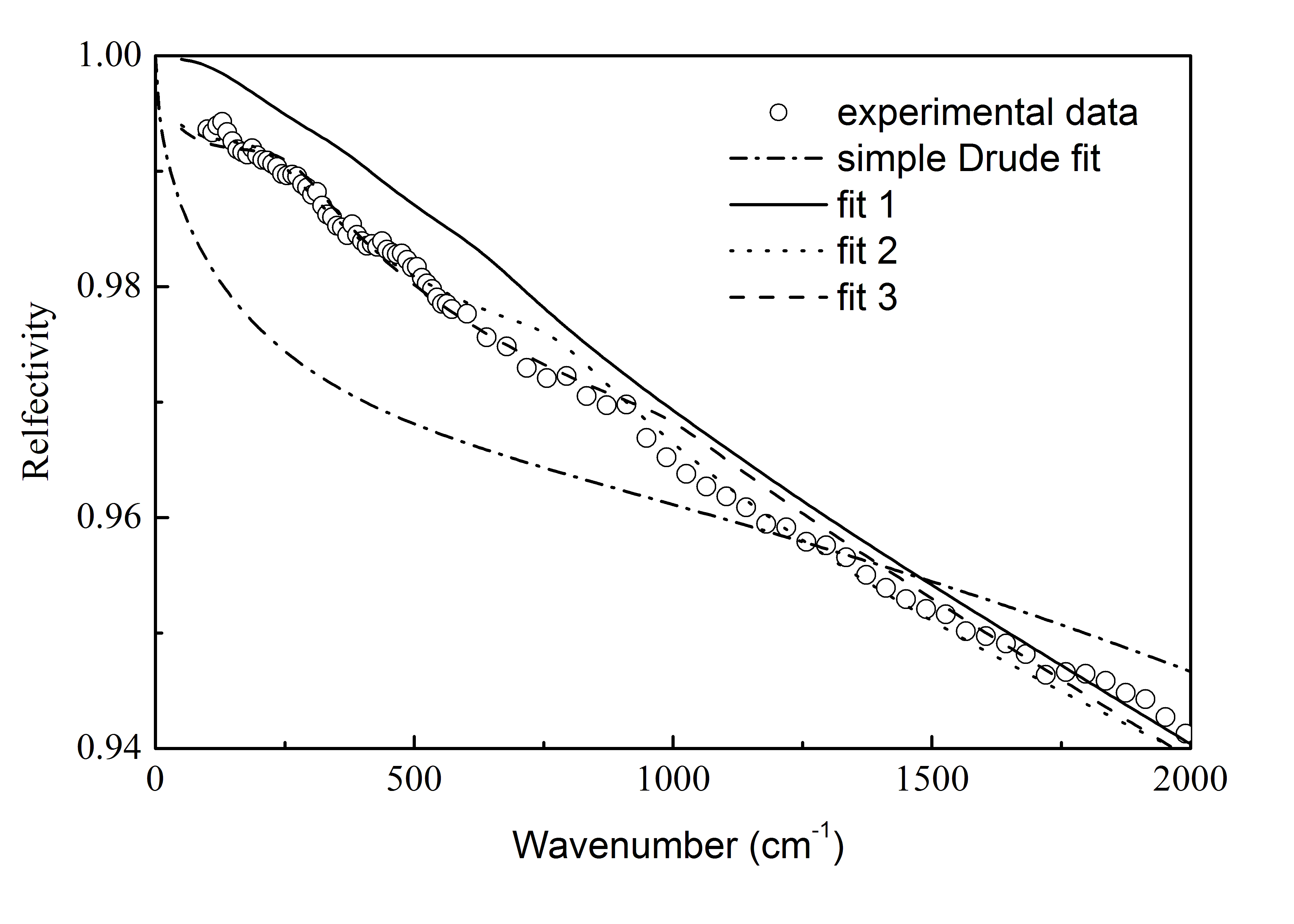}
\caption{\label{ZrB12ref}Reflectivity of ZrB$_{12}$ around 20 K
together with calculations as explained in the text. Figure
adapted from ref. \citep{teyssier}.}
\end{figure}

Electron-phonon coupling is most easily described in the framework
of Migdal-Eliashberg theory. The application of the theory to
optics can be found in the papers by Allen \cite{Allen}. In the
so-called Allen approximation the self-energy in Eq.
(\ref{optcondSigma}) is calculated using,
\begin{equation}\label{SEallenaprox}
\Sigma \left( \omega  \right) =  - 2i\int\limits_0^\infty {d\Omega
\alpha _{tr} ^2 F} (\Omega )K(\frac{\omega }{{2\pi
T}},\frac{\Omega }{{2\pi T}}).
\end{equation}
Here the kernel $K(\frac{\omega }{{2\pi T}},\frac{\Omega }{{2\pi
T}})$ is given by,
\begin{equation}
{\rm K(x}{\rm ,y)} = \frac{{\rm i}}{{\rm y}} + \frac{{y -
x}}{x}\left[ {\Psi (1 - ix + iy) - \Psi (1 + iy)} \right] +
\frac{{y + x}}{x}\left[ {\Psi (1 - ix - iy) + \Psi (1 - iy)}
\right].
\end{equation}
where the $\Psi(x)$ are Digamma functions. The function
$\alpha^{2}_{tr}F(\Omega)$ appearing in Eq. (\ref{SEallenaprox})
is the phonon spectral function. The label "tr" stands for
transport indicating that the spectral function is related to a
transport property. This function is different by a multiplicative
factor from the true $\alpha^{2}F(\Omega)$ as measured by for
instance tunnelling. The electron-phonon coupling strength is
easily calculated from $\alpha^{2}_{tr}F(\Omega)$ by integration,
\begin{equation}\label{lambdatr}
\lambda_{tr}=2\int_{0}^{\infty}\frac{\alpha^{2}_{tr}F(\Omega)}{\Omega}d\Omega.
\end{equation}
This approach was first applied by Timusk and Farnworth in a
comparison of tunnelling and optical measurements on the
superconducting properties of Pb \cite{Timusk}. As an example
we discuss the application of this formalism to the optical
properties of ZrB$_{12}$ \cite{teyssier}. Figure \ref{ZrB12sig}
shows the optical conductivity of ZrB$_{12}$. The spectrum
consists of what appears to be a Drude peak and some interband
contributions. Also shown are the calculated $1/\tau(\omega)$
and $m^{*}(\omega)/m_{b}$. The temperature dependence of
$1/\tau(\omega)$ is what is usually observed for a narrowing of
the Drude peak with decreasing temperature whereas the strong
frequency dependence is suggestive of electron-phonon
interaction. Using the McMillan formula (\ref{lambdatr}) the
coupling strength was estimated to be $\lambda_{tr}\approx0.7$.
In figure \ref{ZrB12ref} the reflectivity of ZrB$_{12}$
together with calculations based on Eq. (\ref{optcondSigma})
and (\ref{SEallenaprox}) is shown. It is clear that a simple
Drude form is not capable of describing the observed
reflectivity. The first fit (fit 1) is a fit where the
$\alpha^{2}_{tr}F(\Omega)$ that was used as input was derived
from specific heat measurements \cite{Lortz}. Although it gives
an improvement over the standard Drude fit there is still some
discrepancy between the data and the fit. To make further
improvements $\alpha^{2}_{tr}F(\Omega)$ was modelled using a
sum of $\delta$-functions. The results of this modelling are
indicated as fit 2 and fit 3. Using Eq. (\ref{lambdatr}) we
find coupling strengths $\lambda_{tr}\approx$ 1 - 1.3. Another
method to roughly estimate $\alpha^{2}_{tr}F(\Omega)$ is due to
Marsiglio \cite{Marsiglio1,Marsiglio2}. It states that a rough
estimate of the shape of $\alpha^{2}_{tr}F(\Omega)$ can be
found by simply differentiating the optical data,
\begin{equation}
\alpha^{2}_{tr}F(\Omega)=\frac{1}{2\pi}\frac{\Omega_{p}^{2}}{4\pi}\frac{d^{2}}{d\omega^{2}}Re(\frac{1}{\sigma(\omega)}),
\end{equation}
where $\Omega_{p}$ is the plasma frequency. The obvious problem
with this method is that it requires the double derivative of the
data. Because of the inevitable noise in the data usually some
form of smoothing is required. Applied to ZrB$_{12}$ the extracted
$\alpha^{2}_{tr}F(\Omega)$ shows peaks at the same positions as
the ones extracted before and a coupling strength
$\lambda_{tr}\approx1.1$. These results indicate a medium to
strong electron-phonon coupling for ZrB$_{12}$.

\section{Polarons}
\begin{figure}[thb]
\includegraphics[width=6 cm]{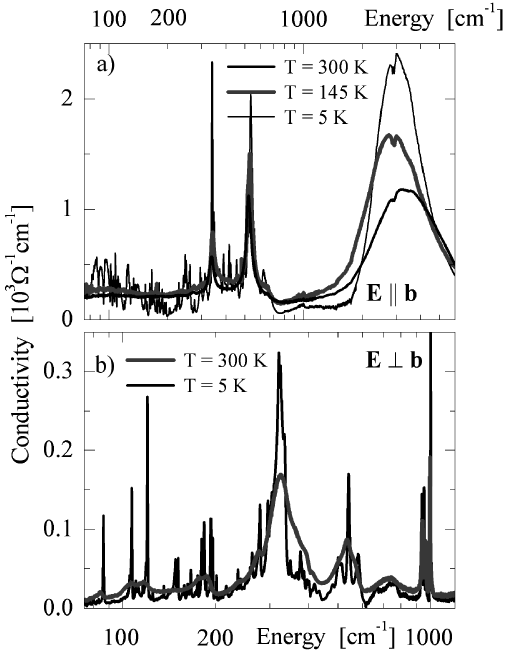}
\caption{\label{NaVOsigma}Optical conductivity of
$\beta$-NaV$_{6}$O$_{15}$ for light polarized along and
perpendicular to the b-axis. Figure adapted from \cite{presura}}
\end{figure}

There exist many definitions of what is a polaron. Electrons
coupled to a phonon have been called polaron as have free
electrons moving around in an insulator. Here we will consider the
Landau-Pekar approximation for a polaron \cite{landau,landau2}.
The idea is that when an electron moves about the crystal it
polarizes the surrounding lattice and this in turn leads to an
attractive potential for the electron. If the interaction between
electron and lattice is sufficiently strong this potential is
capable of trapping the electron and it becomes more or less
localized. The new object, electron plus polarization cloud is
called polaron. This self-trapping of electrons can occur in a
number of different situations and different names are used. For
instance, one talks about small polarons in models where only
short range interactions are considered, because this typically
leads to polaron formation with polarons occupying a single
lattice site. From the Landau-Pekar formalism we can get some
feeling of when polarons form and what their
\begin{figure}[thb]
\includegraphics[width=6 cm]{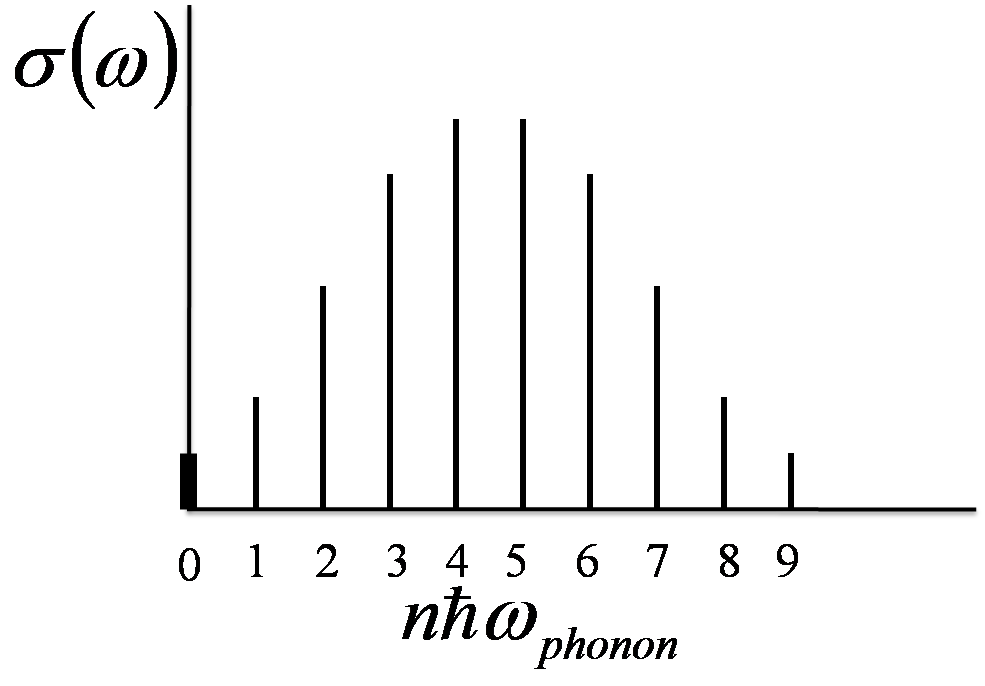}
\caption{\label{polcond} Schematic of the optical conductivity of
electrons interacting with a single Einstein mode.}
\end{figure}
properties will be. First of all, the coupling constant $\alpha$
is given by,
\begin{equation}
\alpha^{2}=\frac{Ry}{\hbar\omega_{0}\tilde{\varepsilon}_{\infty}^{2}}\frac{m_{b}}{m_{e}},
\end{equation}
where $Ry$ stands for the unit Rydberg (1 Rydberg =
$m_{e}e^{4}/2\hbar^{2}$ = 13.6 eV), $\omega_{0}$ is the oscillator
frequency of the (Einstein) phonon mode involved $m_{b}$ and
$m_{e}$ are the band and free electron mass respectively and
$\tilde{\varepsilon}$ is given by,
\begin{equation}
\frac{1}{\tilde{\varepsilon}}=\frac{1}{\varepsilon_{\infty,IR}}-\frac{1}{\varepsilon(0)}.
\end{equation}
For strong coupling (small polarons) the polaron mass is expressed
in terms of the coupling constant as,
\begin{equation}
m_{pol}=m_{b}(1+0.02\alpha^{4}),
\end{equation}
\begin{figure}[bht]
\includegraphics[width=6 cm]{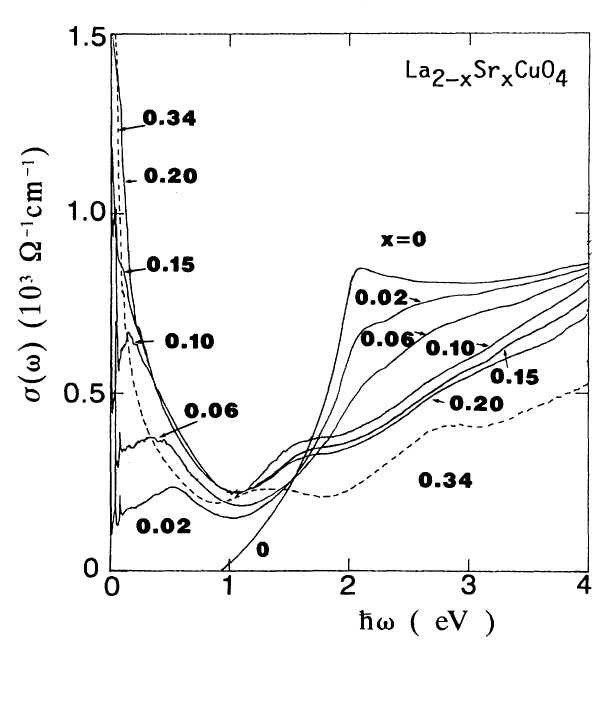}
\caption{\label{LSCOdopdep} Doping dependence of the room
temperature optical conductivity of
La$_{2-x}$Sr$_{x}$CuO$_{4}$. Figure adapted from Uchida
\textit{et al.}, ref. \citep{uchida}.}
\end{figure}
with the polaron binding energy given by,
\begin{equation}
E_{pol}=0.1\frac{Ry}{\tilde{\varepsilon}_{\infty}^{2}}\frac{m_{b}}{m_{e}}.
\end{equation}
The polaron binding energy typically is of the order of a few
100 meV. The polaron mass is typically of the order of 50-100
times the electron mass. The effect of polaron formation on the
optical conductivity can be described by assuming a gas of
non-interacting polarons (i.e. low polaron density). This
results in a spectrum that can be described by a Drude peak and
a so-called Holstein side-band. If we assume that the electrons
interact with a single Einstein mode the spectrum will look as
in figure \ref{polcond}. The spectrum consists of a
zero-phonon, coherent part (n = 0) with a spectral weight
1/(1+0.02$\alpha^{4}$) followed by a series of peaks that
describe the incoherent movement of polarons assisted by
n=1,2,3.. phonons. In real solids the peaks are smeared out due
to the fact that phonons form bands. The real part of the
optical conductivity can thus be described as,
\begin{equation}
{\mathop{\rm Re}\nolimits} \frac{{4\pi }}{{\omega _{_p }^{*2}
}}\sigma \left( \omega  \right) = \frac{1}{{1 + 0.02\alpha ^4
}}\frac{{\pi \delta (\omega )}}{2} + \frac{{0.02\alpha ^4 }}{{1 +
0.02\alpha ^4 }}E_{pol}^{ - 1} \exp \left\{ { - \left(
{\frac{{\omega ^2  - E_{pol}^2 }}{{cE_{pol}^2 }}} \right)^2 }
\right\}.
\end{equation}
The first term in this expression describes the coherent part
of the spectrum, which in real solids will also be smeared out
to finite frequency by other forms of scattering, and an
incoherent term given by the second term which is called the
Holstein band. The shape of the side-band can be qualitatively
understood by imagining how a polaron has to move through the
lattice. In order to move from one site to another the lattice
deformation around the original site has to relax and be
adjusted on the new site. This relaxation process results in
the multi-phonon side-bands of the Drude peak.
\begin{figure}[thb]
\begin{minipage}{6 cm}
\includegraphics[width=4cm]{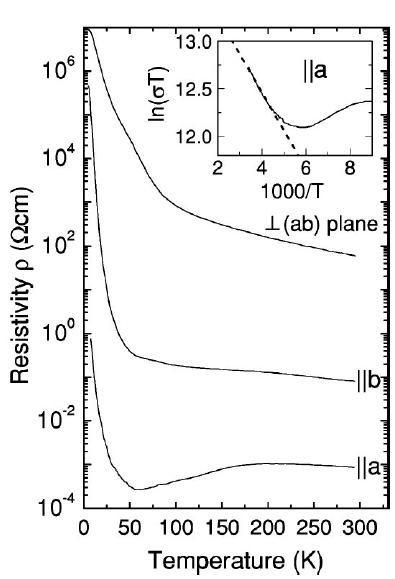}
\end{minipage}\hspace{2pc}%
\begin{minipage}{6cm}
\includegraphics[width=8 cm]{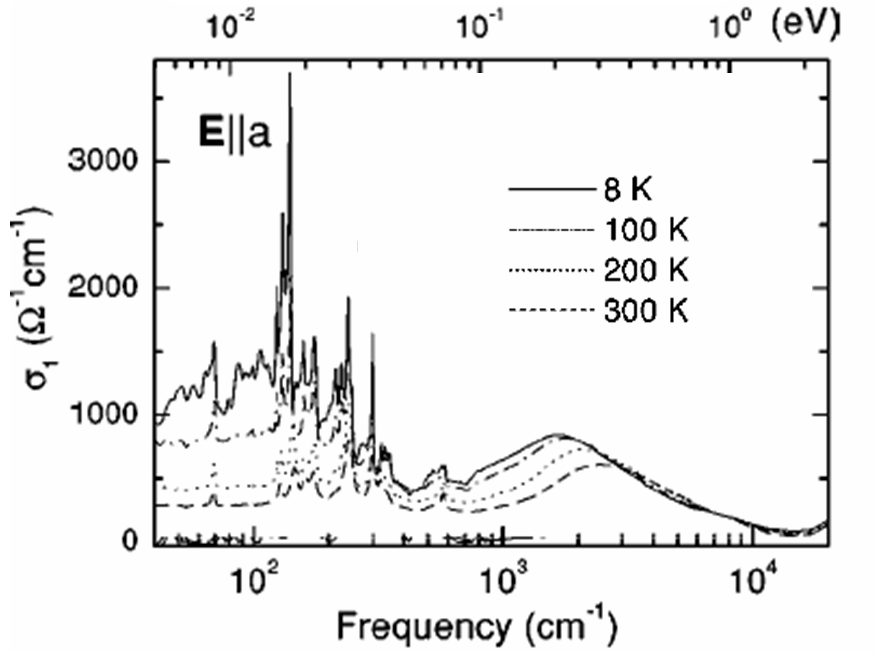}
\end{minipage}
\caption{\label{LaTiO} Left panel: temperature dependent
resistivity of LaTiO$_{3.41}$. Right panel: Optical conductivity
for selected temperatures. Figure adapted from ref.
\citep{kuntscher}.}
\end{figure}
The observation of the Holstein side-band is somewhat
complicated because it is not possible to distinguish between
normal interband transitions and the effects due to polaron
formation. There have been some claims that a band observed in
the mid infrared region ($\approx$ 100 - 500 meV) of the
spectrum of high-T$_{c}$ superconductors is due to polaron
formation but many other interpretations exist. Figure
\ref{LSCOdopdep} shows the doping dependence of
La$_{2-x}$Sr$_{x}$CuO$_{4}$. The peak that occurs around 0.5 eV
for the 0.02 doped sample has been interpreted as the Holstein
side-band. Another example where polarons could play a role is
in LaTiO$_{3.41}$ \cite{kuntscher}. In this material the
resistivity (figure \ref{LaTiO}) shows a quasi one dimensional
behavior with an upturn of the resistivity at lower
temperatures. This could be due to polaron formation but it has
also been interpreted as due to a charge density wave. The
optical conductivity at low temperatures shows that a large
part of the spectral weight is contained in a side-band around
300 meV (see figure \ref{LaTiO}). If this peak would be due to
polarons we expect that when we warm up the system to higher
temperatures its spectral weight should be diminished. This is
because the increased temperature unbinds electrons from their
self-trapping potential and therefore shifts spectral weight
from the Holstein band to the Drude peak. This is also what is
observed and at the same time explains the decrease of
resistivity with increasing temperature.
\begin{figure}[bth]
\includegraphics[width=6 cm]{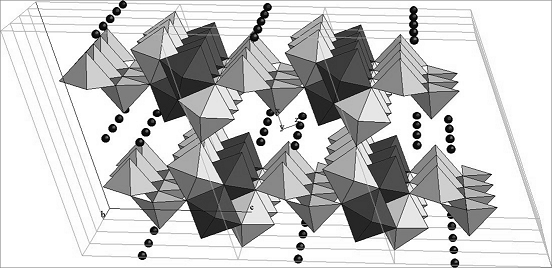}
\caption{\label{NaVOstruct}Crystal structure of
NaV$_{6}$O$_{15}$.}
\end{figure}
The last example we will discuss is NaV$_{6}$O$_{15}$. The
structure of this compound is build up out off octahedra and
tetrahedra of vanadium and oxygen atoms where the tetrahedra
form quasi 1-dimensional zig-zag chains (see figure
\ref{NaVOstruct}). There are 3 different types of vanadium
sites in this structure: 2 of them are ionic with a charge 5+
on the vanadium which has then a $3d^{0}$ configuration. The
third site has half an electron more leading to a charge of
4.5+ on the vanadium atom in a $3d^{1/2}$ configuration.
Because of this we expect a quarter filled band and metallic
behavior. Figure \ref{NaVOsigma} shows the optical conductivity
of $\beta$- NaV$_{6}$O$_{15}$. The chains are along the
direction labelled \textit{b}. At energies around 3000
cm$^{-1}$ we observe a broad peak for light polarized along the
$b$-direction which could be due to polarons although these
transitions also correspond well with the energies predicted by
the Hubbard model for d-d transitions. If we compare the
conductivity with polarization parallel and perpendicular to
the b-axis, we see that the conductivity perpendicular to the
b-axis is insulating whereas the one along the b-axis is
conducting. This conducting behavior is due to the quarter
filled bands.
Are polarons playing an important role in the above examples ? It
is nearly impossible to answer this question experimentally due to
the above mentioned difficulty in separating polaronic behavior
from normal interband transitions. Moreover, in most cases where
polarons are invoked, other theories are also able to reproduce
the experimental results. To close this section we briefly discuss
what happens if the density of polarons becomes larger. Imagine
what happens if we increase the density of polarons such that we
are getting close to a system with one polaron on each site. In
that case the original lattice will almost be completely deformed
and one can wonder wether the electrons are still capable to
self-trap. It seems reasonable that in this limit the polaron
picture no longer applies. Another possibility is the formation of
bipolarons. Since the deformation energy of the lattice is
proportional to the electron charge $E_{pol}\propto -1/2Cq^{2}$,
the binding energy of two polarons is $\propto -Cq^{2}$. The
binding energy of a bipolaron (two electrons trapped by the same
polarization cloud) is twice as large however $E_{bipol}\propto
-1/2C(2q)^{2}$. This binding energy is usually not enough to
overcome the Coulomb repulsion between the electrons.

\section{Spin interactions}\label{spininteraction}
\begin{figure}[thb]
\includegraphics[width=10 cm]{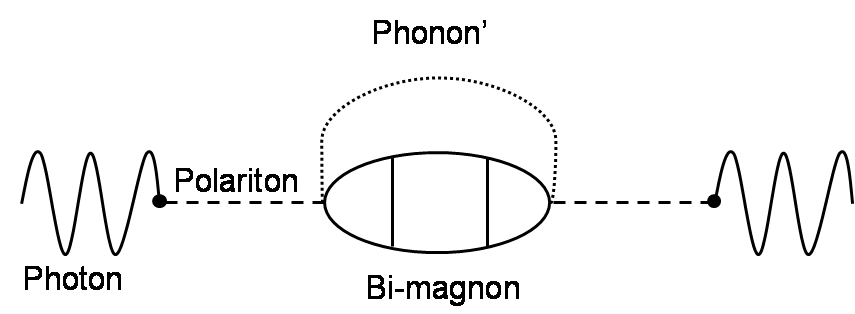}
\caption{\label{spinflip}Interaction diagram for the indirect
interaction of light with spin degrees of freedom.}
\end{figure}

As mentioned in the previous section the signatures for the
presence of polarons can often be interpreted with different
ideas. Most often these models are based on coupling to magnetic
interactions. Consider for example the spectrum of the parent
(undoped) compound YBa$_{2}$Cu$_{3}$O$_{6}$ which is a Mott
insulator (see bottom panel of figure \ref{reftrans}). Below 100
meV we see a series of peaks which are due to phonons. But what
about the structure between 100 meV and 1 eV ? One of the
difficulties in explaining this structure is that light does not
directly couple to spin degrees of freedom. It is however possible
to indirectly make spin flips with photons (see figure
\ref{spinflip}).

For this process to occur we have to include phonon-magnon
interaction. When a photon enters the material it gets dressed
with phonons forming a polariton which is then coupled to the spin
degrees by the phonon-magnon interaction. This leads to the
possibility of so-called phonon assisted absorption of spin-flip
excitations \cite{Lorenzana}. We see from figure \ref{spinflip}
that the polariton creates a bi-magnon. This is because the
intermediate state has to have spin S = 0. The dashed square
represents all magnon-magnon interactions. The coupling constant
for this process was first calculated by Lorenzana and Sawatzky
and is
\begin{equation}
J_{ph-mag}=\frac{1}{2J}<\frac{d^{2}J}{du^{2}}><u^{2}>,
\end{equation}
\begin{figure}[thb]
\includegraphics[width=8.5 cm]{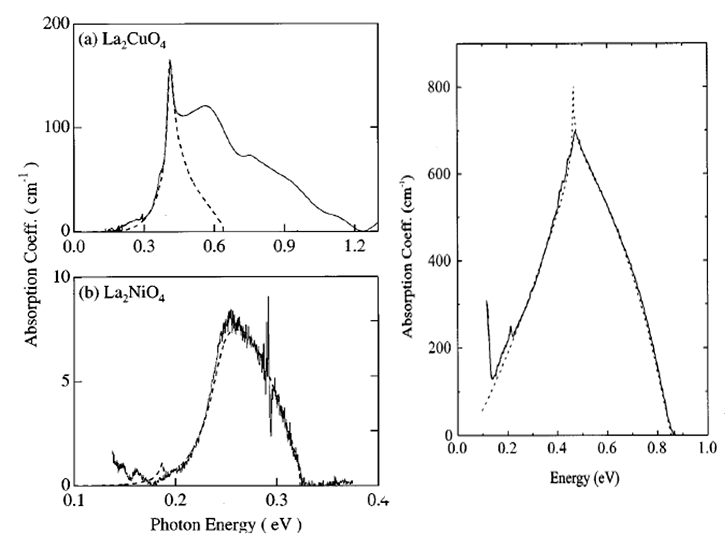}
\caption{\label{spinsigma}Optical conductivity of (a):
La$_{2}$CuO$_{4}$, (b): La$_{2}$NiO$_{4}$ and (c):
Sr$_{2}$CuO$_{4}$. Dashed lines are fits using the
Lorenzana-Sawatzky model.}
\end{figure}
where $J$ is the superexchange constant and $u$ is the atomic
displacement vector. In the process momentum and energy have to be
conserved and this leads to
\begin{equation}
k_{\text{magnon 1}}  + k_{\text{magnon 2}}  + k_{\text{phonon}}  =
k_{\text{photon}} \approx 0.
\end{equation}
and
\begin{equation} \omega _{\text{magnon 1}}  + \omega _{\text{magnon 2}}  +
\omega _{\text{phonon}}  = \omega _{\text{photon}}.
\end{equation}
for the process in figure \ref{spinflip}. This gives constraints
on the possible absorptions. In figure \ref{spinsigma} some
examples are shown of materials in which we believe this process
to play a role. One of the compounds where the predicted optical
conductivity fits the spectrum very well is in the case of
Sr$_{2}$CuO$_{3}$. To make the fit the magnon dispersion as
measured with neutron scattering was used. The reason that this
theory works so well for Sr$_{2}$CuO$_{3}$ is that the conduction
is nearly one dimensional. This gives a good starting point
because the magnon spectrum is completely understood. On the
contrary, the theory is not completely capable of predicting the
spectrum of La$_{2}$CuO$_{4}$. Most likely the peaks around 0.6
and 0.75 eV are due to 4 and 6 magnon absorption. In the case of
YBa$_{2}$Cu$_{3}$O$_{6}$ the situation gets even more complicated
due to the presence of two layers per unit cell. Because of the
doubling of the unit cell, there are now acoustic and optical
magnon branches just as what would happen in the case of phonons.
The effect of this on the optical conductivity was first discussed
by Grueninger \textit{et al.} \cite{Grueninger,Grueninger2}.

\begin{figure}[bht]
\includegraphics[width=8.5 cm]{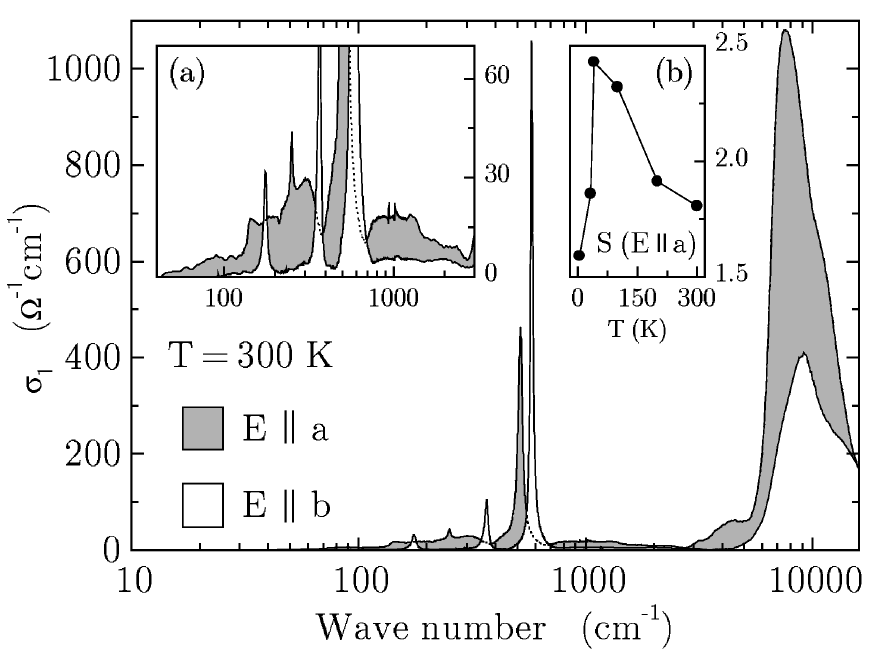}
\caption{\label{alphaNaVO}Optical conductivity of $\alpha$-
NaV$_{2}$O$_{5}$. Inset (a) shows the low energy continuum
attributed to charged bi-magnon excitations and inset (b) shows
the temperature dependence of the spectral weight of this
continuum. Figure adapted from ref. \citep{damascelli3}.}
\end{figure}

Another example of probing of spin excitations occurs in
NaV$_{2}$O$_{5}$. As already discussed in the previous section
this compound has quasi one dimensional chains as shown in figure
\ref{NaVOstruct}. These chains can be seen to form a so-called
ladder structure, with the ladders parallel to the $b$ direction.
\begin{figure}[thb]
\includegraphics[width=3 cm]{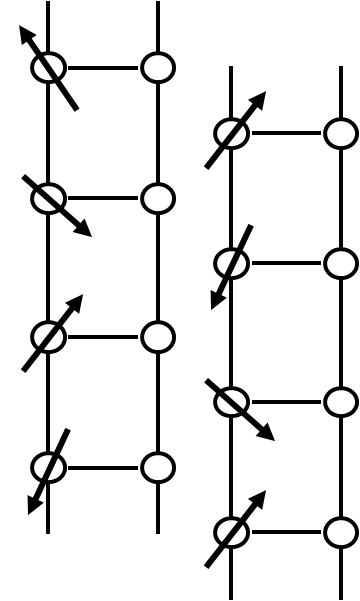}
\caption{\label{NaVOlad}Schematic of the ladder structure of
$\alpha$- NaV$_{2}$O$_{5}$. Arrows indicate the position of the
lectrons and their spin orientation.}
\end{figure}
Each adjacent ladder is shifted with respect to the previous
such that the rungs of one ladder fall in between those of the
next (figure \ref{NaVOlad}). The vanadium atoms that form the
ladders have an average charge of +4.5. It has been claimed
\cite{carpy} that the charge distribution is inhomogeneous with
most of the charge on one side of the ladder as indicated in
figure \ref{NaVOlad}.
\begin{figure}[thb]
\begin{minipage}{7cm}
\includegraphics[width=7cm]{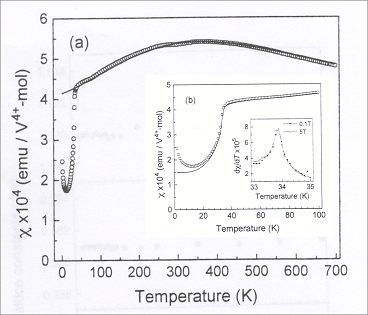}
\caption{\label{NaVOmag}Magnetization of $\alpha$-
NaV$_{2}$O$_{5}$. Figure adapted from ref. \citep{Isobe}.}
\end{minipage}\hspace{2pc}%
\begin{minipage}{7cm}
\includegraphics[width=5cm]{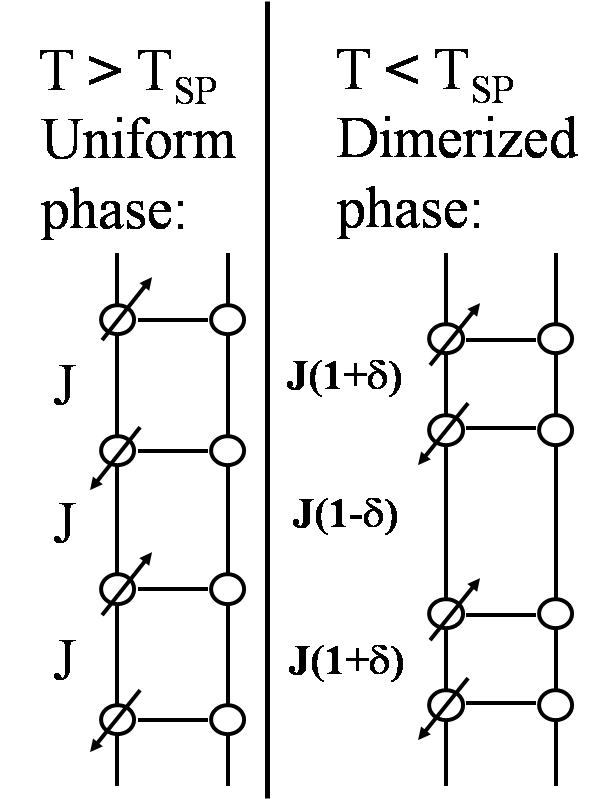}
\caption{\label{NaVOphase}Schematic representation of the low
and high temperature phase of $\alpha$- NaV$_{2}$O$_{5}$.}
\end{minipage}
\end{figure}
The temperature dependence of the magnetic susceptibility can
be modelled pretty well using a Bonner-Fischer model for a
spin-1/2 Heisenberg chain \cite{Isobe} for temperatures higher
then 34 K (see figure \ref{NaVOmag}). Below 34 K, X-ray
analysis shows a doubling of the a- and b- axes and a
quadrupling of the c-axis. It indicates that the new unit cell
consists of 64 vanadium atoms and 32 valence electrons. At the
same temperature the susceptibility shows an abrupt drop. An
explanation for this transition is in terms of a spin-Peierls
transition. In the high temperature phase (T $>$ T$_{SP}$) the
left side of the ladder has a uniform spin distribution, as
indicated in the left panel of figure \ref{NaVOphase}, which is
reasonably well described with an anti-ferromagnetic (AF) S =
1/2 Heisenberg spin chain with uniform exchange coupling $J$.
For T $<$ T$_{SP}$ the system dimerises due to a deformation of
the lattice, leading to an alternation of exchange couplings
(see right panel figure \ref{NaVOphase}). Here we focus on the
high temperature phase. If the charge inhomogeneity is present
it would lead to a breaking of the inversion symmetry which in
turn leads to a non-zero optical matrix element for two magnon
absorption \cite{damascelli3}. The idea is similar to the
Lorenzana-Sawatzky model discussed above. In the latter case
the phonon effectively lowers the symmetry making the process
optically allowed. The optical conductivity of $\alpha$-
NaV$_{2}$O$_{5}$ is shown in figure \ref{alphaNaVO}. We can
model $\alpha$- NaV$_{2}$O$_{5}$ with independent ladders where
the hopping probability along a rung ($t_{\perp}$) is much
larger than that along the ladder ($t_{\parallel}$).
Furthermore we assume a large on-site repulsion U. We can then
model a ladder by independent rungs. Assuming a quarter filled
ladder (one electron per rung) we have a simple two level
problem leading to bonding and anti-bonding levels (see also
the discussion in the section on applications of sum rules). If
we also include a potential energy difference $\Delta$ between
the sites the wavefunctions become asymmetric with higher
probability on the low potential site and one can show that
this again leads to bonding and anti-bonding solutions which
are split by an energy \cite{damascelli3},
\begin{equation}\label{Ect}
E_{CT}=\sqrt{\Delta^{2}+4t^{2}_{\perp}}.
\end{equation}
Transitions from the bonding to anti-bonding band are optically
active and involve charge transfer (CT) from the left side of the
ladder to the right. The large peak seen in figure \ref{alphaNaVO}
around 1 eV is due to these transitions. The energy position of
the peak indirectly gives evidence for the charge inhomogeneity:
band structure calculations and exact diagonalization of finite
clusters give $t_{\perp}\approx$ 0.35 eV, which would put the
charge transfer peak around 0.7 eV. The observed value of 1 eV
thus indicates $\Delta\neq$ 0. The spectral weight of the peak
allows us to make an estimate for $\Delta$. One can show that,
\begin{equation}
\int_{peak}\sigma_{1}(\omega)=\pi
e^{2}Nd^{2}_{\perp}t^{2}_{\perp}\hbar^{-2}E_{CT}^{-1}.
\end{equation}
Using Eq. (\ref{Ect}) we find $t_{\perp}\approx$ 0.3 eV and
$\Delta\approx$ 0.8 eV. Besides the large CT peak seen in
figure \ref{alphaNaVO} there is also a broad continuum in the
infrared region of the spectrum for $E\parallel a$ (see inset).
This part of the spectrum can be understood if we include the
coupling between rungs of the ladder. For parallel spins on
different rungs this coupling would have no effect since the
Pauli principle would forbid hopping between the sites. For an
anti-parallel spin configuration the system can gain some
kinetic energy from virtual hopping of an electron from one
rung to the next, putting two electrons on one rung. For very
large U this electron would occupy the righthand side of the
rung. Starting from an anti-parallel configuration a spin-flip
transition on one rung thus leads to a net dipole displacement
which leads to optical activity of this transition. We note
that because of spin conservation rules we have to make two
spin-flips. These excitations have been dubbed charged
bi-magnon excitations.

\end{document}